\documentclass[prb,twocolumn,superscriptaddress,float,aps]{revtex4-2}
\usepackage{bm,xcolor,amsmath,amssymb,mathrsfs,latexsym,graphicx,psfrag,tabularx,booktabs,amsthm,diagbox,amssymb,bbold}

\usepackage[hidelinks,colorlinks,linkcolor=blue,
citecolor=blue,urlcolor=blue]{hyperref}

\newcommand{\bs}[1]{\boldsymbol{#1}}

\newcommand{\beq}{\begin{equation}}
\newcommand{\eneq}{\end{equation}}

\def\kk{\mathbf{k}}

\def\rr{\mathbf{r}}
\def\GG{\mathbf{G}}

\def\bb{\mathbf{b}}

\def\GG{\mathbf{G}}

\def\bb{\mathbf{b}}

\def\gs{{\bs g}}

\def\ie{{\it i.e.},\ }
\def\eg{{\it e.g.}\ }

\usepackage{etoolbox}

\usepackage{dsfont}

\usepackage{cleveref}
\crefname{appendix}{App.\,}{Apps.\,}
\crefname{equation}{Eq.\,}{Eqs.\,}
\crefname{figure}{Fig.\,}{Figs.\,}
\crefname{table}{Tab.\,}{Tabs.\,}
\crefname{section}{Sec.\,}{Secs.\,}
\creflabelformat{appendix}{[#2#1#3]}

\begin{document}

\title{Symmetry-enforced Moir\'e Topology}

\author{Yunzhe Liu}
\thanks{These authors contributed equally.}
\affiliation{Department of Physics, The Pennsylvania State University, University Park, Pennsylvania 16802, USA}
\author{Ethan Angerhofer}
\thanks{These authors contributed equally.}
\affiliation{Department of Physics, University of Florida, Gainesville, FL, USA}
\author{Kaijie Yang}
\affiliation{Department of Materials Science and Engineering, University of Washington, Seattle, Washington 98195, USA}
\author{Chao-Xing Liu}
\email{cxl56@psu.edu}
\affiliation{Department of Physics, The Pennsylvania State University, University Park,  Pennsylvania 16802, USA}
\author{Jiabin Yu}
\email{yujiabin@ufl.edu}
\affiliation{Department of Physics, University of Florida, Gainesville, FL, USA}
\affiliation{Quantum Theory Project, University of Florida, Gainesville, FL, USA}

\begin{abstract} 
Topological flat bands in two-dimensional (2D) moir\'e materials have emerged as promising platforms for exploring the interplay between topology and correlation effects.
However, realistic calculations of moir\'e band topology using density functional theory (DFT) are computationally inefficient due to the large number of atoms in a single moir\'e unit cell.
In this work, we propose a systematic scheme to predict the topology of moir\'e bands from atomic symmetry data and moir\'e symmetry group, both of which can be efficiently extracted from DFT.
Specifically, for $\Gamma$-valley electron gases, we find that certain combinations of atomic symmetry data and moir\'e symmetry groups can enforce nontrivial band topology in the low-energy moir\'e bands, as long as the moir\'e band gap is smaller than the atomic band splitting at the moir\'e Brillouin zone boundary.
This symmetry-enforced nontrivial moiré topology, including both topological insulators and topological semimetals, is robust against various material-specific details such as the precise form and strength of the moiré potential or the exact twist angle.
By exhaustively scanning all 2D atomic symmetry data and moiré symmetry groups, we identify 197 combinations that can yield symmetry-enforced nontrivial moir\'e topology, and we verify one such combination using a moir\'e model with cubic Rashba spin-orbit coupling. By screening the existing 2D material database, we currently identify 92 monolayer materials with (i) the low-energy bands near $\Gamma$ and (ii)  the atomic symmetry data that belong to those combinations.
Our approach is generalizable to other valleys and provides a useful guideline for experimental efforts to discover and design new topologically nontrivial moir\'e materials.

\end{abstract}

\date{\today}

\maketitle

\section{Introduction}

Moiré materials \cite{cao2018unconventional,bistritzer2011moire} have emerged as one of the most important platforms for exploring exotic strongly correlated physics, primarily because they host energy bands that are (nearly) flat and topological. The most recent prominent example is twisted bilayer MoTe$_2$, which has been extensively studied both experimentally \cite{
cai2023signatures,zeng2023thermodynamic,park2023observation,xu2023observation,ji2024local,redekop2024direct,kang2024evidence,park2025ferromagnetism,an2025observation,jia2025anomalous,xu2025interplay,kang2025time,park2025observation,xu2025signatures,xia2025simulating}
and theoretically \cite{wu2019topological, yu2020giant, pan2020band, zhang2021electronic, 
li2021spontaneous, devakul2021magic, morales2023pressure, 
wang2024fractional, reddy2023fractional, qiu2023interaction, 
dong2023composite, wang2023diverse, goldman2023zero, 
morales2024magic, liu2024gate, xu2024maximally, reddy2023toward, 
song2024phase, wu2024time, yu2024fractional, abouelkomsan2024band, 
li2024electrically, jia2024moire, mao2024transfer, 
zhang2024polarization, wang2023topology, li2024contrasting, 
sheng2024quantum, reddy2024non, xu2025multiple, ahn2024non, 
wang2025higher, shen2024stabilizing, wang2024phase, kwan2024could, 
wu2024quantum, zaklama2025structure, zhang2024universal, 
qiu2025topological, gonccalves2025spinless, liu2025characterization}, where the flat bands with nonzero Chern numbers give rise to fractional Chern insulators \cite{neupert2011fractional,sheng2011fractional,regnault2011fractional,tang2011high,sun2011nearly} at fractional fillings. The small bandwidth of the moir\'e bands in moir\'e materials naturally arises from the large moir\'e unit cell, which folds the non-moir\'e bands (\ie atomic bands) hundreds of times, leading to energy bands that have bandwidths much smaller than their non-moir\'e counterparts. 
However, the nontrivial topology of the flat bands is not necessarily guaranteed.
There is one known example of enforcing the nontrivial topology of the low-energy moiré bands based on the non-moiré band topology---twisted bilayer graphene (and related graphene-based systems) \cite{cao2018unconventional,cao2018correlated,bistritzer2011moire,song2019all,po2019faithful,ahn2019failure,tarnopolsky2019origin, song2021twisted,song2022magic}.
Yet, this case is quite special: one needs \emph{both} the $C_{2}\mathcal{T}$ symmetry (\ie combination of two-fold rotation $C_2$ and time reversal (TR) symmetry $\mathcal{T}$) and an effective normal-state particle-hole symmetry~\cite{song2021twisted}, which, especially the latter, are not common beyond graphene-based systems. Indeed, the nontrivial topology of the flat bands in twisted bilayer MoTe$_2$ is not guaranteed---the Chern numbers of the bands are sensitive to material parameters such as twisted angles~\cite{zhang2024polarization,xu2025multiple,zhang2024universal,li2024electrically}.

\begin{figure}
    \centering
\includegraphics[width=1\columnwidth]{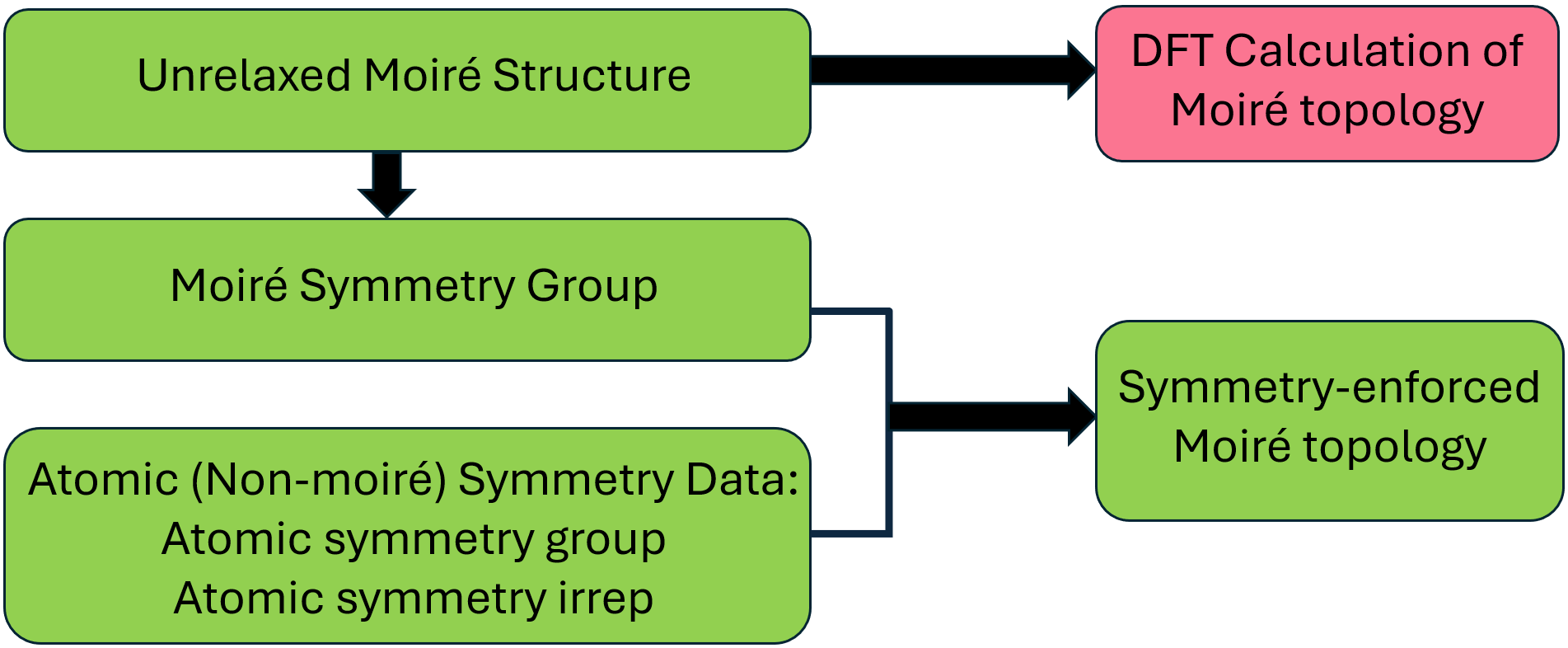}
    \caption{\label{fig:outline}
    {\bf Theoretical Approaches to Identify Moir\'e Topology. }
    All green blocks are easily computationally accessible, while the red block is computationally inefficient. Our work provides an efficient way to indicate moiré topology from the moiré symmetry group and atomic symemtry data.  }
\end{figure}

It would be transformative to develop general principles that can guarantee the nontrivial topology of moiré flat bands based on ubiquitous symmetries, such as crystalline and TR symmetries.
The reason is summarized in Fig.\,\ref{fig:outline} and elaborated in the following.
Given a realistic moir\'e system formed by 2D non-moir\'e layered materials, it is extremely time consuming to directly calculate the moir\'e bands and their band topology (the red box in Fig.\,\ref{fig:outline}), especially when the moir\'e unit cell becomes very large, \eg twist bilayer MoTe$_2$ with a twist angle smaller than $2^\circ$. On the other hand, it is highly efficient to know the symmetry group (crystalline and TR) and symmetry representations of the non-moir\'e layered materials at high-symmetry momenta (which together we refer to as atomic symmetry data), and it takes no time to know the symmetry group of the moir\'e system (which we refer to as moir\'e symmetry group), as depicted by the green boxes in Fig.\,\ref{fig:outline}. Therefore, the principles that can derive moir\'e topology from atomic symmetry data and moir\'e symmetry group can dramatically advance our ability of numerically predicting new topological moir\'e platforms. They are also crucial for experimental studies, as those principles should be robust against sizable variations in material parameters that are often difficult to control in practice---such as the twist angle and moir\'e potential/tunneling strength.

In this work, we successfully establish such a principle. Specifically, for $\Gamma$-valley electron gas subject to weak moir\'e potential, we find that certain combinations of atomic symmetry group, atomic symmetry representation at $\Gamma$ and moir\'e symmetry group can enforce nontrivial topology of the low-energy moir\'e bands, where nontrivial topology includes both topological insulator and topological semi-metal phases. 
Here the low-energy moir\'e bands refer to the moiré bands that are closest to the charge neutrality, \ie conduction bottom bands or valence top bands. We refer to such enforced nontrivial topology as symmetry-enforced moir\'e topology. Weak moir\'e potential means that the strength of the moiré potential is weaker than the atomic band splitting, which is a realistic condition at $\Gamma$ valley satisfied by known examples such as twisted bilayer MoTe$_2$, as discussed in Sec.\ref{sec:discussion}. In other words, for those combinations of atomic symmetry data and moiré symmetry group, the phase diagram of the low-energy bands of the moiré model only includes topological insulator and semi-metal phases---no topologically trivial phases. By systematically analyzing all possible combinations, we identify 197 combinations that can lead to symmetry-enforced nontrivial moir\'e topology, with 16 of them for the topological insulating phase in the phase diagram. 
The general principle is explicitly verified in a moir\'e model with cubic Rashba spin-orbit coupling (SOC). 
We further screen the existing non-moiré 2D material databases \cite{vergniory2019complete, haastrup2018computational,jiang20242d}, focusing on monolayers with plane groups P31m, P3m1, and P3. We identify 92 monolayer materials that have (i)  low-energy bands near $\Gamma$ and (ii)  the atomic symmetry data that belong to those combinations, as discussed in  \cref{sec:material}.

Clearly, our principle does not rely on precisely-tuned values of moir\'e parameters such as twist angle and moir\'e potential strength. It is also independent of the form of moir\'e potential, in contrast to Refs.\cite{crepel2025efficient,Lhachemiarxiv}. 
We note that the moiré topological semi-metal phases may also be a fertile ground for exotic strongly-correlated phases. It is because electron-electron interactions may open a gap between topological metallic bands and drive the system into topologically ordered states. A closely related example is the emergence of fractional Chern insulators in multilayer rhombohedral graphene/hexagonal boron nitride (hBN) heterostructures at various electron fillings \cite{lu2024fractional, xie2025tunable, waters2025chern, lu2025extended, 
aronson2025displacement,zhou2024layer,han2024engineering, xiang2025continuously,wang2025electrical,li2025tunable,xie2025unconventional}, where the moir\'e bands are nearly gapless at the single-particle level \cite{ park2023topological, dong2024theory,zhou2024fractional,dong2024anomalous, herzog2024moire,guo2024fractional,kwan2025moire,soejima2024anomalous,dong2024stability,xie2024integer,kudo2024quantum,crepel2025efficient,huang2024self,yu2024moir,huang2025fractional,das2024thermal, wei2025edge,zhou2025new,zeng2025berry,bernevig2025berry,li2025multiband,uchida2025non}.

\begin{figure}
    \centering
\includegraphics[width=1\columnwidth]{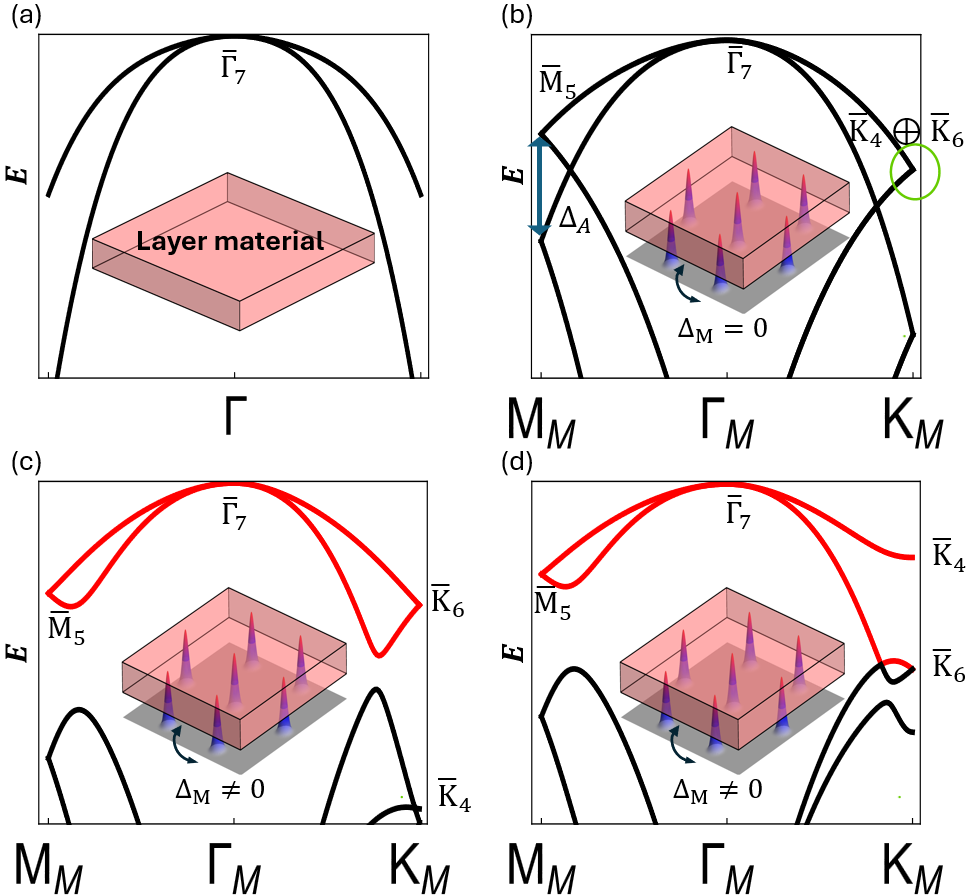}
    \caption{\label{fig:principle}   
    {\bf Schematics of Symmetry-Enforced Moir\'e Topology. }
   (a) The band structure around $\Gamma$ point of a pristine layered material with point group $C_{6v}$. The low-energy band at $\Gamma$ is described by the 2D $\bar{\Gamma}_7$ irrep. (b) Band folding at zero moir\'e potential. At ${\text{K}}_M$, the folded bands form a three-dimensional representation $\bar{\text{K}}_4 \oplus \bar{\text{K}}_6$, while at ${\text{M}}_M$ they form a two-dimensional irrep $\bar{\text{M}}_5$. $\Delta_A$ marks the energy scale of the atomic band splitting at the mBZ boundary.
    (c) One possible moiré band structure in the presence of weak moir\'e potential, \ie the energy splitting generated by $\Delta_M$ is smaller than $\Delta_A$.
    Here the top two bands (red) carry the $\bar{\text{K}}_6$ irrep, while the third top band has the $\bar{\text{K}}_4$ irrep, resulting in nontrivial topology in the isolated top two bands.
     (d) The other possible moir\'e band structure in the presence of weak moir\'e potential.   
    The $\bar{\text{K}}_4$ band has higher energy than the two $\bar{\text{K}}_6$ bands, resulting in a topological semi-metallic phase for the red bands due to the unavoidable touching between the second and third top bands at $\text{K}_M$.
    }
\end{figure}

\section{Illustration of Principle}
\label{sec:principle}
We will first use a specific set of atomic symmetry data and moir\'e symmetry group to illustrate the underlying principle.
Let us consider spin-orbit-coupled electrons around $\Gamma$ point of a non-moir\'e layered material. Suppose the electrons carry total out-of-plane angular momentum $J_z = \pm 3/2$, and have the point group $C_{6v}$ and TR symmetry. $J_z = \pm 3/2$ can be generated by the addition of atomic orbitals ($p$, $d$ or higher) and electron spin. The $C_{6v}$ group is generated by a six-fold rotation along the out-of-plane direction and a mirror reflection about the $y$-axis ($M_y$). Owing to the angular momentum, the electrons at $\Gamma$ form a doublet, furnishing $\bar{\Gamma}_7$ irreducible representation (irrep) of $C_{6v}$. Without loss of generality, we choose both bands to bend downward, and the two bands naturally get split by the SOC away from $\Gamma$, as shown in \cref{fig:principle}(a).

Now we introduce the moir\'e potential that preserves the point group $C_{6v}$ and  TR symmetry. In addition, the moir\'e potential would introduce the discrete moir\'e translation symmetry, leading to a plane group $P6mm$. In order to understand the generation of the moir\'e bands, let us first consider an artificial limit where the moir\'e potential strength $\Delta_M$ vanishes. In this case, the moir\'e bands simply come from the artificial band folding of the $\Gamma$-valley atomic bands (\cref{fig:principle}(a)) into the moiré Brillouin zone (mBZ), as shown in \cref{fig:principle}(b). The $\bar{\Gamma}_7$ irrep survives under the band folding and stays at the moiré $\Gamma_{M}$ point, while the band folding generates a Kramers' pair among the top two bands at the $\text{M}_M$ point, furnishing $\bar{\text{M}}_5$ irrep. Most importantly, the band folding gives rise to three accidentally degenerate states at $\text{K}_M$ point. The degeneracy is accidental because they can be split into a one-dimensional (1D) irrep $\bar{\text{K}}_4$ and a two-dimensional (2D) irrep $\bar{\text{K}}_6$ according to the moir\'e symmetry group discussed below.

Now we resume the nonzero moiré strength $\Delta_M$. We consider the weak moiré region where the energy splitting induced by the moiré potential is smaller than the atomic band splitting ($\Delta_A$) at the mBZ boundary. 
The moiré potential will leave the $\bar{\Gamma}_7$ irrep at $\Gamma_M$ and the $\bar{\text{M}}_5$ irrep at $\text{M}_M$ intact, as protected by symmetry. Nevertheless, the moiré potential will split the $\bar{\text{K}}_4$ and $\bar{\text{K}}_6$ bands at $\text{K}_M$. Focusing on the top two bands, there are two possibilities. One is that the top two bands carry the 2D $\bar{\text{K}}_6$ irrep, and the third band possesses the $\bar{\text{K}}_4$ irrep, resulting in the top two bands being isolated (\cref{fig:principle}(c)). In this case, the top two bands have nontrivial topology as the irreps $\bar{\Gamma}_7$,  $\bar{\text{M}}_5$ and $\bar{\text{K}}_6$ can never appear simultaneously for an isolated set of trivial bands according to topological quantum chemistry \cite{bradlyn2017topological,cano2021band,elcoro2021magnetic} and symmetry indicator theory \cite{po2017symmetry,kruthoff2017topological,fu2007topological,fang2012bulk,po2020symmetry,lenggenhager2022universal}. The other possibility is that the top two bands are from one 1D $\bar{\text{K}}_4$ irrep and one component of the 2D $\bar{\text{K}}_6$ irrep. In this case, the top second band definitely touches the third band as they together form the $\bar{\text{K}}_6$ irrep at $\text{K}_M$. Therefore, regardless of which possibility, the low-energy moir\'e bands are either topologically insulating or topologically semi-metallic. Such nontrivial moir\'e topology is enforced by the atomic symmetry data and moir\'e symmetry group, as long as the moir\'e potential is weak compared to atomic band splitting.

\section{Moir\'e cubic Rashba model}
We now use an specific example to verify the principle. 
We consider a model Hamiltonian describing a spin-orbit-coupled 2D electron gas around $\Gamma$ point subject to a moir\'e superlattice potential. 
As discussed in \cref{sec:principle}, we impose the point group $C_{6v}$ and TR symmetry, and consider $J_z =\pm 3/2$ doublet at $\Gamma$.
Then, the atomic part of the Hamiltonian up to the fourth order of the momentum $\kk$ has the following general form
\begin{align} \label{eq: cubic rashba}
&\hat{H}^{A} = \sum_{\kk n m}c_{n \kk }^\dagger 
 [H^{A}(\kk)]_{n m} c_{m \kk}, \nonumber \\
  &H^{A}(\kk)=\begin{pmatrix}
       \alpha k^2 + \beta  k^4  &  i R_3 k^3_- \\
     -i R_3 k^3_+& \alpha k^2 +  \beta k^4
    \end{pmatrix}.
\end{align}
where $c_{n \kk}$ is the fermion annihilation operator with $n = 1, 2$ indexing $J_z=\pm 3/2$. 
To ensure that both bands of $\hat{H}^A$ bend down, we choose both $\alpha$ and $\beta$ to be negative in the kinetic term $\alpha k^2+\beta  k^4$, where the $k^4$ term is necessary due to the cubic Rashba SOC term, \ie the $R_3$ term in \cref{eq: cubic rashba}.

\begin{figure}
    \centering
\includegraphics[width=1\columnwidth]{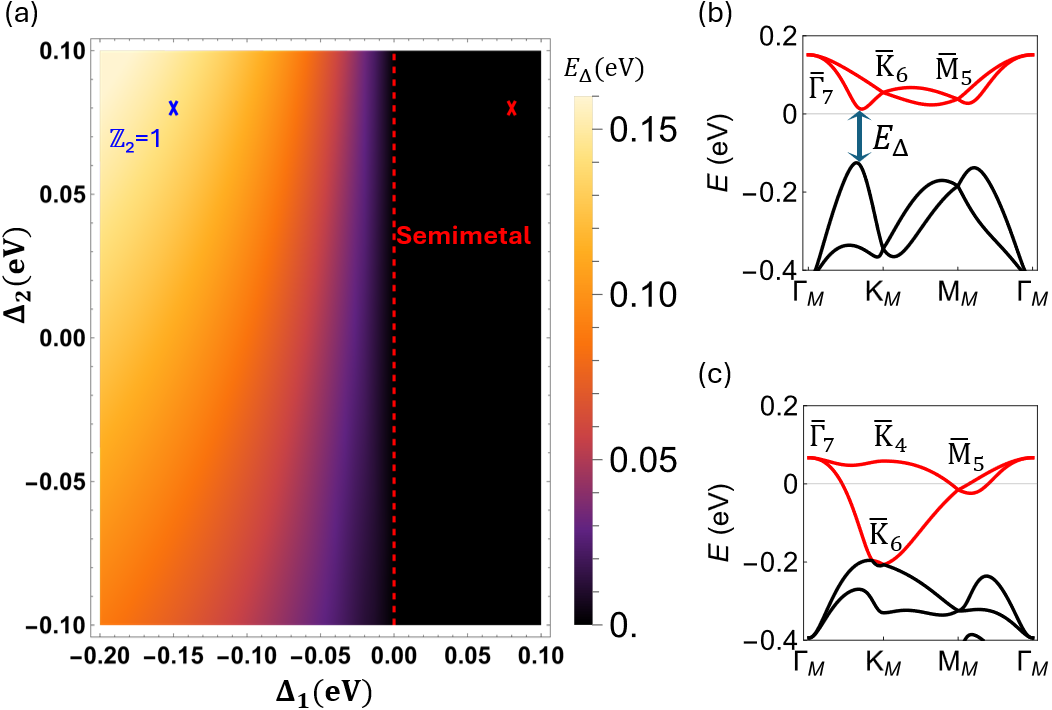}
    \caption{\label{fig:phase diagram}
   {\bf Cubic Rashba SOC model under moir\'e potential. }
    (a) Phase diagram of the topmost two bands of the  moiré cubic Rashaba model in \cref{eq_main:HammC}. $E_\Delta$ denotes the band gap between the second and third topmost bands. The red dashed lines separate the insulator region and semimetal region. (b) The band dispersion and band irreps computed with the parameters labeled by the blue cross point in (a). The top two bands have $\mathbb{Z}_2 = 1$. (c) The band dispersion and band irreps computed with the parameters labeled by the red cross point in (a). The top two bands are connected to the top third band.  
    }
\end{figure}

We consider the moir\'e superlattice potential on a hexagonal moiré lattice, which reads 
\begin{align}
& \hat{H} _ M  = \sum_{\alpha = 1, 2 }\int d^2r  c_{n  \rr }^\dagger H _ \text M (\rr) c_{n  \rr },  \label{eq_main:moirepotential}
\end{align}
with $c_{n  \rr }=\frac{1}{\sqrt{V}}\int_{\mathds{R}^2} d^2k c_{n \kk } e ^{i \kk \cdot \rr}$ and $V$ the system area. 
We include both the first and second harmonic moir\'e supperlattice potential, namely, $H_M(\rr)=\sum_{\gs \in \GG^{\mathds{1}}_M} \Delta_1 e^{i \gs \cdot \rr}+\sum_{\gs \in \GG^{\mathbb{2}}_M} \Delta_2 e^{i \gs \cdot \rr}$, where $\GG^{\mathds{1}}_M$ consists of $\bb^M_1$ and its partners under six-fold rotation symmetries, $\GG^{\mathbb{2}}_M$ consists of $\bb^M_1 + \bb^M_2$ and its partners under six-fold rotation symmetries. $\bb^M_1= 2\pi(1,-1/\sqrt{3})/a_{M}$ and $\bb^M_2 =  2\pi(1,1/\sqrt{3})/a_{M}$ are primitive reciprocal lattice vectors, and $a_{M}$ is the moir\'e lattice constant. The moiré potential in \cref{eq_main:moirepotential} has plane group $P6mm$, generated by the point group $C_{6v}$ and the moiré lattice translation symmetry, and also preserves the TR symmetry. Those symmetries are preserved even after we incorporate the atomic Hamiltonian in \cref{eq: cubic rashba}, resulting in the total Hamiltonian 
\begin{align} 
\hat{H}=\hat{H}^{A}+\hat{H}_M\ .
\label{eq_main:HammC}
\end{align}
To illustrate our principle, we focus on the weak moiré region, where the moiré gap is smaller than the band splitting caused by Rashba SOC. By choosing reasonable parameter values (listed in supplementary material (SM) Sec.C1 \cite{SM}.), we obtain the phase diagram as a function of $\Delta_1$ and $\Delta_2$ in \cref{fig:phase diagram}(a). The color represents the direct band gap $E_\Delta$ between the second and third topmost moir\'e bands (see \cref{fig:phase diagram}(b)) and the red dashed line separates the insulating region from the semi-metallic region for the top two moir\'e bands. (See more details in SM Sec.C1 \cite{SM}.) We find that regardless of the value of the moiré parameter values, there are only two phases for the low-energy physics: (i) topological insulator phase where the top two bands are isolated with band irreps ($\Bar{ \mathbf{\Gamma}}_7$, $\Bar{\textbf{M}}_5$, $\Bar{\textbf{K}}_6$) and have $\mathds{Z}_2 =1 $ in \cref{fig:phase diagram}(b), and (ii) topological semi-metal phase where the top two bands are enforced to be connected to other bands in \cref{fig:phase diagram}(c) as the irrep $\Bar{\textbf{K}}_4$ for the topmost band is 1D, while $\Bar{\textbf{M}}_5$ and $\Bar{\mathbf{\Gamma}}_7$ irreps are 2D. This is consistent with the principle that we discussed in \cref{sec:principle}.

\begin{table}[t]
\centering
\renewcommand{\arraystretch}{1.2}
\begin{tabular}{|c|c|c|c|c| c|c|c|}
\hline
 \diagbox{$G_0$}{$G^A_0$} & $C_{6v}$ & $C_{3v}$ & $C_{6}$ & $C_{3}$  &  $C_{4v}$&  $C_{2v}$ & $C_{s}$
\\
\hline
 $P6mm$ & \textcolor{red}{$\Bar{\Gamma}_7$} &  &  &  & &  &
 \\
  $P3m1$ & \textcolor{red}{$\Bar{\Gamma}_7$} & \textcolor{red}{$\Bar{\Gamma}_4\Bar{\Gamma}_5$} &  &   &    & &
 \\
   $P31m$ & \textcolor{red}{$\Bar{\Gamma}_7$}  & \textcolor{red}{$\Bar{\Gamma}_4\Bar{\Gamma}_5$} &  &   &    & & 
 \\
  $P6$ & \textcolor{red}{$\Bar{\Gamma}_7$}  &  & \textcolor{red}{$\Bar{\Gamma}_7\Bar{\Gamma}_8$} &  &    &  & 
 \\
 $P3$ & \textcolor{red}{$\Bar{\Gamma}_7$}  & \textcolor{red}{$\Bar{\Gamma}_4\Bar{\Gamma}_5$} & \textcolor{red}{$\Bar{\Gamma}_7\Bar{\Gamma}_8$} &  \textcolor{red}{$\Bar{\Gamma}_4\Bar{\Gamma}_4$}  &    &  & 
 \\
  $P4bm$ &   &  &  &    &  $\Bar{\Gamma}_6$, $\Bar{\Gamma}_7$  & & 
 \\
  $Pba2$ & $\Bar{\Gamma}_7$, $\Bar{\Gamma}_8$, $\Bar{\Gamma}_9$  &  &  &    &  $\Bar{\Gamma}_6$, $\Bar{\Gamma}_7$  &$\Bar{\Gamma}_5$  & 
 \\ 
 $Pma2$ &$\Bar{\Gamma}_7$, $\Bar{\Gamma}_8$, $\Bar{\Gamma}_9$ & &  & & $\Bar{\Gamma}_6$, $\Bar{\Gamma}_7$   & $\Bar{\Gamma}_5$ & 
 \\
  $Pb11$ &$\Bar{\Gamma}_7$, $\Bar{\Gamma}_8$, $\Bar{\Gamma}_9$ & $\Bar{\Gamma}_4\Bar{\Gamma}_5$, $\Bar{\Gamma}_6$ &  &    &  $\Bar{\Gamma}_6$, $\Bar{\Gamma}_7$  & $\Bar{\Gamma}_5$ & $\Bar{\Gamma}_3\Bar{\Gamma}_4$
 \\
\hline
\end{tabular}
\caption{   \label{TAB:summary_TR_SOC} {\bf Spinful TR-invariant Cases for Symmetry-enforced Moir\'e Topology. } List of combinations of atomic point group $G_0^A$, atomic $\Gamma$ irrep, and moiré plane gorup $G_0$ that lead to symmetry-enforced nontrivial moiré topology, in the presence of both TR symmetry and SOC. For a given $G_0^A$ and $G_0$, the table entry indicates the required atomic $\Gamma$ irrep to give symmetry-enforced topology. The irrep is highlighted by the red means the low-energy bands can be topological insulating.  }
\end{table}

\begin{table*}[]
\begin{tabular}{|c|c|c|c|c|c|}
\hline
 \diagbox{$G_0$}{$G^A_0$} & $C_{6v}$ & $C_{4v}$ & $C_{3v}$ & $C_{2v}$ & $C_{s}$
\\
\hline
 $P6$ & \textcolor{red}{${\Gamma}_5$, ${\Gamma}_6$} & & & & 
 \\
 $P2$ & \textcolor{red}{${\Gamma}_5$, ${\Gamma}_6$} & \textcolor{red}{${\Gamma}_5$} & & & 
 \\
  $P6mm$ & ${{\Gamma}}_5$, ${{\Gamma}}_6$ & & & & 
 \\
 $P4mm$ &  & ${{\Gamma}}_5$ &  &   &   
 \\
 $P4bm$ &  & ${{\Gamma}}_1$, ${{\Gamma}}_2,{{\Gamma}}_3$, ${{\Gamma}}_4$ &  &   & 
 \\
  $P3m1$ & ${{\Gamma}}_5$, ${{\Gamma}}_6$ &   &  &  &
 \\
 $Pmm2$ & ${{\Gamma}}_5$, ${{\Gamma}}_6$ &${\Gamma}_5$   &  &  &
 \\
  $Cmm2$ & ${{\Gamma}}_5$, ${{\Gamma}}_6$ &${\Gamma}_5$   &  &  &
 \\
 $Pba2$ & ${{\Gamma}}_1$, ${{\Gamma}}_2,{{\Gamma}}_3$, ${{\Gamma}}_4$ & ${{\Gamma}}_1$, ${{\Gamma}}_2,{{\Gamma}}_3$, ${{\Gamma}}_4$  &  & ${{\Gamma}}_1$, ${{\Gamma}}_2,{{\Gamma}}_3$, ${{\Gamma}}_4$    &
 \\
   $Pma2$ & ${{\Gamma}}_1$, ${{\Gamma}}_2,{{\Gamma}}_3$, ${{\Gamma}}_4,{{\Gamma}}_5$, ${{\Gamma}}_6$ &   ${{\Gamma}}_1$, ${{\Gamma}}_2,{{\Gamma}}_3$, ${{\Gamma}}_4$, ${{\Gamma}}_5$& &${{\Gamma}}_1$, ${{\Gamma}}_2,{{\Gamma}}_3$, ${{\Gamma}}_4$&
 \\
   $Pm11$ & ${{\Gamma}}_5$, ${{\Gamma}}_6$ & ${\Gamma}_5$  &   & &
 \\
   $Pb11$ &  ${{\Gamma}}_1$, ${{\Gamma}}_2,{{\Gamma}}_3$, ${{\Gamma}}_4$ & ${{\Gamma}}_1$, ${{\Gamma}}_2,{{\Gamma}}_3$, ${{\Gamma}}_4$ &   ${{\Gamma}}_1$, ${{\Gamma}}_2$  & ${{\Gamma}}_1$, ${{\Gamma}}_2,{{\Gamma}}_3$, ${{\Gamma}}_4$ &${{\Gamma}}_1$, ${{\Gamma}}_2$
 \\
\hline
\end{tabular}
\caption{   \label{TAB:summary_NTR_NSOC}  {\bf Spinless  TR-breaking Cases for Symmetry-enforced Moir\'e Topology. } List of combinations of atomic point group $G_0^A$, atomic $\Gamma$ irrep, and moiré plane gorup $G_0$ that lead to symmetry-enforced nontrivial moiré topology, without TR symmetry and without SOC. The meaning of the table entries is the same as that in \cref{TAB:summary_TR_SOC}.
}
\end{table*}

\section{Symmetry-Enforced Moiré Topology in 2D Plane Group}

The symmetry-enforced moiré topology is not limited to the symmetry data discussed in \cref{sec:principle}. In this section, we will show that there are many other cases where nontrivial moiré topology can be enforced by the atomic symmetry data and the moiré symmetry group. We still focus on the $\Gamma$-valley electron gas (both with and without SOC and TR) in the weak moiré region (\ie the moiré gap is weaker than atomic band splitting if present).

As discussed in \cref{sec:principle}, atomic symmetry data includes the atomic symmetry group $G^A$, and its irrep $\Lambda^A_\Gamma$ furnished by the considered states at $\Gamma$, and we use $G$ to label the moiré symmetry group.
Here $G^A$ is simply the point group at $\Gamma$, $G^A_0$, if there is no TR symmetry, but also includes the TR symmetry, denoted as $\mathcal{T}$,  if it appears. Similarly, $G$ is simply the moiré plane group, $G_0$, if there is no TR symmetry, but also includes the TR symmetry if it appears. We focus on the case that the moiré potential does not cause extra TR breaking. For the specific case in \cref{sec:principle}, we have $G^{A} = G^{A}_0\rtimes \mathcal{T}$ with $G^{A}_0 = C_{6v}$, the irrep $\Lambda^A_\Gamma=\bar{\Gamma}_7$, and $G= G_0 \rtimes \mathcal{T}$ with $G_0 = P6mm$. Although we have the atomic symmetry group to be included by the moiré symmetry group in \cref{sec:principle}, we do not assume such a relation in the general study in this section. The detailed steps for this general search are illustrated in SM Sec.A \cite{SM}.

We went through all possible combinations of $G^{A}$, the irrep $\Lambda^A_\Gamma$, $G$, and the presence/absence of SOC, as elaborated in SM Sec.B \cite{SM}. The results are summarized in \cref{TAB:summary_TR_SOC} 
for the spinful case with TR and in  
\cref{TAB:summary_NTR_NSOC} for the spinless case without TR, where spinful (spinless) simply refers to the case with (without) SOC as indicated by the irrep notation with (without) a bar \cite{aroyo2011crystallography,aroyo2006bilbao,aroyo2006bilbao1}. (The results of spinless cases with TR and spinful cases without TR are summarized in Tab. SI(c) and SIV(b) in the SM Sec.B \cite{SM} respectively. ) 
We find 58 combinations that lead to symmetry-enforced moiré topology in the spinful case, and 139 in the spinless case. Out of the 58 spinful combinations (139 spinless combinations), we find that there are 11 (5) combinations that allow for isolated topological bands at low-energy, while all other combinations only have topological semi-metal phases in the weak moir\'e potential limit. The symmetry-enforced gapped moir\'e topology has the following scenarios: (i) for the TR-preserving case, isolated topological moir\'e bands may carry nonzero $\mathds{Z}_2$ index in addition to other possibilities such as fragile topology;  (ii) for the TR-breaking case, the topological moir\'e bands always have nonzero Chern number.

From our results, we can immediately provide several guiding principles for finding moir\'e topological insulators in $\Gamma$-valley moir\'e systems. First, symmetry-enforced moir\'e gapped topology only happens for doublets at $\Gamma$ for both the TR-preserving and TR-breaking cases. Second, almost all symmetry-enforced moiré topological insulators happen on a triangular/hexagonal lattice, with only three combinations as exceptions. Therefore, to look for moiré topological insulators in $\Gamma$-valley moiré systems, it is favorable to look for low-energy 2D irreps at $\Gamma$ valley subject to triangular/hexagonal moiré potential. Finally, symmetry-enforced moiré topological insulators only happen when SOC and TR symmetry are simultaneously present or simultaneously absent, meaning that the simultaneous presence and absence of TR symmetry and SOC is also favorable for moiré topological insulators in $\Gamma$-valley moiré systems.

On the other hand, symmetry-enforced moiré topological semimetals are much easier to be realized than 
symmetry-enforced moiré topological insulators. There are no particular constraints on the lattice type, or the simultaneous presence/absence of TR symmetry and SOC. Nevertheless, doublets at $\Gamma$ are still required to realize symmetry-enforced moiré topological semimetal, unless the moiré plane group is non-symmorphic \cite{michel1999connectivity,michel2001elementary, watanabe2015filling, watanabe2016filling, vergniory2017graph}.

\section{Material candidates}\label{sec:material}


We now discuss the material candidates.
As our framework only needs information of the monolayers, we can screen through the existing non-moiré 2D material databases \cite{haastrup2018computational,vergniory2019complete, jiang20242d} for suitable monolayer 2D materials for our proposal. 

We first pick out monolayer spin-orbit coupled materials in which (i) VBM or CBM is located at $\Gamma$ and (ii) there are no symmetries that will force the atomic band splitting to  be zero  along at least one line in the BZ.
Particularly, we focus on atomic plane groups P31m, P3m1 and P3, in which 92 monolayer materials are identified with their atomic symmetry data listed in \cref{TAB:summary_TR_SOC}. 
In particular, for 71 of them, the required moiré symmetry group is P3 according to \cref{TAB:summary_TR_SOC}, which can be naturally realized by a small angle twisted homobilayer structure.

\begin{figure}
    \centering
\includegraphics[width=1\columnwidth]{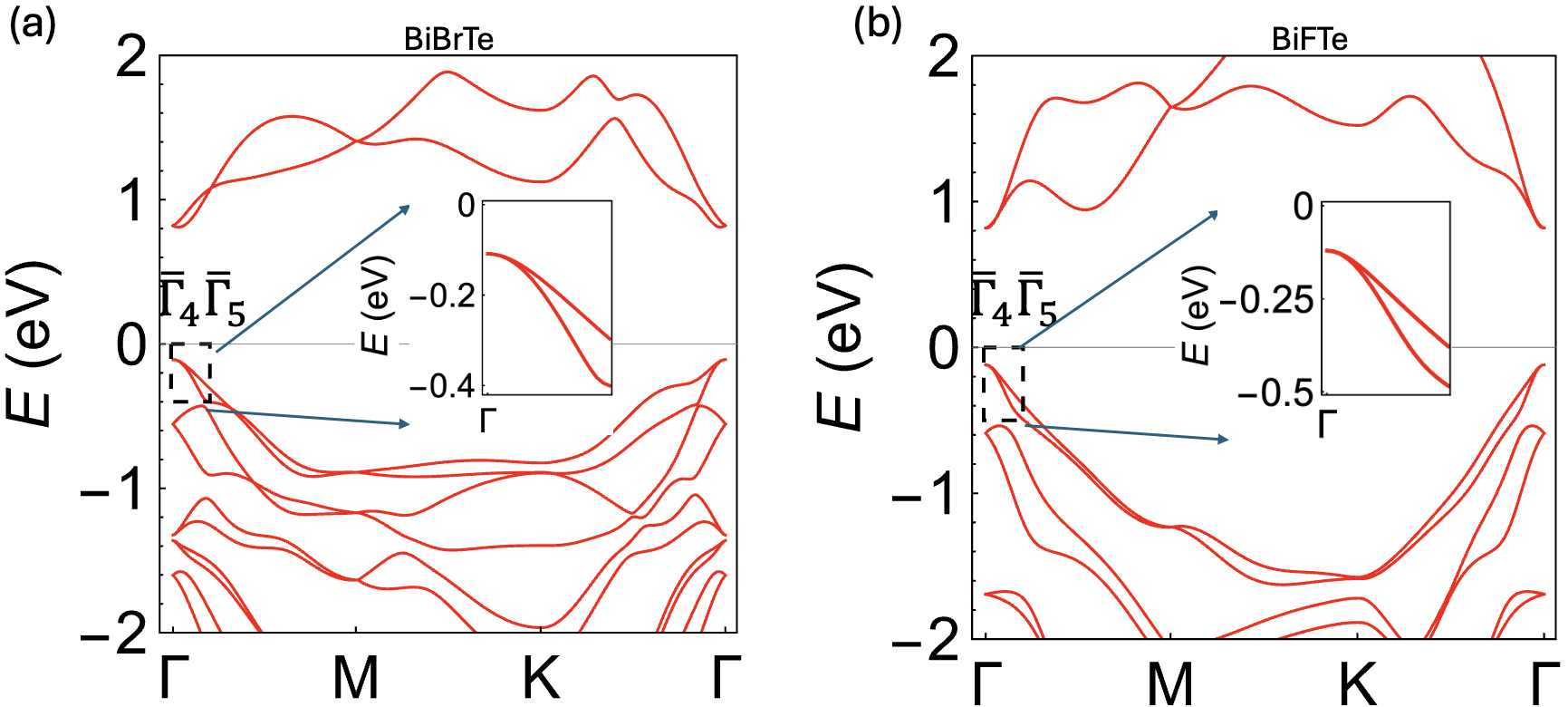}
    \caption{\label{fig: material example}
    (a) Band dispersion of BiBrTe. (b) Band dispersion of BiFTe. The irreps of VBM for both materials are $\bar{\Gamma}_4\bar{\Gamma}_5$. The inset figures show a zoom-in of the regions inside the dashed boxes.
    }
\end{figure}

%
We show two examples, BiBrTe and BiFTe, in \cref{fig: material example}.
For these two examples, the atomic symmetry group at $\Gamma$ is $G_0^A = C_{3v}$ and the irreps of VBM are $\bar{\Gamma}_4\bar{\Gamma}_5$, which belong to \cref{TAB:summary_TR_SOC}.
The required moiré symmetry group P3 naturally arise in the twist homobilayer of either of the two examples (See more discussion in Sec.D in SM \cite{SM}.

\section{Conclusion and Discussion} \label{sec:discussion}
We have proposed the concept of symmetry-enforced moiré topology, and list all combinations of atomic symmetry data and moiré symmetry group to realize it for $\Gamma$-valley electron gas subject to a weak moiré potential. 92 candidate monolayer materials are identified based on a selective search.

We note that the TR-breaking cases with the spinless models in Sec.C2 of SM \cite{SM} can also be applied to topological moir\'e magnon systems \cite{li2020moire,wang2020stacking,ganguli2023visualization}. 
Furthermore, the generation of the moir\'e potential is not limited to twisting two homobilayers---it can also be done (i)  by forming hetero-structure with insulating materials such as boron nitride or transition metal dichalcogenides (TMD)~\cite{yasuda2021stacking, woods2021charge, zhao2021universal, kim2024electrostatic} and (ii) by fabricating patterned hole arrays in dielectric substrate materials~\cite{forsythe2018band, barcons2022engineering}. We note that although the homobilayer structure contains two copies of the $\Gamma$-valley modes from each layer, the low-energy physics can only involve either the bonding combination or the anti-bonding combination between these two copies, owing to the strong uniform interlayer coupling \cite{angeli2021gamma, zhang2021electronic,jia2024moire,liu2025ideal}. As a result, at low energies, homobilayer twist structure effectively has only one set of the $\Gamma$-valley modes. This is exactly the case in twisted homobilayer TMD  at $\Gamma$ valley~\cite{angeli2021gamma, zhang2021electronic,jia2024moire}. While our current study focuses on layered materials with low-energy $\Gamma$-valley modes, the methodology should be generalizable to other valleys, such as $K$ or $M$ valleys \cite{cualuguaru2025moire}, which we leave for the future work.

\section{Acknowledgment}
We acknowledge the helpful discussion with Jennifer Cano and Lei Chen. 
J. Y.'s work is supported by startup funds at University of Florida. Y.L. and C.L. acknowledge the support from the Penn State Materials Research Science and Engineering Center for Nanoscale Science under National Science Foundation award DMR-2011839. This work was performed in part at Aspen Center for Physics, which is supported by National Science Foundation grant PHY-2210452.

\bibliography{ref}

\end{document}


\setcitestyle{numbers,square}
\title{Supplemental Materials for "Symmetry-enforced Moir\'e Topology"}

\author{Yunzhe Liu$^{1, \ddag}$}
\author{Ethan Angerhofer$^{2,\ddag}$}
\author{Kaijie Yang$^{2}$}
\author{Chao-Xing Liu$^{1, }$}
\author{Jiabin Yu$^{3,\dagger}$}

\affiliation{$^{1}$Department of Physics, The Pennsylvania State University, University Park, Pennsylvania 16802, USA}
\affiliation{$^{2}$Department of Materials Science and Engineering, University of Washington, Seattle, Washington 98195, USA}
\affiliation{$^{3}$Department of Physics, University of Florida, Gainesville, FL, USA}

\date{\today}

\maketitle

\begingroup
\renewcommand\thefootnote{\fnsymbol{footnote}}
\footnotetext[1]{\texttt{cxl56@psu.edu}}
\footnotetext[2]{\texttt{yujiabin@ufl.edu}}
\footnotetext[3]{\texttt{These authors contributed equally.}}
\endgroup

\section{General formalism for Topological properties of moir\'e systems based on topological quantum chemistry}\label{sec:moire TQC}

We consider a two-dimensional (2D) layered material, in which the conduction band minimum (CBM) or valence band maximum (VBM) is near the $\mathbf{\Gamma}$ point in the atomic Brillouin zone (ABZ).  We assume the low-energy physics of this 2D material occurs around $\mathbf{\Gamma}$ and can be described by an effective Hamiltonian $\hat{H}^{A}$. 
We further introduce a moir\'e superlattice potential, leading to the following full Hamiltonian
\begin{eqnarray}
    \hat{H}= \hat{H}^A + \hat{H}_M,
\end{eqnarray}
where $\hat{H}_M$ describes the weak moir\'e superlattice potential. The low-energy moir\'e bands refer to conduction bottom bands or valence top bands, as they are closest to charge neutrality. Here we will show how the low-energy moir\'e bands can be guaranteed to be topologically non-trivial by atomic symmetry data and moiré symmetry group ($G$), as long as the moiré potential is weaker than the atomic band splitting---the weak moiré condition is satisfied. The atomic symmetry data include the atomic symmetry group of $\hat{H}^A$, labeled as $G^A$, and the symmetry representation of $G^A$, denoted as $\Lambda^A_\Gamma$, furnished by the basis of $\hat{H}^A$ at $\Gamma$. Throughout the paper, we always make sure the basis of $\hat{H}^A$ at $\Gamma$ furnishes an irreducible representation (irrep).

To address this question, the key step is to determine all the possible symmetry representations of the low-energy moiré bands at high symmetry momenta in the moir\'e Brillouin zone (mBZ) given a fixed combination of atomic symmetry data and moiré symmetry group. Once the moiré symmetry representations are obtained, we can directly use the theoretical formalism of topological quantum chemistry (TQC) \cite{bradlyn2017topological,cano2021band,elcoro2021magnetic} and symmetry indicators \cite{fu2007topological,fang2012bulk,kruthoff2017topological,po2017symmetry,po2020symmetry,lenggenhager2022universal} to determine the moir\'e topology.

We first consider the case where the full Hamiltonian $\hat{H}$ does not have time reversal (TR) symmetry. In this case, the atomic symmetry group $G^{A}=G_0^{A}$ is a point group. The effective atomic model also has the continuous translation $T$, in addition to $G^{A}=G_0^{A}$. Let us denote the plane group of the moir\'e potential $\hat{H}_M$ as $G^M_0$, which contains the moiré translations $T_M$. 
Then, the symmetry group of the full Hamiltonian $\hat{H}$ is $G = G_0 = G^M_0 \cap (G^A_0 \rtimes T)$. Below we always use $\kk_0$ to denote the momentum in the ABZ and ${\kk}{_M}$ to denote the momentum in the mBZ, e.g. $\kk_0 = {\kk}{_M} +\mathbf g$ for a certain moir\'e reciprocal lattice vector $\mathbf g$. Given a moiré momentum ${\kk}{_M}$, the group of operations in $G^A_0\rtimes T$ and $G_0$ that leaves $\kk_M$ invariant are denoted as
 \begin{eqnarray}
&&G_{\kk_M}^{A} = \{ \mathcal S  \vert  \mathcal S\in G^A_0\rtimes T \And \mathcal S \kk_M = \kk_M+ \exists\gs \},\label{eq:little group}\\
&&G_{{\kk}{_M}} = \{ \mathcal S  \vert  \mathcal S\in G_0 \And \mathcal S {\kk}{_M} = {\kk}{_M} + \exists\mathbf g \} \label{eq:momentum shell T},
\end{eqnarray} respectively, where $\mathbf g$ is a moir\'e reciprocal lattice vector. $G_{\kk_M}$ is a subgroup of $G^A_{\kk_M}$, because $G_0$ is a subgroup of $G^A_0\rtimes T$. We note that as $T$ is the group of continuous translation, $G $ is allowed to be nonsymmorphic.

We label the dimension of the irrep $\Lambda^{A}_\mathbf{\Gamma}$ as $n_d$. Owing to the irreducible nature, all eigen-states of $\hat{H}^A$ are degenerate at $\mathbf{\Gamma}$ with the degeneracy $n_d$  for $n_d>1$. The highest dimension of  $\Lambda^{A}_\mathbf{\Gamma}$ for all possible $G^{A}$ for 2D materials is two \cite{fu2024symmetry}, so that we only need to consider $n_d\leq2$. We refer to $\Lambda^{A}_\mathbf{\Gamma}$ together with $G^A$ as atomic symmetry data.  For the momentum $\kk$ away from $\mathbf{\Gamma}$, the band splitting of these $n_d$ -fold degenerate states will follow the compatibility relations \cite{bradlyn2017topological,cano2021band,elcoro2021magnetic,dresselhaus2007group,evarestov2012site,po2017symmetry,po2020symmetry}. The moir\'e potential $\hat{H}_M$ can further lower the space group symmetry, thus inducing additional band splitting. Here we focus on the \textit{weak moir\'e potential limit}, which means the band splitting induced by the moir\'e potential $\hat{H}_M$ is smaller than the intrinsic band splitting of $\hat{H}^A$ at the mBZ boundary. 

\begin{figure}
    \centering
    \includegraphics[width=1\linewidth]{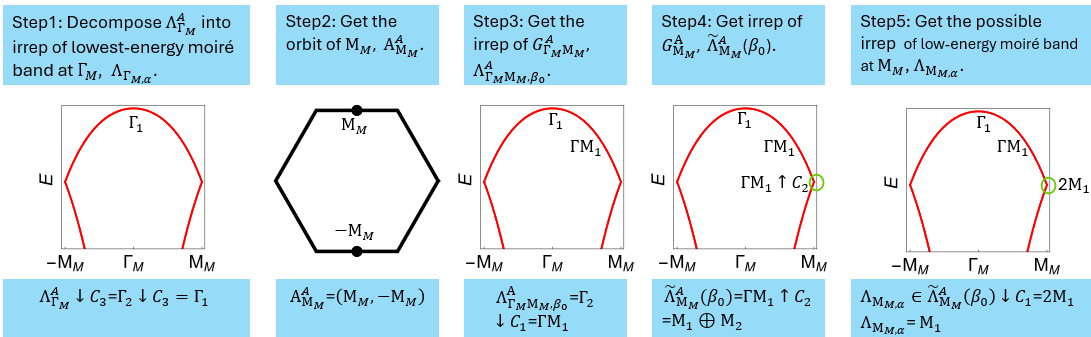}
    \caption{\justifying   The workflow chart illustrating how to determine the irreps 
at high-symmetry points in the MBZ. We take the combination of $G^A_0=C_{6}$,  $\Lambda^A_\Gamma=\Gamma_2$ and $G_0=P3$ as an example. In this example,  $G^A_{\mathbf{\Gamma}_M\mathbf{M}_M}=C_1$, $G^A_{\mathbf{M}_M}=C_2$, $G_{\mathbf{\Gamma}_M}=C_3$, and $G_{\mathbf{M}_M}=C_1$.  We get the irrep of $G_{\mathbf{\Gamma}_M}$ at  $\mathbf{\Gamma}$ in the step 1, which is $\mathbf{\Gamma}_1$.  We obtain the representation of $G^A_{\mathbf{M}_M}$, and subsequently that of 
$G_{\mathbf{M}_M}$, for the folded bands through Steps 2 to 5. 
Specifically, $\mathbf{M}_1 \oplus \mathbf{M}_2$ denotes the representation of 
$G^A_{\mathbf{M}_M}$ for the folded bands, whereas $2\mathbf{M}_1$ denotes 
the representation of $G_{\mathbf{M}_M}$ for the folded bands. Therefore, when moir\'e potential is included, the irrep $\Lambda_{\mathbf{M}_M,\alpha}$ of topmost band at $\mathbf{M}_M$ is $\mathbf{M}_1$. }
    \label{fig:set up}
\end{figure}

Based on the above notation, we next will identify all the possible symmetry irreps of the low-energy moir\'e bands in the mBZ, namely, the moiré symmetry data. The moiré symmetry data allow us to straightforwardly extract the topological information of these moir\'e bands according to the TQC \cite{bradlyn2017topological,cano2021band,elcoro2021magnetic} and  symmetry indicator \cite{fu2007topological,fang2012bulk,kruthoff2017topological,po2017symmetry,po2020symmetry,lenggenhager2022universal}. Given the atomic symmetry data and moiré symmetry group, the moiré symmetry data can be obtained in five steps, as illustrated in the Fig. \ref{fig:set up}. In the following, we consider the case where the atomic Hamiltonian comes from VBM, without loss of generality. We focus on the $n_d$  lowest energy moiré bands of $\hat{H}$, and discuss these five steps in detail.

\textbf{Step 1}: The irrep of the $n_d$ moiré bands  at $\mathbf{\Gamma}_M$ in the mBZ is inherited directly from the irrep $\Lambda^{A}_\mathbf{\Gamma}$ of $G^A$ for $\hat{H}^A$ as  $\mathbf{\Gamma}_M$=$\mathbf{\Gamma}$ in the ABZ. In particular, $\Lambda^{A}_\mathbf{\Gamma}$ can be decomposed into the irreps of the little group $G_{\mathbf{\Gamma}_M}$ at $\mathbf{\Gamma}_M$ in the mBZ as
\begin{eqnarray} \label{eq: irrep at Gamma}
 \Lambda^{A}_{\mathbf{\Gamma}} \downarrow  G_{\mathbf{\Gamma}_M}= \oplus_{\alpha} c_\alpha \Lambda_{\mathbf{\Gamma}_M, \alpha},
\end{eqnarray}
where $\Lambda_{\mathbf{\Gamma}_M, \alpha}$ is the $\alpha$th irrep of $G_{\mathbf{\Gamma}_M}$ that is contained in $ \Lambda^{A}_{\mathbf{\Gamma}}$ (\ie, restricted irrep of $ \Lambda^{A}_{\mathbf{\Gamma}}$), 
$c_\alpha $ is the multiplicity of this decomposition, and $\downarrow$ means the restricted representation of  $G_{\mathbf{\Gamma}_M}$. Eq.(\ref{eq: irrep at Gamma}) determines the band splitting and band irreps of $n_d$ moiré bands at $\mathbf{\Gamma}_M$ in the mBZ from the atomic irrep $\Lambda^{A}_\mathbf{\Gamma}$.

\textbf{Step 2}: We now start addressing the possible irreps at other high symmetry points at the mBZ boundary, which come from the band folding. In this step, we list the atomic momenta that fold into the same moiré momentum. For a high symmetry momentum $\kk_{M}$, we label the set of the momenta in the ABZ that are folded into $\kk_{M}$ as
\begin{eqnarray}
  A_{\kk_{M}} =\{\kk_0 \vert \kk_0=\kk_{M} + \mathbf g\}
\end{eqnarray}
with the moir\'e reciprocal lattice vector $\mathbf g$.  Any element of $G^A_{\kk_{M}}$, defined in Eq.~(\ref{eq:little group}), leaves $A_{\kk_{M}}$ invariant. The orbit of $\kk_M$, labelled as $A^{A}_{\kk_M}$, is particularly important and defined by applying all symmetry operators in $G^A_{\kk_{M}}$ to $\kk_M$, namely, 
\begin{equation}\label{eq:momentum shell N0}
A^{A}_{\kk_M} = \{ S {\kk}_M \vert  S\in G^A_{\kk_{M}}\}\ ,
\end{equation} where all elements are guaranteed to lie within the first mBZ or 
on the boundary of the first mBZ.

As any two momenta in $A^{A}_{\kk_{M}}$ are related by symmetry, the eigen-energy spectrum of $\hat{H}^{A}(\kk)$ is the same for any $\kk \in A^{A}_{\kk_{M}}$. With a zero moir\'e potential, all the non-moiré (\ie  atomic) bands at $\kk \in A^{A}_{\kk_{M}}$ will be folded into $\kk_{M}$.

\textbf{Step 3}: 
Clearly, the $n_d$ atomic bands at $\kk_{M}$ may not have the same energy, leading to different sets of moiré bands at $\kk_{M}$ with different energies. It comes from the fact that the symmetry along the path from $\mathbf{\Gamma}_M$ to $\kk_{M}$, which we label as  ${\mathbf{\Gamma}_M \kk_{M}}$, can be lower than that at $\mathbf{\Gamma}_M$, leading to the atomic band splitting. To account for this, we use $G^A_{\mathbf{\Gamma}_M \kk_{M}}$ to label the little group of $G^A\rtimes T$ at any momentum on ${\mathbf{\Gamma}_M \kk_{M}}$, which is given by the union of all $G^A_{\kk}$ with $\kk$ ranging over $\mathbf{\Gamma}_M \kk_{M}$. (Here, $G^A_{\kk}$ is simply given by replacing $\kk_M$ in  {Eq.\,\eqref{eq:little group}} by $\kk$.) Since there are momenta on ${\mathbf{\Gamma}_M \kk_{M}}$ that are not on the mBZ boundary, all elements of $G^A_{\mathbf{\Gamma}_M \kk_{M}}$ simply leave ${\mathbf{\Gamma}_M \kk_{M}}$ invariant without being up to moiré reciprocal lattice vectors.

Naturally, $G^A_{\mathbf{\Gamma}_M \kk_{M}}$ is a subgroup of $G^A_{\mathbf{\Gamma}_{M}}$, and thus we can decompose 
 $\Lambda^{A}_{\mathbf{\Gamma}}$ into the irreps $\Lambda^A_{\mathbf{\Gamma}_M\kk_{M}}$ of the little group $G^A_{\mathbf{\Gamma}_M \kk_{M}}$ as
\begin{eqnarray} \label{eq:irrep high symmetry line}
\Lambda^{A}_{\mathbf{\Gamma}_M} \downarrow G^A_{\mathbf{\Gamma}_M \kk_{M}}= \oplus_{\beta} \Lambda^{A}_{ \mathbf{\Gamma}_M\kk_{M},\beta}\ ,
\end{eqnarray}
where $\beta = (\alpha,\mu)$, $\alpha$ labels the type of irrep, and $\mu$ labels the copies of the same irrep---different copies are furnished by different physical states. Clearly, it is possible that $\Lambda^{A}_{ \mathbf{\Gamma}_M\kk_{M},\beta}$ and $\Lambda^{A}_{ \mathbf{\Gamma}_M\kk_{M},\beta'}$ are the same irreps for $\beta\neq \beta'$, but they are always furnished by different physical states. There are no symmetries that relate states that belong to different $\beta$, and thus it is natural to expect different $\Lambda^{A}_{ \mathbf{\Gamma}_M\kk_{M},\beta}$ correspond to different energies.

\textbf{Step 4}: 
For each copy of irrep $\Lambda^A_{ \mathbf{\Gamma}_M\kk_{M},\beta}$, we can construct the corresponding induced representation $\tilde \Lambda^A_{{\kk}_M}(\beta)$ of $G^A_{\kk_{M}}$ as
\begin{eqnarray}
\label{eq: representation after MBZ folding2}
   \tilde \Lambda^A_{{\kk}_M}(\beta) =\Lambda^A_{ \mathbf{\Gamma}_M\kk_{M},\beta} \uparrow G^A_{\kk_{M}}
\end{eqnarray}
where $\uparrow$ denotes the induction of the representation $\Lambda^A_{ \mathbf{\Gamma}_M\kk_{M},\beta}$ in the little group $G^A_{{\kk}_M}$ defined in Eq.\,\eqref{eq:little group}. This induction process is nothing but the band folding---all moiré bands that belong to
$\tilde{\Lambda}^{A}_{{\kk}_M}(\beta)$ have the same energy as the $\Lambda^A_{ \mathbf{\Gamma}_M\kk_{M},\beta}$ atomic band at $\kk_M$. For any $\beta$ and any high symmetry ${\kk}_M$, the dimension of $\tilde{\Lambda}^{A}_{{\kk}_M}(\beta)$ is 
\begin{eqnarray} \label{eq: dimension of part folding band}
    N[ \tilde \Lambda^{A}_{{\kk}_M,\beta} ]= 
 N[\Lambda^{A}_{ \mathbf{\Gamma}_M\kk_{M},\beta} ]\times 
N[A^{A}_{\kk_M}]\ ,
\end{eqnarray}
where $N[...]$ is the dimension if ``...'' is a representation or the number of elements if ``...'' is a set. If $N[ \tilde \Lambda^{A}_{{\kk}_M,\beta} ] \geq n_d$, we can just choose $\tilde \Lambda^{A}_{{\kk}_M,\beta}$ that corresponds to the lowest energy, which we labeled as $\beta_0$. Here the lowest energy refers to the energy of the topmost valence band  or the bottommost conduction band. If $N[ \tilde \Lambda^{A}_{{\kk}_M,\beta} ] < n_d$, we  just should choose the lowest energy two irreps, $\beta_0$ and $\beta_1$.

\textbf{Step 5}: 
At last, the possible irreps of the low-energy $n_d$ moiré bands at $\kk_M$ are contained in 
\begin{equation}
\label{eq: representation decomposition nt0}
\oplus_{\beta} (\tilde \Lambda^A_{{\kk}_M}(\beta)\downarrow G_{\kk_M} )  = \oplus_{\beta} \oplus_{\alpha} c_{\kk_M,\beta\alpha} \Lambda_{\kk_M, \alpha},
\end{equation}
where $\beta$ only ranges over $\beta_0$ if $N[ \tilde \Lambda^{A}_{{\kk}_M,\beta} ] \geq n_d$ or $\beta_0$ and $\beta_1$ if $N[ \tilde \Lambda^{A}_{{\kk}_M,\beta} ] < n_d$. Here, $\Lambda_{\kk_M, \alpha}$ is the $\alpha$th irrep of $G_{\kk_M}$, and $c_{\kk_M,\alpha}$ is the multiplicity.

Here,we need to emphasize one more point that the irreps at two \textit{different} high symmetry momenta $\kk_{M}$ and $\kk_{M}'$ may be related by $G^{A}$. If there is a symmetry operator $\mathcal{S}_{\kk_{M}}\in G^{A}$ such that  $\mathcal{S}_{\kk_{M}}\kk_{M}={\kk'}_M$, the bands of $\hat{H}^A$ at $\kk_{M}$ and ${\kk'}_M$ have the same energy under zero moir\'e potential and are related by symmetry. This leads to a relationship for the representations of folding bands at $\kk_{M}$ and ${\kk'}_M$, given by
\begin{eqnarray}
     \chi_{\tilde \Lambda^A_{{\kk}_M}(\beta)}( \mathcal{G}_{{\kk}_M})=\chi_{\tilde \Lambda^A_{{ \kk'}_M}(\beta)}( \mathcal{S}_{\kk_{M}} \mathcal{G}_{\kk_{M}} \mathcal{S}^{-1}_{\kk_{M}}) \label{eq: representation after MBZ folding32_1}, 
\end{eqnarray}
where $\mathcal{G_{{\kk}_M}}$ and $\mathcal{S}_{\kk_{M}} \mathcal{G}_{\kk_{M}} \mathcal{S}^{-1}_{\kk_{M}}$ are  symmetry operators in group $G_{  { \kk}_{M}}$ and $G_{  { \kk}'_{M}}$, respectively. One example is the combination of  $G^{A}_{0}=C_6$ and  $G_0=P3$, two distinct high symmetry points $\textbf{K}_M$ and $\textbf{KA}_M$ ($-\textbf{K}_M$) are related by the symmetry $C_{2z}$ in $G^{A}_{0}$. 
The representation formed by the folded band at $\textbf{KA}_M$, 
 $\tilde{\Lambda}^A_{\textbf{KA}_M}(\beta)$, 
is related to the representation at $\textbf{K}_M$, 
 $\tilde{\Lambda}^A_{\textbf{K}_M}(\beta)$, by the Eq.\,\eqref{eq: representation after MBZ folding32_1}.

With the above steps, we can derive all possible moiré symmetry data of the low-energy $n_d$ moiré bands, as long as the moiré potential strength is weaker than the atomic band splitting. We can then compare the moiré symmetry data to the EBR tables in Bilbao Crystallographic Server \cite{aroyo2011crystallography,aroyo2006bilbao,aroyo2006bilbao1} and indicate the topology of the low-energy $n_d$ moiré bands, based on  the TQC \cite{bradlyn2017topological,cano2021band,elcoro2021magnetic} and  symmetry indicator \cite{fu2007topological,fang2012bulk,kruthoff2017topological,po2017symmetry,po2020symmetry,lenggenhager2022universal}.

In the above, we only focus on the space group symmetry and assume $\hat{H}^A$ has no TR symmetry. In a TR invariant system, $G^{A}= \mathcal{T}\rtimes G^{A}_0$ and $G= \mathcal{T}\rtimes G_0$, where $\mathcal{T}$ represents TR symmetry and $G_0$ represents plane group symmetry of $\hat{H}$. With TR, we can still follow exactly step 1 to step 5 to determine the moiré symmetry data of the low-energy $n_d$ moiré bands. Of course, the extra TR symmetry can enlarge the irreps.

\section{Symmetry-Enforced Moiré Topology} \label{sec: Classification}

In this section, we list all combinations of atomic symmetry data and moiré symmetry group that enforce nontrivial topology of low-energy moiré bands, using the method developed in Sec.\ref{sec:moire TQC}.

\subsection{Cases with TR symmetry} \label{sec: Cases with TR}

We start with the cases with TR symmetry. Table I in the main text summarizes all the combinations of $G^A_0$, $\Lambda^A_{\Gamma}$ and $G_0$ that lead to symmetry-enforced moir\'e topology in a spinful TR-invariant system. Table \ref{Tab:Symmetry enfored topology  TR} summarizes all the combinations of $G^A_0$, $\Lambda^A_{\Gamma}$ and $G_0$ that lead to symmetry-enforced moir\'e topology in a spinless TR-invariant system. In spinless systems, all symmetry-enforced topological moiré bands are semimetallic. We find 34 combinations that lead to symmetry-enforced moir\'e topology in the spinful case, and 73  in the spinless case, when there is TR symmetry. Among the 34 spinful cases, we find that there are 11 cases that allow the existence of low-energy isolated topological moir\'e bands in the phase diagram, while all other cases are completely semimetallic. We notice that most non-trivial cases are for 2D irreps. For a 1D spinless irrep $\Lambda^{A}_{\Gamma}$, we can only get a symmetry-enforced topological semimetal when $G_0$ is a nonsymmorphic plane group, in which a single isolated moir\'e band is not allowed \cite{michel1999connectivity,michel2001elementary, watanabe2015filling, vergniory2017graph, watanabe2016filling}.

\begin{figure}
\centering
\includegraphics[width=0.7\textwidth]{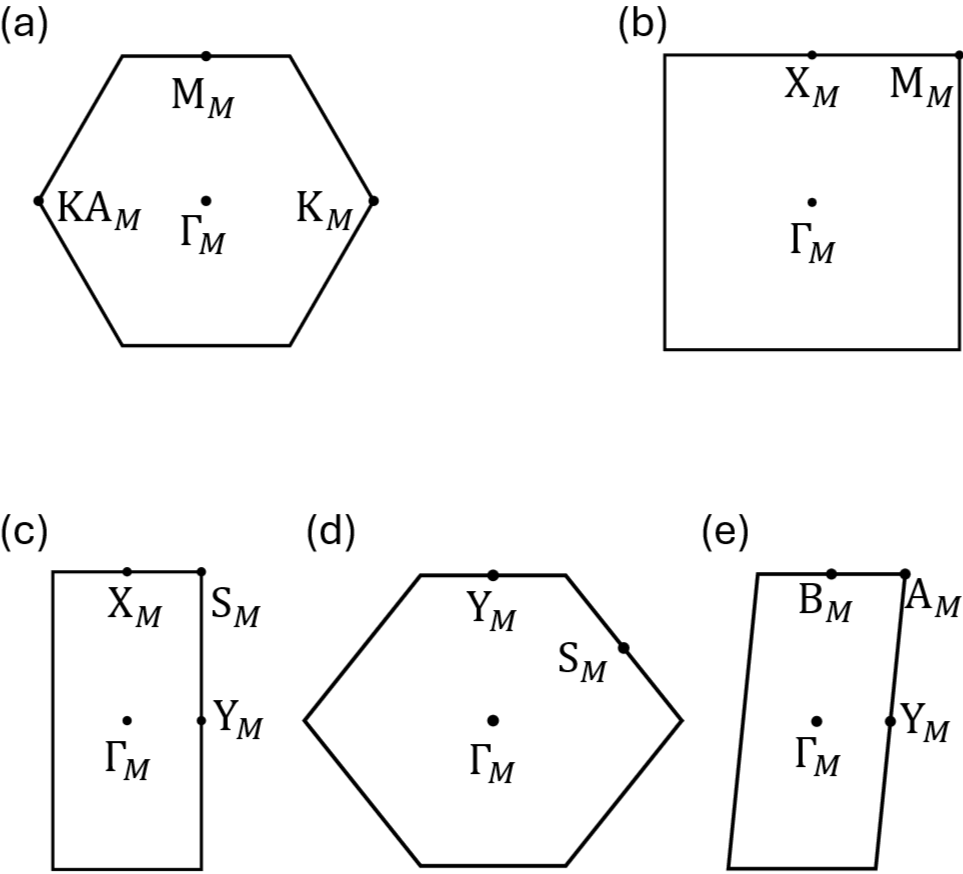}
\caption{\justifying\label{fig:BZ Figure} The mBZs for different plane groups and high symmetry momenta.
(a) shows the mBZs for plane group group $P3$, $P3m1$, $P31m$, $P6$ and $P6mm$. The Bravais lattices are hexagonal. (b) shows the mBZs for plane group $P4$, $P4mm$ and $P4bm$. Bravais lattices of these groups are square (c) shows the mBZ for plane group $pmm2$, $pma2$, $pmb2$, $pb11$ and $pm11$. Bravais lattices of these groups are rectangular. (d) shows the mBZs for plane group $Cmm2$ and $Cmm1$. Bravais lattices of these groups are centered rectangular (e) shows the mBZ for plane group $P2$ and $P1$.  }

\end{figure}

\begin{figure}
\centering
\includegraphics[width=0.35\textwidth]{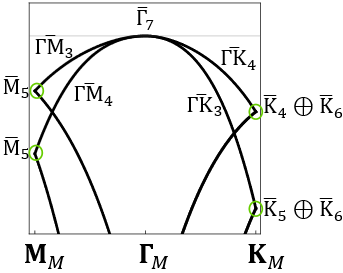}
\caption{\justifying\label{fig:TR_symmetry_enforce} 
(a) The schematic figure of band folding, and band irreps are labeled.  }

\end{figure}

Tables \ref{tab:C6v TR TS}-\ref{tab:C3 TR T} show all the representations at high-symmetry momenta  in the mBZ for a certain $\Lambda^{A}_\mathbf{\Gamma}$, which are constructed by applying the general approach developed in  Sec.\ref{sec:moire TQC} to all combinations of $G^A_0$ and $G_0$ listed in 
Table I in the main text and Table \ref{tab:summary of TR}. We organize these tables according to $G^A_0$, namely, Tables \ref{tab:C6v TR TS}-\ref{tab:C6v TR S} are for $G^A_0=C_{6v}$, Tables \ref{tab:C3V TR TS}-\ref{tab:C3v TR S} are for $G^A_0=C_{3v}$, Tables \ref{tab:C4v TR S0}-\ref{tab:C4v TR S} are for $G^A_0=C_{4v}$, Table \ref{tab:C2v TR S} is for $G^A_0=C_{2v}$, Table \ref{tab:Cs TR S} is for $G^A_0=C_{s}$, Table \ref{tab: C6 TR T} is for $G^A_0=C_{6}$, Table \ref{tab:C4 TR T} is for $G^A_0=C_{4}$ and Table \ref{tab:C3 TR T} is for $G^A_0=C_{3}$.  The charter tables for above $G^A_0$ are shown in Tables \ref{tab: character table point group} and \ref{tab: character table point group rotation}.

As an example to illustrate how to identify the nontrivial cases from these tables, let us look at Table~\ref{tab:C6v TR TS}(a), where $G^A_0 = C_{6v}$ and $G_0 = P6mm$, and the corresponding character tables for all high-symmetry points are shown in Table~\ref{tab: character table P6mm}. In the first row of Table~\ref{tab:C6v TR TS}(a), the first and second entries show the symmetry groups of $G^A_0$ and $G_0$, respectively. One should note that TR also exists for this table, so $G^A=\mathcal{T}\rtimes G^A_0$. Starting from the second row, the first column lists the irreps $\Lambda^{A}_\mathbf{\Gamma}$.
The irreps with a bar correspond to spinful systems, while those without a bar represent spinless systems \cite{aroyo2011crystallography,aroyo2006bilbao,aroyo2006bilbao1}. For irreps with dimension greater than one, we explicitly label their dimension in the bracket. The second column shows the irrep of {$G_{\bf \Gamma_M}$} in a TR-invariant system after taking into account the moir\'e potential, following {\bf Step 1} in Sec.\ref{sec:moire TQC}. The third and fourth columns show the possible representations of the folded bands at the high-symmetry momenta $\mathbf{M}_M$ and $\mathbf{K}_M$ on the mBZ boundary, respectively, obtained by following \textbf{Steps 2--4} in Sec.\ref{sec:moire TQC}. The locations of high-symmetry momenta ${\bf M_M}$ and ${\bf K_M}$ in the mBZ are shown in Fig. \ref{fig:BZ Figure}(a). The irrep $\Lambda^{A}_\mathbf{\Gamma}$ labeled with a star ($\bigstar$) indicates the case with isolated symmetry-enforced topological moir\'e bands,  while the irreps $\Lambda^{A}_\mathbf{\Gamma}$ labeled with a triangle ($\blacktriangle$) 
indicate the cases where the symmetry-enforced topological  moir\'e bands must be semimetallic.

To be more specific, we examine the case with  $\Lambda^{A}_\mathbf{\Gamma}=\bar{\mathbf{\Gamma}}_7$, which corresponds to the basis functions with the out-of-plane total angular momentum $J_z=\pm\frac{3}{2}$. The irrep $\Lambda^{A}_{\mathbf{ \Gamma}_M,\alpha}$ of low-energy moir\'e bands for the little group $G_\Gamma$ at $\mathbf{\Gamma}_M$ only takes one value for $\alpha$ and directly equals to $\Lambda^{A}_\mathbf{\Gamma}$---$\bar{\mathbf\Gamma}_7$ in the second column of Table \ref{tab:C6v TR TS}(a). Along the $\mathbf{\Gamma}_M \textbf{M}_M$ line (with the little group $G^{A}_{\mathbf{\Gamma}_M \mathbf{M}_M}=C_{s}$ and character table in Table~\ref{tab: character table Cs}), the two-fold degenerate $\bar{\mathbf\Gamma}_7$ bands are split into two bands, which belong to irreps $\bar{\mathbf{\Gamma}\mathbf{M}}_3$ and $\bar{\mathbf{\Gamma}\mathbf{M}}_4$  with opposite mirror $M_x$ (flipping $x$)  eigenvalues. Therefore $\Lambda^A_{\mathbf{\Gamma}_M\mathbf{M}_M,\beta_0}$ and $\Lambda^A_{\mathbf{\Gamma}_M\mathbf{M}_M,\beta_1}$ take values from $\bar{\mathbf{\Gamma}\mathbf{M}}_3$ and $\bar{\mathbf{\Gamma}\mathbf{M}}_4$. The bands of $\hat{H}^A$ at $\kk_0$=$\textbf{M}_M$ and -$\textbf{M}_M$ in the ABZ are folded into $\textbf{M}_M$ in the mBZ. According to Eqs.\,\eqref{eq: representation after MBZ folding2} and \eqref{eq: representation decomposition nt0}, regardless of $\Lambda^A_{\mathbf{\Gamma}_M\mathbf{M}_M,\beta_0}$, the irrep formed by the folded band is the 2D irrep $\bar{\mathbf{M}}_5$, as shown in the third column of Table \ref{tab:C6v TR TS}(a). Therefore, we don’t need to consider $\Lambda^A_{\mathbf{\Gamma}_M\mathbf{M}_M,\beta_1}$. Similarly, along the $\mathbf{\Gamma}_M\mathbf{K}_M$ line, the two-fold degenerate $\bar{\mathbf\Gamma}_7$ bands are also split into two bands in this case. Along the $\mathbf{\Gamma}_M\mathbf{K}_M$ line (with the little group again $G^{A}_{\mathbf{\Gamma}_M \mathbf{K}_M}$=$C_{s}$), these two bands belong to irreps $\bar{\mathbf{\Gamma}\mathbf{K}}_3$ and $\bar{\mathbf{\Gamma}\mathbf{K}}_4$  with opposite mirror $M_y$ (flipping y) values. Therefore, $\Lambda^A_{\mathbf{\Gamma}_M\mathbf{K}_M,\beta_0}$ and $\Lambda^A_{\mathbf{\Gamma}_M\mathbf{K}_M,\beta_1}$ take values from $\bar{\mathbf{\Gamma}\mathbf{K}}_3$ and $\bar{\mathbf{\Gamma}\mathbf{K}}_4$. There are three equivalent $\mathbf{K}_M$, denoted as $\kk_0$=$\mathbf{K}_M$, $C_{3z} \mathbf{K}_M$, and $C^{-1}_{3z} \mathbf{K}_M$ in the ABZ, related by the three-fold rotation $C_{3z}$ around the out-of-plane axis. The bands of $\hat{H}^A$ at these three momenta are folded into $\mathbf{K}_M$ in the mBZ and form a three dimensional (3D) representation. Thus, again we don't need to consider $\Lambda^A_{\mathbf{\Gamma}_M\mathbf{K}_M,\beta_1}$. The 3D representation can be decomposed into one 1D irrep and one 2D irrep. 
According to Eqs.\,\eqref{eq: representation after MBZ folding2} and \eqref{eq: representation decomposition nt0}, the decomposition gives rise to $\bar{\mathbf{K}}_4 \oplus \bar{\mathbf{K}}_6$ when $\Lambda^A_{\mathbf{\Gamma}_M\mathbf{K}_M,\beta_0} = \bar{\mathbf{\Gamma}\mathbf{K}}_4$, and yields $\bar{\mathbf{K}}_5 \oplus \bar{\mathbf{K}}_6$ when $\Lambda^A_{\mathbf{\Gamma}_M\mathbf{K}_M,\beta_0} = \bar{\mathbf{\Gamma}\mathbf{K}}_3$, where $\bar{\mathbf{K}}_4$ and $\bar{\mathbf{K}}_5$ are 1D irreps, while $\bar{\mathbf{K}}_6$ is a 2D irrep. 
No matter which case, $\bar{\mathbf{K}}_5 \oplus \bar{\mathbf{K}}_6$ and $\bar{\mathbf{K}}_4 \oplus \bar{\mathbf{K}}_6$ have different energies caused by the atomic band splitting, even if there is no moiré potential. The moir\'e potential will further split the three-fold degenerate states of $\bar{\mathbf{K}}_5 \oplus \bar{\mathbf{K}}_6$ ($\bar{\mathbf{K}}_4 \oplus \bar{\mathbf{K}}_6$), but this moir\'e energy scale splitting is  smaller than the atomic energy scale. Thus, the low-energy two bands should be from either $\bar{\mathbf{K}}_5 \oplus \bar{\mathbf{K}}_6$ or $\bar{\mathbf{K}}_4 \oplus \bar{\mathbf{K}}_6$. 

Without loss of generality, we consider the low-energy moir\'e bands to be from the $\bar{\mathbf{K}}_4 \oplus \bar{\mathbf{K}}_6$ bands, as schematically shown in Fig. \ref{fig:TR_symmetry_enforce} for zero moir\'e potential after band folding. In Section~II of the main text, we demonstrated that under the weak moir\'e condition, the low-energy two bands must be topological. Indeed, if the $\bar{\mathbf{K}}_6$ irrep is kept for the low-energy two bands, we can find they form a topological insulator by comparing with all the 2D EBRs in Table \ref{TAB:EBR} \cite{aroyo2011crystallography,aroyo2006bilbao,aroyo2006bilbao1}, according to the theory of topological quantum chemistry \cite{bradlyn2017topological,cano2021band,elcoro2021magnetic} and symmetry indicator \cite{po2017symmetry,po2020symmetry,kruthoff2017topological}. 
In contrast, if the energy of $\bar{\text{K}}_4$ band is closer to the band gap, two low-energy bands are from one 1D $\bar{\text{K}}_4$ irrep and one component of the 2D $\bar{\text{K}}_6$ irrep. In this case, the second low-energy band must connect to another band at $\mathbf{K}_M$, resulting in a topological semimetal. We can reach the exactly same conclusion if the low-energy moir\'e bands are from the $\bar{\mathbf{K}}_5 \oplus \bar{\mathbf{K}}_6$ bands. In conclusion, when incorporating TR with the combination $G^A_0 = C_{6v}$, $\Lambda^A_{\Gamma} = \bar{\mathbf{\Gamma}}_7$, and $G_0 = P6mm$, the low-energy two moir\'e  bands in this case must be symmetry-enforced topological under the weak moiré condition.

From Table \ref{tab:summary of TR}(a), all isolated symmetry-enforced topological moir\'e bands exhibit two common characteristics. Firstly, the basis states of $\hat{H}^A$ carry the total out-of-plane angular momentum $J_{z}=\pm\frac{3}{2}$. Secondly, the moir\'e  potential maintains $C_{3z}$ symmetry, so that the mBZ is hexagonal. 

\begin{table}[H]
  
\centering
\renewcommand{\arraystretch}{1.2}

\renewcommand{\arraystretch}{1.2}

\centering

\centering
\begin{tabular}{|c|c|c|c|c|c|c|c|c|}
\hline
 \diagbox{$G_0$}{$G^A_0$} & $C_{6v}$ & $C_{4v}$ & $C_{3v}$ & $C_{2v}$ & $C_{s}$ & $C_6$ & $C_4$
\\
\hline
  $P6mm$ & ${\mathbf{\Gamma}}_5$, ${\mathbf{\Gamma}}_6$ & & & & & &
 \\
 $P4mm$ &  & $\mathbf{\Gamma}_5$  &  & & & &
 \\
 $P4bm$ &  & ${\mathbf{\Gamma}}_1$, ${\mathbf{\Gamma}}_2,{\mathbf{\Gamma}}_3$, ${\mathbf{\Gamma}}_4$ &  & & & &
 \\
  $P3m1$ & ${\mathbf{\Gamma}}_5$, ${\mathbf{\Gamma}}_6$ & & ${\mathbf{\Gamma}}_3$ &  & & &
 \\
 $Pmm2$ & ${\mathbf{\Gamma}}_5$, ${\mathbf{\Gamma}}_6$ &$\mathbf{\Gamma}_5$ &  &  & & &
 \\
  $Cmm2$ & ${\mathbf{\Gamma}}_5$, ${\mathbf{\Gamma}}_6$ &$\mathbf{\Gamma}_5$ &  &  & & &
 \\
 $Pba2$ & ${\mathbf{\Gamma}}_1$, ${\mathbf{\Gamma}}_2,{\mathbf{\Gamma}}_3$, ${\mathbf{\Gamma}}_4$ & ${\mathbf{\Gamma}}_1$, ${\mathbf{\Gamma}}_2,{\mathbf{\Gamma}}_3$, ${\mathbf{\Gamma}}_4$  &  & ${\mathbf{\Gamma}}_1$, ${\mathbf{\Gamma}}_2,{\mathbf{\Gamma}}_3$, ${\mathbf{\Gamma}}_4$ &
& &  \\
 $Pma2$ & ${\mathbf{\Gamma}}_1$, ${\mathbf{\Gamma}}_2,{\mathbf{\Gamma}}_3$, ${\mathbf{\Gamma}}_4,{\mathbf{\Gamma}}_5$, ${\mathbf{\Gamma}}_6$ & ${\mathbf{\Gamma}}_1$, ${\mathbf{\Gamma}}_2,{\mathbf{\Gamma}}_3$, ${\mathbf{\Gamma}}_4$, ${\mathbf{\Gamma}}_5$& &${\mathbf{\Gamma}}_1$, ${\mathbf{\Gamma}}_2,{\mathbf{\Gamma}}_3$, ${\mathbf{\Gamma}}_4$& & &
 \\
 $Pm11$ & ${\mathbf{\Gamma}}_5$, ${\mathbf{\Gamma}}_6$ & $\mathbf{\Gamma}_5$  &  ${\mathbf{\Gamma}}_3$ & && &
 \\
 $Pb11$ &  ${\mathbf{\Gamma}}_1$, ${\mathbf{\Gamma}}_2,{\mathbf{\Gamma}}_3$, ${\mathbf{\Gamma}}_4$ & ${\mathbf{\Gamma}}_1$, ${\mathbf{\Gamma}}_2,{\mathbf{\Gamma}}_3$, ${\mathbf{\Gamma}}_4$ & ${\mathbf{\Gamma}}_1$, ${\mathbf{\Gamma}}_2$  & ${\mathbf{\Gamma}}_1$, ${\mathbf{\Gamma}}_2,{\mathbf{\Gamma}}_3$, ${\mathbf{\Gamma}}_4$ &${\mathbf{\Gamma}}_1$, ${\mathbf{\Gamma}}_2$ & &
 \\
$P6$ &  ${\mathbf{\Gamma}}_5$, ${\mathbf{\Gamma}}_6$ & & & &  & $\mathbf{\Gamma}_3\mathbf{\Gamma}_5$, $\mathbf{\Gamma}_4\mathbf{\Gamma}_6$ &  
 \\
$P2$ &  ${\mathbf{\Gamma}}_5$, ${\mathbf{\Gamma}}_6$ & ${\mathbf{\Gamma}}_5$ & & &  & $\mathbf{\Gamma}_3\mathbf{\Gamma}_5$, $\mathbf{\Gamma}_4\mathbf{\Gamma}_6$ & $\mathbf{\Gamma}_3\mathbf{\Gamma}_4$
 \\
  
\hline
\end{tabular}

\caption{\justifying \textbf{Spinless TR-invariant cases for symmetry-enforced moir\'e topology.}
List of combinations of atomic point group $G_0^A$, atomic $\Gamma$ irrep, and moiré plane group $G_0$ that lead to symmetry-enforced nontrivial moiré topology, in the presence of TR symmetry and without SOC. For a given $G_0^A$ and $G_0$, the table entry indicates the required atomic $\Gamma$ irrep to give symmetry-enforced topology.  The low-energy bands are all semimetallic in this case.  \label{Tab:Symmetry enfored topology TR}
}
  \label{tab:summary of TR}
\end{table}

\begin{table}[H]
\begin{subtable}{0.49\textwidth}
\renewcommand{\arraystretch}{1.2}
\centering
\renewcommand{\arraystretch}{1.2}
\begin{tabular}{c|c|c|c}
\hline
$G_0^A = C_{6v}$ & \multicolumn{3}{c}{$G_0 = P6mm$} \\
\hline
$\mathbf{\Gamma}$ & $\mathbf{\Gamma}_M$ & $\textbf{M}_M$ & $\textbf{K}_M$ \\
\hline
$\mathbf{\Gamma}_{1}$ & $\mathbf{\Gamma}_{1}$ & $\textbf{M}_{1} \oplus \textbf{M}_4$ & $\textbf{K}_{1} \oplus \textbf{K}_{3}(2)$ \\
\hline
$\mathbf{\Gamma}_{2}$ & $\mathbf{\Gamma}_{2}$ & $\textbf{M}_{2} \oplus \textbf{M}_3$ & $\textbf{K}_{2} \oplus \textbf{K}_{3}(2)$ \\
\hline
$\mathbf{\Gamma}_{3}$ & $\mathbf{\Gamma}_{3}$ & $\textbf{M}_{2} \oplus \textbf{M}_3$ & $\textbf{K}_{1} \oplus \textbf{K}_{3}(2)$ \\
\hline$
\mathbf{\Gamma}_{4}$ & $\mathbf{\Gamma}_{4}$ & $\textbf{M}_{1} \oplus \textbf{M}_4$ & $\textbf{K}_{2} \oplus \textbf{K}_{3}(2)$ \\
\hline
$\blacktriangle \mathbf{\Gamma}_{5}(2)$ & $\mathbf{\Gamma}_{5}(2)$ & $\textbf{M}_{1} \oplus \textbf{M}_4$ & $\textbf{K}_{1} \oplus \textbf{K}_{3}(2)$ \\
 &  & $\textbf{M}_{2} \oplus \textbf{M}_3$ & $\textbf{K}_{2} \oplus \textbf{K}_{3}(2)$ \\
\hline
  $\blacktriangle \mathbf{\Gamma}_{6}(2)$ & $\mathbf{\Gamma}_{6}(2)$ & $\textbf{M}_{1} \oplus \textbf{M}_4$ & $\textbf{K}_{1} \oplus \textbf{K}_{3}(2)$ \\
 &  & $\textbf{M}_{2} \oplus \textbf{M}_3$ & $\textbf{K}_{2} \oplus \textbf{K}_{3}(2)$ \\
\hline
  $\bigstar \bar{\mathbf{\Gamma}}_{7}(2)$ & $\bar{\mathbf{\Gamma}}_{7}(2)$ & $\bar{\textbf{M}}_{5}(2)$ & $\bar{\textbf{K}}_{4} \oplus \bar{\textbf{K}}_{6}(2)$ \\
 &  &  & $\bar{\textbf{K}}_{5} \oplus \bar{\textbf{K}}_{6}(2)$ \\
\hline
  $\bar{\mathbf{\Gamma}}_{8}(2)$ & $\bar{\mathbf{\Gamma}}_{8}(2)$ & $\bar{\textbf{M}}_{5}(2)$ & $\bar{\textbf{K}}_{4} \oplus \bar{\textbf{K}}_{6}(2)$ \\
 &  &  & $\bar{\textbf{K}}_{5} \oplus \bar{\textbf{K}}_{6}(2)$ \\
\hline
  $\bar{\mathbf{\Gamma}}_{9}(2)$ & $\bar{\mathbf{\Gamma}}_{9}(2)$ & $\bar{\textbf{M}}_{5}(2)$ & $\bar{\textbf{K}}_{4} \oplus \bar{\textbf{K}}_{6}(2)$ \\
 &  &  & $\bar{\textbf{K}}_{5} \oplus \bar{\textbf{K}}_{6}(2)$ \\
\hline
\end{tabular}
\caption{$G_0^A = C_{6v}$ and $G_0 = P6mm$}
\end{subtable}
\begin{subtable}{0.49\textwidth}
\renewcommand{\arraystretch}{1.2}
\centering
\begin{tabular}{c|c|c|c}
\hline
$G_0^A = C_{6v}$ & \multicolumn{3}{c}{$G_0 = P3m1$} \\
\hline

$\mathbf{\Gamma}$ & $\mathbf{\Gamma}_M$ & $\textbf{M}_M$ & $\textbf{K}_M$ \\
\hline
$\mathbf{\Gamma}_{1}$ & $\mathbf{\Gamma}_{1}$ & $2\textbf{M}_{1} $ & $\textbf{K}_{1} \oplus \textbf{K}_{2}\oplus \textbf{K}_{3}$ \\
\hline
$\mathbf{\Gamma}_{2}$ & $\mathbf{\Gamma}_{2}$ & $2\textbf{M}_{2} $ & $\textbf{K}_{1} \oplus \textbf{K}_{2}\oplus \textbf{K}_{3}$ \\
\hline
$\mathbf{\Gamma}_{3}$ & $\mathbf{\Gamma}_{2}$ & $2\textbf{M}_{2} $ & $\textbf{K}_{1} \oplus \textbf{K}_{2}\oplus \textbf{K}_{3}$ \\
\hline

$
\mathbf{\Gamma}_{4}$ & $\mathbf{\Gamma}_{1}$ & $2\textbf{M}_{1} $ & $\textbf{K}_{1} \oplus \textbf{K}_{2}\oplus \textbf{K}_{3}$ \\

\hline
  $\blacktriangle \mathbf{\Gamma}_{5}(2)$ & $\mathbf{\Gamma}_{3}(2)$ & $2\textbf{M}_{1} $ & $\textbf{K}_{1} \oplus \textbf{K}_{2} \oplus \textbf{K}_{3}$ \\
  & & $2\textbf{M}_{2} $ & \\
\hline
  $\blacktriangle \mathbf{\Gamma}_{6}(2)$ & $\mathbf{\Gamma}_{3}(2)$ & $2\textbf{M}_{1} $ & $\textbf{K}_{1} \oplus \textbf{K}_{2} \oplus \textbf{K}_{3}$ \\
  & & $2\textbf{M}_{2} $ & \\
\hline
  $\bigstar \bar{\mathbf{\Gamma}}_{7}(2)$ & $\bar{\mathbf{\Gamma}}_{4}  \bar{\mathbf{\Gamma}}_{5}(2)$ & $\bar{\textbf{M}}_{3} \bar{\textbf{M}}_{4}(2)$ & $\bar{\textbf{K}}_{4} \oplus \bar{\textbf{K}}_{5} \oplus \bar{\textbf{K}}_{6}$ \\
\hline
  $\bar{\mathbf{\Gamma}}_{8}(2)$ & $\bar{\mathbf{\Gamma}}_{6}(2)$ & $\bar{\textbf{M}}_{3} \bar{\textbf{M}}_{4}(2)$ & $\bar{\textbf{K}}_{4} \oplus \bar{\textbf{K}}_{5} \oplus \bar{\textbf{K}}_{6}$ \\
\hline
  $\bar{\mathbf{\Gamma}}_{9}(2)$ & $\bar{\mathbf{\Gamma}}_{6}(2)$ & $\bar{\textbf{M}}_{3} \bar{\textbf{M}}_{4}(2)$ & $\bar{\textbf{K}}_{4} \oplus \bar{\textbf{K}}_{5} \oplus \bar{\textbf{K}}_{6}$ \\
\hline
\end{tabular}
\caption{$G_0^A = C_{6v}$ and $G_0 = P3m1$}
\end{subtable}
\hfill
\begin{subtable}{0.49\textwidth}
\renewcommand{\arraystretch}{1.2}
\centering
\begin{tabular}{c|c|c|c}
\hline
$G_0^A = C_{6v}$ & \multicolumn{3}{c}{$G_0 = P6$} \\
\hline
$\mathbf{\Gamma}$ & $\mathbf{\Gamma}_M$ & $\textbf{M}_M$ & $\textbf{K}_M$ \\
\hline
$\mathbf{\Gamma}_{1}$ & $\mathbf{\Gamma}_{1}$ & $\textbf{M}_{1} \oplus \textbf{M}_{2}$ & $\textbf{K}_{1} \oplus \textbf{K}_{2}\oplus \textbf{K}_{3}$ \\
\hline
$\mathbf{\Gamma}_{2}$ & $\mathbf{\Gamma}_{1}$ & $\textbf{M}_{1} \oplus \textbf{M}_{2}$ & $\textbf{K}_{1} \oplus \textbf{K}_{2}\oplus \textbf{K}_{3}$ \\
\hline
$\mathbf{\Gamma}_{3}$ & $\mathbf{\Gamma}_{2}$ & $\textbf{M}_{1} \oplus \textbf{M}_{2}$ & $\textbf{K}_{1} \oplus \textbf{K}_{2}\oplus \textbf{K}_{3}$ \\
\hline

$
\mathbf{\Gamma}_{4}$ & $\mathbf{\Gamma}_{2}$ & $\textbf{M}_{1} \oplus \textbf{M}_{2}$& $\textbf{K}_{1} \oplus \textbf{K}_{2}\oplus \textbf{K}_{3}$ \\ \hline
  $ \blacktriangle\mathbf{\Gamma}_{5}(2)$ & $\mathbf{\Gamma}_{3} \mathbf{\Gamma}_{5}(2)$ & $\textbf{M}_{1} \oplus \textbf{M}_{2}$ & $\textbf{K}_{1} \oplus \textbf{K}_{2} \oplus \textbf{K}_{3}$ \\ \hline
  $ \blacktriangle\mathbf{\Gamma}_{6}(2)$ & $\mathbf{\Gamma}_{4} \mathbf{\Gamma}_{6}(2)$ & $\textbf{M}_{1} \oplus \textbf{M}_{2}$ & $\textbf{K}_{1} \oplus \textbf{K}_{2} \oplus \textbf{K}_{3}$ \\ \hline
  $\bigstar \bar{\mathbf{\Gamma}}_{7}(2)$ & $\bar{\mathbf{\Gamma}}_{7} \bar{\mathbf{\Gamma}}_{8}(2)$ & $\bar{\textbf{M}}_{3} \bar{\textbf{M}}_{4}(2)$ & $\bar{\textbf{K}}_{4} \oplus \bar{\textbf{K}}_{5} \oplus \bar{\textbf{K}}_{6}$ \\ \hline
  $\bar{\mathbf{\Gamma}}_{8}(2)$ & $\bar{\mathbf{\Gamma}}_{9}  \bar{\mathbf{\Gamma}}_{12}(2)$ & $\bar{\textbf{M}}_{3} \bar{\textbf{M}}_{4}(2)$ & $\bar{\textbf{K}}_{4} \oplus \bar{\textbf{K}}_{5} \oplus \bar{\textbf{K}}_{6}$ \\ \hline
  $\bar{\mathbf{\Gamma}}_{9}(2)$ & $\bar{\mathbf{\Gamma}}_{10} \bar{\mathbf{\Gamma}}_{11}(2)$ & $\bar{\textbf{M}}_{3} \bar{\textbf{M}}_{4}(2)$ & $\bar{\textbf{K}}_{4} \oplus \bar{\textbf{K}}_{5} \oplus \bar{\textbf{K}}_{6}$ \\ 
\hline
\end{tabular}
\caption{$G_0^A = C_{6v}$ and $G_0 = P6$}
\end{subtable}
\caption{\justifying 
With TR and $G^A_0=C_{6v}$, for the space groups $G_0$ of the moir\'e system listed above, there are two possible situations with nontrivial moir\'e topology: 
(i) for a certain irrep of $G^A_0$, symmetry-enforced moir\'e topology must occur  and the low-energy moir\'e bands can be topologically insulating; 
(ii) for a certain irrep, the low-energy $n_d$ moiré bands are necessarily symmetry-enforced semimetallic.  Irreps denoted with a bar (e.g., $\Bar{\mathbf{\Gamma}}_7$) are spinful, while those without a bar are spinless. The number 2 in the bracket after the irrep indicates the dimension of 2 for the irrep, otherwise they are 1D. The first column lists the irreps of the atomic symmetry group $G^A$ at the $\mathbf{\Gamma}$ point. The corresponding entries show the possible representation from band folding at high-symmetry momenta of space group $G_0$ with TR. If a cell contains multiple rows, it indicates that multiple representations are allowed for the folded bands at that high-symmetry momentum. For example, for the irrep $\Bar{\mathbf{\Gamma}}_7$ of $G^A$ and $G_0 = P6mm$, the representation of folded bands at the $\mathbf{K}_M$ point can be $\bar{\textbf{K}}_5 \oplus \bar{\textbf{K}}_6(2)$ or $\bar{\textbf{K}}_4 \oplus \bar{\textbf{K}}_6(2)$.  The star  ($\bigstar$) label indicates the topologically insulating low-energy moir\'e bands, while the triangle ($\blacktriangle$) label indicates guaranteed semi-metallic low-energy moir\'e bands.
\label{tab:C6v TR TS} }
\end{table}
\begin{table}[H]
\centering
\begin{subtable}{0.49\textwidth}
\centering
\renewcommand{\arraystretch}{1.2}
\begin{tabular}{c|c|c|c}
\hline
$G_0^A = C_{6v}$ & \multicolumn{3}{c}{$G_0 = P31m$} \\
\hline
$\mathbf{\Gamma}$ & $\mathbf{\Gamma}_M$ & $\textbf{M}_M$ & $\textbf{K}_M(\textbf{KA}_M)$ \\
\hline
$\mathbf{\Gamma}_{1}$ & $\mathbf{\Gamma}_{1}$ & $\textbf{M}_{1}\oplus\textbf{M}_{2} $ & $\textbf{K}_{1}\oplus \textbf{K}_{3}(2)$ $(\textbf{KA}_{1}\oplus \textbf{KA}_{3}(2))$ \\
\hline
$\mathbf{\Gamma}_{2}$ & $\mathbf{\Gamma}_{2}$ & $\textbf{M}_{1}\oplus\textbf{M}_{2} $ & $\textbf{K}_{2}\oplus \textbf{K}_{3}(2)$ $(\textbf{KA}_{2}\oplus \textbf{KA}_{3}(2))$ \\
\hline
$\mathbf{\Gamma}_{3}$ & $\mathbf{\Gamma}_{1}$ & $\textbf{M}_{1}\oplus\textbf{M}_{2} $ & $\textbf{K}_{1}\oplus \textbf{K}_{3}(2)$ $(\textbf{KA}_{1}\oplus \textbf{KA}_{3}(2))$ \\
\hline

$
\mathbf{\Gamma}_{4}$ & $\mathbf{\Gamma}_{2}$ & $\textbf{M}_{1}\oplus\textbf{M}_{2} $ & $\textbf{K}_{2}\oplus \textbf{K}_{3}(2)$ $(\textbf{KA}_{2}\oplus \textbf{KA}_{3}(2))$ \\
\hline
  $\mathbf{\Gamma}_{5}(2)$ & $\mathbf{\Gamma}_{3}(2)$ & $\textbf{M}_{1}\oplus \textbf{M}_{2}$ & $\textbf{K}_{1}\oplus \textbf{K}_{3}(2)$ $(\textbf{KA}_{1}\oplus \textbf{KA}_{3}(2))$ \\
  & & & $\textbf{K}_{2}\oplus \textbf{K}_{3}(2)$ $(\textbf{KA}_{2}\oplus \textbf{KA}_{3}(2))$ \\
\hline
  $\mathbf{\Gamma}_{6}(2)$ & $\mathbf{\Gamma}_{3}(2)$ & $\textbf{M}_{1}\oplus \textbf{M}_{2}$ & $\textbf{K}_{1}\oplus \textbf{K}_{3}(2)$ $(\textbf{KA}_{1}\oplus \textbf{KA}_{3}(2))$ \\
  & & & $\textbf{K}_{2}\oplus \textbf{K}_{3}(2)$ $(\textbf{KA}_{2}\oplus \textbf{KA}_{3}(2))$ \\
\hline
  $\bigstar \bar{\mathbf{\Gamma}}_{7}(2)$ & $\bar{\mathbf{\Gamma}}_{4}\bar{\mathbf{\Gamma}}_{5}(2)$ & $\bar{\textbf{M}}_{3}\bar{\textbf{M}}_{4}(2)$ & $\bar{\textbf{K}}_{4}\oplus \bar{\textbf{K}}_{6}(2)$ $(\bar{\textbf{KA}}_{4}\oplus \bar{\textbf{KA}}_{6}(2))$ \\
  & & & $\bar{\textbf{K}}_{5}\oplus \bar{\textbf{K}}_{6}(2)$ $(\bar{\textbf{KA}}_{5}\oplus \bar{\textbf{KA}}_{6}(2))$ \\
\hline
  $\bar{\mathbf{\Gamma}}_{8}(2)$ & $\bar{\mathbf{\Gamma}}_{6}(2)$ & $\bar{\textbf{M}}_{3}\bar{\textbf{M}}_{4}(2)$ & $\bar{\textbf{K}}_{4}\oplus \bar{\textbf{K}}_{6}(2)$ $(\bar{\textbf{KA}}_{4}\oplus \bar{\textbf{KA}}_{6}(2))$ \\
  & & & $\bar{\textbf{K}}_{5}\oplus \bar{\textbf{K}}_{6}(2)$ $(\bar{\textbf{KA}}_{5}\oplus \bar{\textbf{KA}}_{6}(2))$ \\
\hline
  $\bar{\mathbf{\Gamma}}_{9}(2)$ & $\bar{\mathbf{\Gamma}}_{6}(2)$ & $\bar{\textbf{M}}_{3}\bar{\textbf{M}}_{4}(2)$ & $\bar{\textbf{K}}_{4}\oplus \bar{\textbf{K}}_{6}(2)$ $(\bar{\textbf{KA}}_{4}\oplus \bar{\textbf{KA}}_{6}(2))$ \\
  & & & $\bar{\textbf{K}}_{5}\oplus \bar{\textbf{K}}_{6}(2)$ $(\bar{\textbf{KA}}_{5}\oplus \bar{\textbf{KA}}_{6}(2))$ \\
\hline
\end{tabular}
\caption{$G_0^A = C_{6v}$ and $G_0 = P31m$}
\end{subtable}
\centering
\begin{subtable}{1\textwidth}
\renewcommand{\arraystretch}{1.2}
\centering
\begin{tabular}{c|c|c|c}
\hline
$G_0^A = C_{6v}$ & \multicolumn{3}{c}{$G_0 = P3$} \\
\hline
$\mathbf{\Gamma}$ & $\mathbf{\Gamma}_M$ & $\textbf{M}_M$ & $\textbf{K}_M(\textbf{KA}_M)$ \\
\hline
$\mathbf{\Gamma}_{1}$ & $\mathbf{\Gamma}_{1}$ & $2\textbf{M}_{1} $ & $\textbf{K}_{1}\oplus \textbf{K}_{2}\oplus \textbf{K}_{3}$ $(\textbf{KA}_{1}\oplus \textbf{KA}_{2}\oplus \textbf{KA}_{3})$  \\
\hline
$\mathbf{\Gamma}_{2}$ & $\mathbf{\Gamma}_{1}$ & $2\textbf{M}_{1} $  & $\textbf{K}_{1}\oplus \textbf{K}_{2}\oplus \textbf{K}_{3}$ $(\textbf{KA}_{1}\oplus \textbf{KA}_{2}\oplus \textbf{KA}_{3})$  \\
\hline
$\mathbf{\Gamma}_{3}$ & $\mathbf{\Gamma}_{1}$ & $2\textbf{M}_{1} $  & $\textbf{K}_{1}\oplus \textbf{K}_{2}\oplus \textbf{K}_{3}$ $(\textbf{KA}_{1}\oplus \textbf{KA}_{2}\oplus \textbf{KA}_{3})$  \\
\hline

$
\mathbf{\Gamma}_{4}$ & $\mathbf{\Gamma}_{1}$ & $2\textbf{M}_{1} $ & $\textbf{K}_{1}\oplus \textbf{K}_{2}\oplus \textbf{K}_{3}$ $(\textbf{KA}_{1}\oplus \textbf{KA}_{2}\oplus \textbf{KA}_{3})$  \\
\hline
  $\mathbf{\Gamma}_{5}(2)$ & $\mathbf{\Gamma}_{2}\mathbf{\Gamma}_{3}(2)$ & $2\textbf{M}_{1} $  & $\textbf{K}_{1}\oplus \textbf{K}_{2}\oplus \textbf{K}_{3}$ $(\textbf{KA}_{1}\oplus \textbf{KA}_{2}\oplus \textbf{KA}_{3})$  \\
\hline
  $\mathbf{\Gamma}_{6}(2)$ & $\mathbf{\Gamma}_{2}\mathbf{\Gamma}_{3}(2)$ & $2\textbf{M}_{1} $ & $\textbf{K}_{1}\oplus \textbf{K}_{2}\oplus \textbf{K}_{3}$ $(\textbf{KA}_{1}\oplus \textbf{KA}_{2}\oplus \textbf{KA}_{3})$ \\
\hline
  $\bigstar \bar{\mathbf{\Gamma}}_{7}(2)$ & $\bar{\mathbf{\Gamma}}_{4}\bar{\mathbf{\Gamma}}_{4}(2)$ & $\bar{\textbf{M}}_{2}\bar{\textbf{M}}_{2}(2)$ & $\bar{\textbf{K}}_{4}\oplus \bar{\textbf{K}}_{5}\oplus \bar{\textbf{K}}_{6}$ $(\bar{\textbf{KA}}_{4}\oplus \bar{\textbf{KA}}_{5}\oplus \bar{\textbf{KA}}_{6})$ \\
\hline
  $\bar{\mathbf{\Gamma}}_{8}(2)$ & $\bar{\mathbf{\Gamma}}_{5}\bar{\mathbf{\Gamma}}_{6}(2)$ & $\bar{\textbf{M}}_{2}\bar{\textbf{M}}_{2}(2)$ & $\bar{\textbf{K}}_{4}\oplus \bar{\textbf{K}}_{5}\oplus \bar{\textbf{K}}_{6}$ $(\bar{\textbf{KA}}_{4}\oplus \bar{\textbf{KA}}_{5}\oplus \bar{\textbf{KA}}_{6})$ \\
\hline
  $\bar{\mathbf{\Gamma}}_{9}(2)$ & $\bar{\mathbf{\Gamma}}_{5}\bar{\mathbf{\Gamma}}_{6}(2)$ & $\bar{\textbf{M}}_{2}\bar{\textbf{M}}_{2}(2)$ & $\bar{\textbf{K}}_{4}\oplus \bar{\textbf{K}}_{5}\oplus \bar{\textbf{K}}_{6}$ $(\bar{\textbf{KA}}_{4}\oplus \bar{\textbf{KA}}_{5}\oplus \bar{\textbf{KA}}_{6})$ \\
\hline
\end{tabular}
\caption{$G_0^A = C_{6v} $ and $G_0 = P3$}
\end{subtable}

\caption{\justifying With TR and $G^A_0 = C_{6v}$, for the space groups $G_0$ listed above, there exists at least one irrep for symmetry-enforced topologically insulating moir\'e bands. In the last columns of panels (a) and (b), we indicate $\textbf{KA}_M$ in parentheses after $\textbf{K}_M$ to highlight that the $\mathbf{K}_M$ and $\mathbf{KA}_M$ points are related by symmetry. For example, for the irrep $\Bar{\mathbf{\Gamma}}_7$ of $G^A$ and $G_0 = P31m$, if the representation of folded bands at $\mathbf{K}_M$ is $\bar{\mathbf{K}}_5 \oplus \bar{\mathbf{K}}_6(2)$, the representation at $\mathbf{KA}_M$ is $\bar{\mathbf{KA}}_5 \oplus \bar{\mathbf{KA}}_6(2)$.
 \label{tab:C6v TR T} }
\end{table}

\begin{table}[H]

\begin{subtable}{0.49\textwidth}
\centering
\renewcommand{\arraystretch}{1.2}
\begin{tabular}{c|c|c|c|c}
\hline
$G_0^A = C_{6v}$ & \multicolumn{4}{c}{$G_0 = Pmm2$} \\
\hline
$\mathbf{\Gamma}$ & $\mathbf{\Gamma}_M$ & $\textbf{Y}_M$ & $\textbf{X}_M$ & $\textbf{S}_M$ \\
\hline
$ \mathbf{\Gamma}_{1}$ & $\mathbf{\Gamma}_{1}$ & $\textbf{Y}_{1}\oplus \textbf{Y}_{3}$ & $\textbf{X}_{1}\oplus \textbf{X}_{4}$ & $\textbf{S}_{1} \oplus \textbf{S}_{2} \oplus \textbf{S}_{3} \oplus \textbf{S}_{4}$  \\
\hline
 $\mathbf{\Gamma}_{2}$& $\mathbf{\Gamma}_{2}$ & $\textbf{Y}_{2}\oplus \textbf{Y}_{4}$ & $\textbf{X}_{2}\oplus \textbf{X}_{3}$ & $\textbf{S}_{1} \oplus \textbf{S}_{2} \oplus \textbf{S}_{3} \oplus \textbf{S}_{4}$  \\
\hline
 $\mathbf{\Gamma}_{3}$& $\mathbf{\Gamma}_{3}$ & $\textbf{Y}_{1}\oplus \textbf{Y}_{3}$ & $\textbf{X}_{1}\oplus \textbf{X}_{4}$ & $\textbf{S}_{1} \oplus \textbf{S}_{2} \oplus \textbf{S}_{3} \oplus \textbf{S}_{4}$  \\
\hline
 $\mathbf{\Gamma}_{4}$& $\mathbf{\Gamma}_{4}$ & $\textbf{Y}_{2}\oplus \textbf{Y}_{4}$ & $\textbf{X}_{2}\oplus \textbf{X}_{3}$ & $\textbf{S}_{1} \oplus \textbf{S}_{2} \oplus \textbf{S}_{3} \oplus \textbf{S}_{4}$  \\
\hline
  $\blacktriangle \mathbf{\Gamma}_{5}(2)$ & $\mathbf{\Gamma}_{1}\oplus\mathbf{\Gamma}_{2}$ & $\textbf{Y}_{1} \oplus \textbf{Y}_{3}$ & $\textbf{X}_{1} \oplus \textbf{X}_{4}$ & $\textbf{S}_{1} \oplus \textbf{S}_{2} \oplus \textbf{S}_{3} \oplus \textbf{S}_{4}$ \\
  & & $\textbf{Y}_{2} \oplus \textbf{Y}_{4}$ & $\textbf{X}_{2} \oplus \textbf{X}_{3}$ & \\
\hline
\end{tabular}
\caption{$G_0^A = C_{6v}$ and $G_0 = Pmm2$}
\end{subtable}
\begin{subtable}{0.49\textwidth}
\centering
\renewcommand{\arraystretch}{1.2}
\begin{tabular}{c|c|c|c|c}
\hline
$G_0^A = C_{6v}$ & \multicolumn{4}{c}{$G_0 = Pm11$} \\
\hline
$\mathbf{\Gamma}$ & $\mathbf{\Gamma}_M$ & $\textbf{Y}_M$ & $\textbf{X}_M$ & $\textbf{S}_M$ \\
\hline
$ \mathbf{\Gamma}_{1}$ & $\mathbf{\Gamma}_{1}$& $\textbf{Y}_{1} \oplus \textbf{Y}_{2}$ & $2\textbf{X}_{1}$ & $2\textbf{S}_{1} \oplus 2\textbf{S}_{2}$\\
\hline
 $\mathbf{\Gamma}_{2}$ & $\mathbf{\Gamma}_{2}$& $\textbf{Y}_{1} \oplus \textbf{Y}_{2}$ & $2\textbf{X}_{2}$ & $2\textbf{S}_{1} \oplus 2\textbf{S}_{2}$\\
\hline
 $\mathbf{\Gamma}_{3}$ & $\mathbf{\Gamma}_{1}$ & $\textbf{Y}_{1} \oplus \textbf{Y}_{2}$ & $2\textbf{X}_{1}$ & $2\textbf{S}_{1} \oplus 2\textbf{S}_{2}$ \\
\hline
 $\mathbf{\Gamma}_{4}$ & $\mathbf{\Gamma}_{2}$ & $\textbf{Y}_{1} \oplus \textbf{Y}_{2}$ & $2\textbf{X}_{2}$ & $2\textbf{S}_{1} \oplus 2\textbf{S}_{2}$ \\
\hline
  $\blacktriangle \mathbf{\Gamma}_{5}(2)$ & $\mathbf{\Gamma}_{1}\oplus\mathbf{\Gamma}_{2}$ & $\textbf{Y}_{1} \oplus \textbf{Y}_{2}$ & $2\textbf{X}_{1}$ & $2\textbf{S}_{1} \oplus 2\textbf{S}_{2}$ \\
  & & & $2\textbf{X}_{2}$ & \\
\hline
\end{tabular}
\caption{$G_0^A = C_{6v}$ and $G_0 = Pm11$}
\end{subtable} 
\begin{subtable}{0.49\textwidth}
\centering
\renewcommand{\arraystretch}{1.2}
\begin{tabular}{c|c|c|c}
\hline
$G_0^A = C_{6v}$ &\multicolumn{3}{c}{ $G = C2mm$} \\
\hline
$\mathbf{\Gamma}$ & $\mathbf{\Gamma}_M$ & $\textbf{Y}_M$ & $\textbf{S}_M$ \\
\hline
$ \mathbf{\Gamma}_{1}$ & $\mathbf{\Gamma}_{1}$ & $\textbf{Y}_{1}\oplus \textbf{Y}_{4}$  & $\textbf{S}_{1} \oplus \textbf{S}_{2}$  \\
\hline
 $\mathbf{\Gamma}_{2}$& $\mathbf{\Gamma}_{2}$ & $\textbf{Y}_{2}\oplus \textbf{Y}_{3}$  & $\textbf{S}_{1} \oplus \textbf{S}_{2} $  \\
\hline
 $\mathbf{\Gamma}_{3}$& $\mathbf{\Gamma}_{3}$ & $\textbf{Y}_{2}\oplus \textbf{Y}_{3}$  & $\textbf{S}_{1} \oplus \textbf{S}_{2} $  \\
\hline
 $\mathbf{\Gamma}_{4}$& $\mathbf{\Gamma}_{4}$ & $\textbf{Y}_{1}\oplus \textbf{Y}_{4}$  & $\textbf{S}_{1} \oplus \textbf{S}_{2}$  \\
\hline
  $\blacktriangle \mathbf{\Gamma}_{5}(2)$ & $\mathbf{\Gamma}_{1}\oplus\mathbf{\Gamma}_{2}$ & $\textbf{Y}_{1}\oplus \textbf{Y}_{4}$ & $\textbf{S}_{1}\oplus \textbf{S}_{2}$ \\
  & & $\textbf{Y}_{2}\oplus \textbf{Y}_{3}$ & \\
\hline
  $\blacktriangle \mathbf{\Gamma}_{6}(2)$ & $\mathbf{\Gamma}_{3}\oplus\mathbf{\Gamma}_{4}$ & $\textbf{Y}_{2}\oplus \textbf{Y}_{3}$ & $\textbf{S}_{1}\oplus \textbf{S}_{2}$ \\
  & & $\textbf{Y}_{1}\oplus \textbf{Y}_{4}$ & \\
\hline
  $\mathbf{\Bar{\mathbf{\Gamma}}}_{7}(2)$ & $\mathbf{\Bar{\mathbf{\Gamma}}}_{5}(2)$ & $\bar{\textbf{Y}}_{5}(2)$ & $\bar{\textbf{S}}_{3} \bar{\textbf{S}}_{4}(2)$ \\
\hline
  $\mathbf{\Bar{\mathbf{\Gamma}}}_{8}(2)$ & $\mathbf{\Bar{\mathbf{\Gamma}}}_{5}(2)$ & $\bar{\textbf{Y}}_{5}(2)$ & $\bar{\textbf{S}}_{3}\bar{\textbf{S}}_{4}(2)$ \\
\hline
  $\mathbf{\Bar{\mathbf{\Gamma}}}_{9}(2)$ & $\mathbf{\Bar{\mathbf{\Gamma}}}_{5}(2)$ & $\bar{\textbf{Y}}_{5}(2)$ & $\bar{\textbf{S}}_{3}\bar{\textbf{S}}_{4}(2)$ \\
\hline
\end{tabular}
\caption{$G_0^A = C_{6v}$ and $G_0 = C2mm$}
\end{subtable}
\begin{subtable}{0.49\textwidth}
\centering
\renewcommand{\arraystretch}{1.2}
\begin{tabular}{c|c|c|c|c}
\hline
$G_0^A = C_{6v}$ & \multicolumn{4}{c}{$G_0 = P2$} \\
\hline
$\mathbf{\Gamma}$ & $\mathbf{\Gamma}_M$ & $\textbf{Y}_M$ & $\textbf{A}_M$& $\textbf{B}_M$ \\
\hline
$\mathbf{\Gamma}_{1}$ & $\mathbf{\Gamma}_{1}$  & $\textbf{Y}_{1} \oplus \textbf{Y}_2$ & $\textbf{A}_{1} \oplus \textbf{A}_2$ & $\textbf{B}_{1} \oplus \textbf{B}_2$ \\
\hline
$\mathbf{\Gamma}_{2}$ & $\mathbf{\Gamma}_{1}$ & $\textbf{Y}_{1} \oplus \textbf{Y}_2$ & $\textbf{A}_{1} \oplus \textbf{A}_2$ & $\textbf{B}_{1} \oplus \textbf{B}_2$ \\
\hline
$\mathbf{\Gamma}_{3}$ & $\mathbf{\Gamma}_{2}$ & $\textbf{Y}_{1} \oplus \textbf{Y}_2$ & $\textbf{A}_{1} \oplus \textbf{A}_2$ & $\textbf{B}_{1} \oplus \textbf{B}_2$ \\
\hline
$
\mathbf{\Gamma}_{4}$ & $\mathbf{\Gamma}_{2}$& $\textbf{Y}_{1} \oplus \textbf{Y}_2$ & $\textbf{A}_{1} \oplus \textbf{A}_2$ & $\textbf{B}_{1} \oplus \textbf{B}_2$ \\ \hline
  $\blacktriangle  \mathbf{\Gamma}_{5}(2)$ & $2\mathbf{\Gamma}_{1}$ & $\textbf{Y}_{1} \oplus \textbf{Y}_2$ & $\textbf{A}_{1} \oplus \textbf{A}_2$ & $\textbf{B}_{1} \oplus \textbf{B}_2$  \\
\hline
  $\blacktriangle \mathbf{\Gamma}_{6}(2)$  & $2\mathbf{\Gamma}_{2}$ & $\textbf{Y}_{1} \oplus \textbf{Y}_2$ & $\textbf{A}_{1} \oplus \textbf{A}_2$ & $\textbf{B}_{1} \oplus \textbf{B}_2$  \\
\hline
\end{tabular}
\caption{$G_0^A = C_{6v}$ and $G_0 = P2$}
\end{subtable}
\caption{\justifying With TR and $G^A_0 = C_{6v}$,  for the symmorphic space groups $G_0$ listed above, there exists at least one irrep for which the low-energy  moiré bands are necessarily symmetry-enforced semimetallic.
 \label{tab:C6v TR S0} }
\end{table}
\begin{table}[H]
\begin{subtable}{0.49\textwidth}
\centering
\renewcommand{\arraystretch}{1.2}
\begin{tabular}{c|c|c|c|c}
\hline
$G_0^A = C_{6v}$ & \multicolumn{4}{c}{$G_0 = Pba2$} \\
\hline
$\mathbf{\Gamma}$ & $\mathbf{\Gamma}_M$ & $\textbf{Y}_M$ & $\textbf{X}_M$ & $\textbf{S}_M$ \\
\hline
$\blacktriangle \mathbf{\Gamma}_{1}$ & $\mathbf{\Gamma}_{1}$ & $\textbf{Y}_{1}(2)$ & $\textbf{X}_{1}(2)$ & $\textbf{S}_{1}\textbf{S}_{2}(2) \oplus \textbf{S}_{3}\textbf{S}_{4}(2)$  \\
\hline
 $\blacktriangle \mathbf{\Gamma}_{2}$& $\mathbf{\Gamma}_{2}$ & $\textbf{Y}_{1}(2)$ & $\textbf{X}_{1}(2)$ &  $\textbf{S}_{1}\textbf{S}_{2}(2) \oplus \textbf{S}_{3}\textbf{S}_{4}(2)$  \\
\hline
 $\blacktriangle \mathbf{\Gamma}_{3}$& $\mathbf{\Gamma}_{3}$ & $\textbf{Y}_{1}(2)$ & $\textbf{X}_{1}(2)$ &  $\textbf{S}_{1}\textbf{S}_{2}(2) \oplus \textbf{S}_{3}\textbf{S}_{4}(2)$  \\
\hline
 $\blacktriangle \mathbf{\Gamma}_{4}$& $\mathbf{\Gamma}_{4}$ & $\textbf{Y}_{1}(2)$ & $\textbf{X}_{1}(2)$ &  $\textbf{S}_{1}\textbf{S}_{2}(2) \oplus \textbf{S}_{3}\textbf{S}_{4}(2)$  \\
\hline
  $\mathbf{\Gamma}_{5}(2)$ & $\mathbf{\Gamma}_{1}\oplus\mathbf{\Gamma}_{2}$ & $\textbf{Y}_{1}(2)$ & $\textbf{X}_{1}(2)$ & $\textbf{S}_{1}\textbf{S}_{2}(2) \oplus \textbf{S}_{3}\textbf{S}_{4}(2)$ \\
\hline
  $\mathbf{\Gamma}_{6}(2)$ & $\mathbf{\Gamma}_{3}\oplus\mathbf{\Gamma}_{4}$ & $\textbf{Y}_{1}(2)$ & $\textbf{X}_{1}(2)$ & $\textbf{S}_{1}\textbf{S}_{2}(2) \oplus \textbf{S}_{3}\textbf{S}_{4}(2)$ \\
\hline
\end{tabular}
\caption{$G_0^A = C_{6v}$ and $G_0 = Pba2$}
\end{subtable}
\begin{subtable}{0.49\textwidth}
\centering
\renewcommand{\arraystretch}{1.2}
\begin{tabular}{c|c|c|c|c}
\hline
$G_0^A = C_{6v}$ & \multicolumn{4}{c}{$G_0 = Pma2$} \\
\hline
$\mathbf{\Gamma}$ & $\mathbf{\Gamma}_M$ & $\textbf{Y}_M$ & $\textbf{X}_M$ & $\textbf{S}_M$ \\
\hline
$\blacktriangle \mathbf{\Gamma}_{1}$ & $\mathbf{\Gamma}_{1}$& $\textbf{Y}_{1} \oplus \textbf{Y}_{3}$ & $\textbf{X}_{1}(2)$ & $2\textbf{S}_{1}(2)$\\
\hline
 $\blacktriangle  \mathbf{\Gamma}_{2}$ & $\mathbf{\Gamma}_{2}$& $\textbf{Y}_{2} \oplus \textbf{Y}_{4}$ & $\textbf{X}_{1}(2)$ & $2\textbf{S}_{1}(2)$\\
\hline
 $\blacktriangle  \mathbf{\Gamma}_{3}$ & $\mathbf{\Gamma}_{3}$ & $\textbf{Y}_{1} \oplus \textbf{Y}_{3}$ & $\textbf{X}_{1}(2)$ & $2\textbf{S}_{1}(2)$ \\
\hline
 $\blacktriangle  \mathbf{\Gamma}_{4}$ & $\mathbf{\Gamma}_{4}$ & $\textbf{Y}_{2} \oplus \textbf{Y}_{4}$ & $\textbf{X}_{1}(2)$ & $2\textbf{S}_{1}(2)$ \\
\hline
  $\blacktriangle \mathbf{\Gamma}_{5}(2)$ & $\mathbf{\Gamma}_{1}\oplus\mathbf{\Gamma}_{2}$ & $\textbf{Y}_{1} \oplus \textbf{Y}_{3}$ & $\textbf{X}_{1}(2)$ & $2\textbf{S}_{1}(2)$ \\
  & & $\textbf{Y}_{2} \oplus \textbf{Y}_{4}$ & & \\
\hline
\end{tabular}
\caption{$G_0^A = C_{6v}$ and $G_0 = Pma2$}
\end{subtable}
 
\begin{subtable}{0.49\textwidth}
\centering
\renewcommand{\arraystretch}{1.2}
\begin{tabular}{c|c|c|c|c}
\hline
$G_0^A = C_{6v}$ & \multicolumn{4}{c}{$G_0 = Pb11$} \\
\hline
$\mathbf{\Gamma}$ & $\mathbf{\Gamma}_M$ & $\textbf{Y}_M$ & $\textbf{X}_M$ & $\textbf{S}_M$ \\
\hline
$ \blacktriangle \mathbf{\Gamma}_{1}$ & $\mathbf{\Gamma}_{1}$ & $\textbf{Y}_{1} \oplus \textbf{Y}_{2}$ & $\textbf{X}_{1}\textbf{X}_{2}(2)$ & $ 2\textbf{S}_{1}\textbf{S}_{2}(2)$  \\
\hline
 $ \blacktriangle \mathbf{\Gamma}_{2}$& $\mathbf{\Gamma}_{2}$ & $\textbf{Y}_{1} \oplus \textbf{Y}_{2}$ & $\textbf{X}_{1}\textbf{X}_{2}(2)$ & $ 2\textbf{S}_{1}\textbf{S}_{2}(2)$ \\
\hline
 $ \blacktriangle \mathbf{\Gamma}_{3}$& $\mathbf{\Gamma}_{1}$ & $\textbf{Y}_{1} \oplus \textbf{Y}_{2}$ & $\textbf{X}_{1}\textbf{X}_{2}(2)$ & $ 2\textbf{S}_{1}\textbf{S}_{2}(2)$  \\
\hline
 $\blacktriangle \mathbf{\Gamma}_{4}$& $\mathbf{\Gamma}_{2}$ & $\textbf{Y}_{1} \oplus \textbf{Y}_{2}$ & $\textbf{X}_{1}\textbf{X}_{2}(2)$ & $ 2\textbf{S}_{1}\textbf{S}_{2}(2)$  \\
\hline
  $ \mathbf{\Gamma}_{5}(2)$ & $\mathbf{\Gamma}_{1}\oplus\mathbf{\Gamma}_{2}$ & $\textbf{Y}_{1} \oplus \textbf{Y}_{2}$ & $\textbf{X}_{1}\textbf{X}_{2}(2)$ & $ 2\textbf{S}_{1}\textbf{S}_{2}(2)$ \\
\hline
  $\mathbf{\Gamma}_{6}(2)$ & $\mathbf{\Gamma}_{1}\oplus\mathbf{\Gamma}_{2}$ & $\textbf{Y}_{1} \oplus \textbf{Y}_{2}$ & $\textbf{X}_{1}\textbf{X}_{2}(2)$ & $ 2\textbf{S}_{1}\textbf{S}_{2}(2)$ \\
\hline
\end{tabular}
\caption{$G_0^A = C_{6v}$ and $G_0 = Pb11$}
\end{subtable}
\caption{\justifying With TR and $G^A_0 = C_{6v}$, for the nonsymmorphic space groups $G_0$ listed above, there exists at least one irrep for which the low-energy  moiré bands are necessarily symmetry-enforced semimetallic. Irreps with $(2)$ are 2D, those with $(4)$ are 4 dimensional (4D), and otherwise they are 1D.
 \label{tab:C6v TR S} }
\end{table}

\begin{table}[H]
\centering
\renewcommand{\arraystretch}{1.2}
\begin{tabular}{c|c|c|c}
\hline
$G_0^A = C_{3v}$&\multicolumn{3}{c}{ $G = P3m1$}\\
\hline
$\mathbf{\Gamma}$ & $\mathbf{\Gamma}_M$ & $\textbf{M}_M$ & $\textbf{K}_M$ \\
\hline
$\mathbf{\Gamma}_{1}$ & $\mathbf{\Gamma}_{1}$ & $2\textbf{M}_{1} $ & $\textbf{K}_{1} \oplus \textbf{K}_{2}\oplus \textbf{K}_{3}$ \\
\hline
$\mathbf{\Gamma}_{2}$ & $\mathbf{\Gamma}_{2}$ & $2\textbf{M}_{2} $ & $\textbf{K}_{1} \oplus \textbf{K}_{2}\oplus \textbf{K}_{3}$ \\
\hline
  $\blacktriangle \mathbf{\Gamma}_{3} (2)$ & $\mathbf{\Gamma}_{3}  (2)$ & $2\textbf{M}_{1}$ & $\textbf{K}_{1}\oplus \textbf{K}_{2}\oplus \textbf{K}_{3}$ \\
  & & $2\textbf{M}_{2}$ & \\
\hline
  $\bigstar \mathbf{\Bar{\mathbf{\Gamma}}}_{4}\mathbf{\Bar{\mathbf{\Gamma}}}_{5}(2)$ & $\mathbf{\Bar{\mathbf{\Gamma}}}_{4}\mathbf{\Bar{\mathbf{\Gamma}}}_{5}(2)$ & $\bar{\textbf{M}}_{3}\bar{\textbf{M}}_{4}(2)$ & $\bar{\textbf{K}}_{4}\oplus \bar{\textbf{K}}_{5}\oplus \bar{\textbf{K}}_{6}$ \\
\hline
  $\Bar{\mathbf{\Gamma}}_{6}(2)$ & $\Bar{\mathbf{\Gamma}}_{6}(2)$ & $\bar{\textbf{M}}_{3}\bar{\textbf{M}}_{4}(2)$ & $\bar{\textbf{K}}_{4}\oplus \bar{\textbf{K}}_{5}\oplus \bar{\textbf{K}}_{6}$ \\
\hline
\end{tabular}

\caption{\justifying With TR and $G^A_0=C_{3v}$, for the space group $G_0$ listed above, there are two possible situations: 
(i) for a certain irrep of $G^A_0$, the symmetry-enforced nontrivial moir\'e topology occur and low-energy moiré bands can be topologically insulating; 
(ii) for a certain irrep, the  low-energy $n_d$ moiré bands are necessarily symmetry-enforced semimetallic. \label{tab:C3V TR TS} }
\end{table}
\begin{table}[H]
\centering
 \begin{subtable}{1\textwidth}
\centering
\renewcommand{\arraystretch}{1.2}
\begin{tabular}{c|c|c|c}
\hline
$G_0^A = C_{3v}$&\multicolumn{3}{c}{ $G = P31m$}\\
\hline
$\mathbf{\Gamma}$& $\mathbf{\Gamma}_M$ & $\textbf{M}_M$ & $\textbf{K}_M(\textbf{KA})_M$ \\
\hline
$\mathbf{\Gamma}_{1}$ & $\mathbf{\Gamma}_{1}$ & $\textbf{M}_{1}\oplus\textbf{M}_{2} $ & $\textbf{K}_{1}\oplus \textbf{K}_{3}(2)$ $(\textbf{KA}_{1}\oplus \textbf{KA}_{3}(2))$ \\
\hline
$\mathbf{\Gamma}_{2}$ & $\mathbf{\Gamma}_{2}$ & $\textbf{M}_{1}\oplus\textbf{M}_{2} $ & $\textbf{K}_{2}\oplus \textbf{K}_{3}(2)$ $(\textbf{KA}_{2}\oplus \textbf{KA}_{3}(2))$ \\
\hline
  $\mathbf{\Gamma}_{3}(2)$ & $\mathbf{\Gamma}_{3}(2)$ & $\textbf{M}_{1}\oplus \textbf{M}_{2}$ & $\textbf{K}_{1}\oplus \textbf{K}_{3}(2)$ $(\textbf{KA}_{1}\oplus \textbf{KA}_{3}(2))$ \\
  & & & $\textbf{K}_{2}\oplus \textbf{K}_{3}(2)$ $(\textbf{KA}_{2}\oplus \textbf{KA}_{3}(2))$ \\
\hline
  $\bigstar \mathbf{\Bar{\mathbf{\Gamma}}}_{4}\mathbf{\Bar{\mathbf{\Gamma}}}_{5}(2)$ & $\mathbf{\Bar{\mathbf{\Gamma}}}_{4}\mathbf{\Bar{\mathbf{\Gamma}}}_{5}(2)$ & $\bar{\textbf{M}}_{3}\bar{\textbf{M}}_{4}(2)$ & $\bar{\textbf{K}}_{4}\oplus \bar{\textbf{K}}_{6}(2)$ $(\bar{\textbf{KA}}_{4}\oplus \bar{\textbf{KA}}_{6}(2))$ \\
  & & & $\bar{\textbf{K}}_{5}\oplus \bar{\textbf{K}}_{6}(2)$ $(\bar{\textbf{KA}}_{5}\oplus \bar{\textbf{KA}}_{6}(2))$ \\
\hline
  $\Bar{\mathbf{\Gamma}}_{6}(2)$ & $\Bar{\mathbf{\Gamma}}_{6}(2)$ & $\bar{\textbf{M}}_{3}\bar{\textbf{M}}_{4}(2)$ & $\bar{\textbf{K}}_{4}\oplus \bar{\textbf{K}}_{6}(2)$ $(\bar{\textbf{KA}}_{4}\oplus \bar{\textbf{KA}}_{6}(2))$ \\
  & & & $\bar{\textbf{K}}_{5}\oplus \bar{\textbf{K}}_{6}(2)$ $(\bar{\textbf{KA}}_{5}\oplus \bar{\textbf{KA}}_{6}(2))$ \\
\hline
\end{tabular}
\caption{$G_0^A = C_{3v}$ and $G_0 = P31m$}
\end{subtable}\\
\centering
\begin{subtable}{1\textwidth}
\centering
\renewcommand{\arraystretch}{1.2}
\begin{tabular}{c|c|c|c}
\hline
$G_0^A = C_{3v}$&\multicolumn{3}{c}{ $G = P3$}\\
\hline
$\mathbf{\Gamma}$& $\mathbf{\Gamma}_M$ & $\textbf{M}_M$ & $\textbf{K}_M(\textbf{KA})_M$ \\
\hline
$\mathbf{\Gamma}_{1}$ & $\mathbf{\Gamma}_{1}$ & $2\textbf{M}_{1} $ & $\textbf{K}_{1}\oplus \textbf{K}_{2}\oplus \textbf{K}_{3}$ $(\textbf{KA}_{1}\oplus \textbf{KA}_{2}\oplus \textbf{KA}_{3})$  \\
\hline
$\mathbf{\Gamma}_{2}$ & $\mathbf{\Gamma}_{1}$ & $2\textbf{M}_{1} $  & $\textbf{K}_{1}\oplus \textbf{K}_{2}\oplus \textbf{K}_{3}$ $(\textbf{KA}_{1}\oplus \textbf{KA}_{2}\oplus \textbf{KA}_{3})$  \\
\hline
  $\mathbf{\Gamma}_{3}(2)$ & $\mathbf{\Gamma}_{2}\mathbf{\Gamma}_{3}(2)$ & $2\textbf{M}_{1}$ & $\textbf{K}_{1}\oplus \textbf{K}_{2}\oplus \textbf{K}_{3}$ $(\textbf{KA}_{1}\oplus \textbf{KA}_{2}\oplus \textbf{KA}_{3})$ \\
\hline
  $\bigstar \mathbf{\Bar{\mathbf{\Gamma}}}_{4}\mathbf{\Bar{\mathbf{\Gamma}}}_{5}(2)$ & $\Bar{\mathbf{\Gamma}}_{4}\Bar{\mathbf{\Gamma}}_{4}(2)$ & $\bar{\textbf{M}}_{2}\bar{\textbf{M}}_{2}(2)$ & $\bar{\textbf{K}}_{4}\oplus \bar{\textbf{K}}_{5}\oplus \bar{\textbf{K}}_{6}$ $(\bar{\textbf{KA}}_{4}\oplus \bar{\textbf{KA}}_{5}\oplus \bar{\textbf{KA}}_{6})$ \\
\hline
  $\Bar{\mathbf{\Gamma}}_{6}(2)$ & $\mathbf{\Bar{\mathbf{\Gamma}}}_{5}\mathbf{\Bar{\mathbf{\Gamma}}}_{6}(2)$ & $\bar{\textbf{M}}_{2}\bar{\textbf{M}}_{2}(2)$ & $\bar{\textbf{K}}_{4}\oplus \bar{\textbf{K}}_{5}\oplus \bar{\textbf{K}}_{6}$ $(\bar{\textbf{KA}}_{4}\oplus \bar{\textbf{KA}}_{5}\oplus \bar{\textbf{KA}}_{6})$ \\
\hline
\end{tabular}
\caption{$G_0^A = C_{3v}$ and $G_0 = P3$}
\end{subtable}
 \caption{\justifying With TR and $G^A_0 = C_{3v}$, for the space groups $G_0$ listed above, there exists at least one irrep for symmetry-enforced topologically insulating moir\'e bands.
 \label{tab:C3v TR T} }
\end{table}

\begin{table}[H]
\centering
\begin{subtable}{0.49\textwidth}
\centering
\renewcommand{\arraystretch}{1.2}
\begin{tabular}{c|c|c|c|c}
\hline
$G_0^A = C_{3v}$ & \multicolumn{4}{c}{$G_0 = Pb11$} \\
\hline
$\mathbf{\Gamma}$ & $\mathbf{\Gamma}_M$ & $\textbf{Y}_M$ & $\textbf{X}_M$ & $\textbf{S}_M$ \\
\hline
$ \blacktriangle \mathbf{\Gamma}_{1}$ & $\mathbf{\Gamma}_{1}$ & $\textbf{Y}_{1} \oplus \textbf{Y}_{2}$ & $\textbf{X}_{1}\textbf{X}_{2}(2)$ & $ 2\textbf{S}_{1}\textbf{S}_{2}(2)$  \\
\hline
 $ \blacktriangle \mathbf{\Gamma}_{2}$& $\mathbf{\Gamma}_{2}$ & $\textbf{Y}_{1} \oplus \textbf{Y}_{2}$ & $\textbf{X}_{1}\textbf{X}_{2}(2)$ & $ 2\textbf{S}_{1}\textbf{S}_{2}(2)$ \\
\hline
  $\mathbf{\Gamma}_{3}(2)$ & $\mathbf{\Gamma}_{1} \oplus \mathbf{\Gamma}_{2}$ & $\textbf{Y}_{1} \oplus \textbf{Y}_{2}$ & $\textbf{X}_{1} \textbf{X}_{2}(2)$ & $2\textbf{S}_{1} \textbf{S}_{2}(2) $ \\
\hline
  $\blacktriangle \Bar{\mathbf{\Gamma}}_{4} \Bar{\mathbf{\Gamma}}_{5}(2)$ & $\Bar{\mathbf{\Gamma}}_{3} \Bar{\mathbf{\Gamma}}_{4}(2)$ & $\bar{\textbf{Y}}_{3} \bar{\textbf{Y}}_{4}(2)$ & $\bar{\textbf{X}}_{3} \bar{\textbf{X}}_{3}(2)$ & $\bar{\textbf{S}}_{3} \bar{\textbf{S}}_{3} (2)\oplus \bar{\textbf{S}}_{4} \bar{\textbf{S}}_{4}(2)$ \\
  & & & $\bar{\textbf{X}}_{4} \bar{\textbf{X}}_{4}(2)$ & \\
\hline
  $\blacktriangle \Bar{\mathbf{\Gamma}}_{6}(2)$ & $\Bar{\mathbf{\Gamma}}_{3} \Bar{\mathbf{\Gamma}}_{4}(2)$ & $\bar{\textbf{Y}}_{3} \bar{\textbf{Y}}_{4}(2)$ & $\bar{\textbf{X}}_{3} \bar{\textbf{X}}_{3}(2)$ & $\bar{\textbf{S}}_{3} \bar{\textbf{S}}_{3} (2)\oplus \bar{\textbf{S}}_{4} \bar{\textbf{S}}_{4}(2)$ \\
  & & & $\bar{\textbf{X}}_{4} \bar{\textbf{X}}_{4}(2)$ & \\
\hline
\end{tabular}
\caption{$G_0^A = C_{3v}$ and $G_0 = Pb11$}
\end{subtable}
\begin{subtable}{0.49\textwidth}
\centering
\renewcommand{\arraystretch}{1.2}
\begin{tabular}{c|c|c|c|c}
\hline
$G_0^A = C_{3v}$ & \multicolumn{4}{c}{$G_0 = Pm11$} \\
\hline
$\mathbf{\Gamma}$ & $\mathbf{\Gamma}_M$ & $\textbf{Y}_M$ & $\textbf{X}_M$ & $\textbf{S}_M$ \\
\hline
$ \mathbf{\Gamma}_{1}$ & $\mathbf{\Gamma}_{1}$& $\textbf{Y}_{1} \oplus \textbf{Y}_{2}$ & $2\textbf{X}_{1}$ & $2\textbf{S}_{1} \oplus 2\textbf{S}_{2}$\\
\hline
 $\mathbf{\Gamma}_{2}$ & $\mathbf{\Gamma}_{2}$& $\textbf{Y}_{1} \oplus \textbf{Y}_{2}$ & $2\textbf{X}_{2}$ & $2\textbf{S}_{1} \oplus 2\textbf{S}_{2}$\\
\hline
  $\blacktriangle \mathbf{\Gamma}_{3}(2)$ & $\mathbf{\Gamma}_{1} \oplus \mathbf{\Gamma}_{2}$ & $\textbf{Y}_{1} \oplus \textbf{Y}_{2}$ & $2\textbf{X}_{1}$ & $2\textbf{S}_{1} \oplus 2\textbf{S}_{2} $ \\
  & & & $2\textbf{X}_{2}$ & \\
\hline
  $\Bar{\mathbf{\Gamma}}_{4} \Bar{\mathbf{\Gamma}}_{5}(2)$ & $\Bar{\mathbf{\Gamma}}_{3} \Bar{\mathbf{\Gamma}}_{4}(2)$ & $\bar{\textbf{Y}}_{3} \bar{\textbf{Y}}_{4}(2)$ & $\bar{\textbf{X}}_{3} \bar{\textbf{X}}_{4}(2)$ & $\bar{\textbf{S}}_{3} \bar{\textbf{S}}_{4}(2) \oplus \bar{\textbf{S}}_{3} \bar{\textbf{S}}_{4}(2)$ \\
\hline
  $\Bar{\mathbf{\Gamma}}_{6}(2)$ & $\Bar{\mathbf{\Gamma}}_{3} \Bar{\mathbf{\Gamma}}_{4}(2)$ & $\bar{\textbf{Y}}_{3} \bar{\textbf{Y}}_{4}(2)$ & $\bar{\textbf{X}}_{3} \bar{\textbf{X}}_{4}(2)$ & $\bar{\textbf{S}}_{3} \bar{\textbf{S}}_{4}(2) \oplus \bar{\textbf{S}}_{3} \bar{\textbf{S}}_{4}(2)$ \\
\hline
\end{tabular}
\caption{$G_0^A = C_{3v}$ and $G_0 = Pm11$}
\end{subtable}
\caption{\justifying With TR and $G^A_0 = C_{3v}$, for the  space groups $G_0$ listed above, there exists at least one irrep for which the  low-energy $n_d$ moiré bands are necessarily symmetry-enforced semimetallic.
 \label{tab:C3v TR S} }
\end{table}
\begin{table}[H]
\centering
\begin{subtable}{0.49\textwidth}
\centering
\renewcommand{\arraystretch}{1.2}
\begin{tabular}{c|c|c|c}
\hline
$G_0^A = C_{4v}$ & \multicolumn{3}{c}{$G_0 = P4mm$} \\
\hline
$\mathbf{\Gamma}$ & $\mathbf{\Gamma}_M$ & $\textbf{M}_M$ & $\textbf{X}_M$ \\
\hline
  $ \mathbf{\Gamma}_{1}$ & $\mathbf{\Gamma}_{1}$ & $\textbf{M}_{1} \oplus \textbf{M}_{3} \oplus \textbf{M}_{5}(2)$ & $\textbf{X}_{1} \oplus \textbf{X}_{3}$ \\
  \hline
  $ \mathbf{\Gamma}_{2}$ & $\mathbf{\Gamma}_{2}$ & $\textbf{M}_{2} \oplus \textbf{M}_{4} \oplus \textbf{M}_{5}(2)$ & $\textbf{X}_{1} \oplus \textbf{X}_{3}$ \\
  \hline
  $\mathbf{\Gamma}_{3}$ & $\mathbf{\Gamma}_{3}$ & $\textbf{M}_{1} \oplus \textbf{M}_{3} \oplus \textbf{M}_{5}(2)$ & $\textbf{X}_{2} \oplus \textbf{X}_{4}$ \\
  \hline
  $\mathbf{\Gamma}_{4}$ & $\mathbf{\Gamma}_{4}$ & $\textbf{M}_{2} \oplus \textbf{M}_{4} \oplus \textbf{M}_{5}(2)$ & $\textbf{X}_{2} \oplus \textbf{X}_{4}$ \\
\hline
  $\blacktriangle \mathbf{\Gamma}_{5}(2)$ & $\mathbf{\Gamma}_{5}(2)$ & $\textbf{M}_{1} \oplus \textbf{M}_{3} \oplus \textbf{M}_{5}(2)$ & $\textbf{X}_{1} \oplus \textbf{X}_{3}$ \\
  & & $\textbf{M}_{2} \oplus \textbf{M}_{4} \oplus \textbf{M}_{5}(2)$ & $\textbf{X}_{2} \oplus \textbf{X}_{4}$ \\
\hline
  $\Bar{\mathbf{\Gamma}}_{6}(2)$ & $\Bar{\mathbf{\Gamma}}_{6}(2)$ & $\bar{\textbf{M}}_{6}(2) \oplus \bar{\textbf{M}}_{7}(2)$ & $\bar{\textbf{X}}_{5}(2)$ \\
\hline
  $\Bar{\mathbf{\Gamma}}_{7}(2)$ & $\Bar{\mathbf{\Gamma}}_{7}(2)$ & $\bar{\textbf{M}}_{6}(2) \oplus \bar{\textbf{M}}_{7}(2)$ & $\bar{\textbf{X}}_{5}(2)$ \\
\hline
\end{tabular}
\caption{$G_0^A = C_{4v}$ and $G_0 = P4mm$}
\end{subtable}
\begin{subtable}{0.49\textwidth}
\centering
\renewcommand{\arraystretch}{1.2}
\begin{tabular}{c|c|c|c|c}
\hline
$G_0^A = C_{4v}$ & \multicolumn{4}{c}{$G_0 = Pmm2$} \\
\hline
$\mathbf{\Gamma}$ & $\mathbf{\Gamma}_M$ & $\textbf{Y}_M$ & $\textbf{X}_M$ & $\textbf{S}_M$ \\
\hline
  $ \mathbf{\Gamma}_{1}$ & $ \mathbf{\Gamma}_{1}$ & $\textbf{Y}_{1} \oplus \textbf{Y}_{3}$ & $\textbf{X}_{1} \oplus \textbf{X}_{4}$& $\textbf{S}_{1} \oplus \textbf{S}_{2} \oplus \textbf{S}_{3} \oplus \textbf{S}_{4}$ \\
  \hline
  $ \mathbf{\Gamma}_{2}$ & $\mathbf{\Gamma}_{1}$ & $\textbf{Y}_{1} \oplus \textbf{Y}_{3}$ &  $\textbf{X}_{2} \oplus \textbf{X}_{3}$& $\textbf{S}_{1} \oplus \textbf{S}_{2} \oplus \textbf{S}_{3} \oplus \textbf{S}_{4}$\\
  \hline
  $\mathbf{\Gamma}_{3}$ & $\mathbf{\Gamma}_{2}$ & $\textbf{Y}_{2} \oplus \textbf{Y}_{4}$ &  $\textbf{X}_{2} \oplus \textbf{X}_{3}$ & $\textbf{S}_{1} \oplus \textbf{S}_{2} \oplus \textbf{S}_{3} \oplus \textbf{S}_{4}$\\
  \hline
  $\mathbf{\Gamma}_{4}$ & $\mathbf{\Gamma}_{2}$ & $\textbf{Y}_{2} \oplus \textbf{Y}_{4}$ & $\textbf{X}_{1} \oplus \textbf{X}_{4}$ & $\textbf{S}_{1} \oplus \textbf{S}_{2} \oplus \textbf{S}_{3} \oplus \textbf{S}_{4}$\\
\hline
  $\blacktriangle \mathbf{\Gamma}_{5}(2)$ & $\mathbf{\Gamma}_{3} \oplus \mathbf{\Gamma}_{4}$ & $\textbf{Y}_{1} \oplus \textbf{Y}_{3}$ & $\textbf{X}_{1} \oplus \textbf{X}_{4}$ & $\textbf{S}_{1} \oplus \textbf{S}_{2} \oplus \textbf{S}_{3} \oplus \textbf{S}_{4}$ \\
  & & $\textbf{Y}_{2} \oplus \textbf{Y}_{4}$ & $\textbf{X}_{2} \oplus \textbf{X}_{3}$ & \\
\hline
  $\Bar{\mathbf{\Gamma}}_{6}(2)$ & $\Bar{\mathbf{\Gamma}}_{5}(2)$ & $\bar{\textbf{Y}}_{5}(2)$ & $\bar{\textbf{X}}_{5}(2)$ & $\bar{\textbf{S}}_{5} (2)\oplus \bar{\textbf{S}}_{5}(2)$ \\
\hline
  $\Bar{\mathbf{\Gamma}}_{7}(2)$ & $\Bar{\mathbf{\Gamma}}_{5}(2)$ & $\bar{\textbf{Y}}_{5}(2)$ & $\bar{\textbf{X}}_{5}(2)$ & $\bar{\textbf{S}}_{5} (2)\oplus \bar{\textbf{S}}_{5}(2)$ \\
\hline
\end{tabular}
\caption{$G_0^A = C_{4v}$ and $G_0 = Pmm2$}
\end{subtable}

\begin{subtable}{0.49\textwidth}
\centering
\renewcommand{\arraystretch}{1.2}
\begin{tabular}{c|c|c|c|c}
\hline
$G_0^A = C_{4v}$ & \multicolumn{4}{c}{$G_0 = Pm11$} \\
\hline
$\mathbf{\Gamma}$ & $\mathbf{\Gamma}$ & \textbf{Y} &\textbf{X} & \textbf{S} \\
\hline
  $ \mathbf{\Gamma}_{1}$ & $ \mathbf{\Gamma}_{1}$ & $\textbf{Y}_{1} \oplus \textbf{Y}_{2}$ & 2$\textbf{X}_{1}$& $2\textbf{S}_{1} \oplus 2\textbf{S}_{2}$ \\
  \hline
  $ \mathbf{\Gamma}_{2}$ & $\mathbf{\Gamma}_{1}$ & $\textbf{Y}_{1} \oplus \textbf{Y}_{2}$ &  2$\textbf{X}_{2}$& $2\textbf{S}_{1} \oplus 2\textbf{S}_{2}$\\
  \hline
  $\mathbf{\Gamma}_{3}$ & $\mathbf{\Gamma}_{2}$ & $\textbf{Y}_{1} \oplus \textbf{Y}_{2}$ &  2$\textbf{X}_{1}$ & $2\textbf{S}_{1} \oplus 2\textbf{S}_{2}$\\
  \hline
  $\mathbf{\Gamma}_{4}$ & $\mathbf{\Gamma}_{2}$ & $\textbf{Y}_{1} \oplus \textbf{Y}_{2}$ & 2$\textbf{X}_{2}$ & $2\textbf{S}_{1} \oplus 2\textbf{S}_{2}$\\
\hline
  $\blacktriangle \mathbf{\Gamma}_{5}(2)$ & $\mathbf{\Gamma}_{1} \oplus \mathbf{\Gamma}_{2}$ & $\textbf{Y}_{1} \oplus \textbf{Y}_{2}$ &$2\textbf{X}_{1}$ & $2\textbf{S}_{1} \oplus 2\textbf{S}_{2}$ \\
  & & $\textbf{Y}_{1} \oplus \textbf{Y}_{2}$ &$2\textbf{X}_{2}$ & \\
\hline
  $\bar{\mathbf{\Gamma}}_{6}(2)$ & $\bar{\mathbf{\Gamma}}_{3} \bar{\mathbf{\Gamma}}_{4}(2)$ & $\bar{\textbf{Y}}_{3} \bar{\textbf{Y}}_{4}(2)$ &$\bar{\textbf{X}}_{3} \bar{\textbf{X}}_{4}(2)$ & $\bar{\textbf{S}}_{3} \bar{\textbf{S}}_{4}(2) \oplus \bar{\textbf{S}}_{3} \bar{\textbf{S}}_{4}(2)$ \\
\hline
  $\bar{\mathbf{\Gamma}}_{7}(2)$ & $\bar{\mathbf{\Gamma}}_{3} \bar{\mathbf{\Gamma}}_{4}(2)$ & $\bar{\textbf{Y}}_{3} \bar{\textbf{Y}}_{4}(2)$ &$\bar{\textbf{X}}_{3} \bar{\textbf{X}}_{4}(2)$ & $\bar{\textbf{S}}_{3} \bar{\textbf{S}}_{4}(2) \oplus \bar{\textbf{S}}_{3} \bar{\textbf{S}}_{4}(2)$ \\
\hline
\end{tabular}
\caption{$G_0^A = C_{4v}$ and $G_0 = Pm11$}
\end{subtable}
\begin{subtable}{0.49\textwidth}
\centering
\renewcommand{\arraystretch}{1.2}
\begin{tabular}{c|c|c|c}
\hline
$G_0^A = C_{4v}$ & \multicolumn{3}{c}{$G_0 = C2mm$} \\
\hline
$\mathbf{\Gamma}$ & $\mathbf{\Gamma}_M$ & $\textbf{Y}_M$ & $\textbf{S}_M$ \\
\hline
  $ \mathbf{\Gamma}_{1}$ & $ \mathbf{\Gamma}_{1}$ & $\textbf{Y}_{1} \oplus \textbf{Y}_{4}$ & $\textbf{S}_{1} \oplus \textbf{S}_{2}$ \\
  \hline
  $ \mathbf{\Gamma}_{2}$ & $\mathbf{\Gamma}_{1}$ & $\textbf{Y}_{1} \oplus \textbf{Y}_{4}$ & $\textbf{S}_{1} \oplus \textbf{S}_{2}$\\
  \hline
  $\mathbf{\Gamma}_{3}$ & $\mathbf{\Gamma}_{2}$ & $\textbf{Y}_{2} \oplus \textbf{Y}_{3}$ & $\textbf{S}_{1} \oplus \textbf{S}_{2}$\\
  \hline
  $\mathbf{\Gamma}_{4}$ & $\mathbf{\Gamma}_{2}$ & $\textbf{Y}_{2} \oplus \textbf{Y}_{3}$ &  $\textbf{S}_{1} \oplus \textbf{S}_{2}$\\
\hline
  $\blacktriangle \mathbf{\Gamma}_{5}(2)$ & $\mathbf{\Gamma}_{3} \oplus \mathbf{\Gamma}_{4}$ & $\textbf{Y}_{2} \oplus \textbf{Y}_{3}$ & $\textbf{S}_{1} \oplus \textbf{S}_{2}$ \\
  & & $\textbf{Y}_{1} \oplus \textbf{Y}_{4}$ & \\
\hline
  $\Bar{\mathbf{\Gamma}}_{6}(2)$ & $\Bar{\mathbf{\Gamma}}_{5}(2)$ & $\bar{\textbf{Y}}_{5}(2)$ & $\bar{\textbf{S}}_{3} \bar{\textbf{S}}_{4}(2)$ \\
\hline
  $\Bar{\mathbf{\Gamma}}_{7}(2)$ & $\Bar{\mathbf{\Gamma}}_{5}(2)$ & $\bar{\textbf{Y}}_{5}(2)$ & $\bar{\textbf{S}}_{3} \bar{\textbf{S}}_{4}(2)$ \\
\hline
\end{tabular}
\caption{$G_0^A = C_{4v}$ and $G_0 = C2mm$}
\end{subtable}
\RaggedRight
\begin{subtable}{0.49\textwidth}
\centering
\renewcommand{\arraystretch}{1.2}
\begin{tabular}{c|c|c|c|c}
\hline
$G_0^A = C_{4v}$ & \multicolumn{4}{c}{$G_0 = P2$} \\
\hline
$\mathbf{\Gamma}$ & $\mathbf{\Gamma}_M$ & $\textbf{Y}_M$ & $\textbf{A}_M$& $\textbf{B}_M$ \\
\hline
$\mathbf{\Gamma}_{1}$ & $\mathbf{\Gamma}_{1}$  & $\textbf{Y}_{1} \oplus \textbf{Y}_2$ & $\textbf{A}_{1} \oplus \textbf{A}_2$ & $\textbf{B}_{1} \oplus \textbf{B}_2$ \\
\hline
$\mathbf{\Gamma}_{2}$ & $\mathbf{\Gamma}_{1}$ & $\textbf{Y}_{1} \oplus \textbf{Y}_2$ & $\textbf{A}_{1} \oplus \textbf{A}_2$ & $\textbf{B}_{1} \oplus \textbf{B}_2$ \\
\hline
$\mathbf{\Gamma}_{3}$ & $\mathbf{\Gamma}_{2}$ & $\textbf{Y}_{1} \oplus \textbf{Y}_2$ & $\textbf{A}_{1} \oplus \textbf{A}_2$ & $\textbf{B}_{1} \oplus \textbf{B}_2$ \\
\hline
$
\mathbf{\Gamma}_{4}$ & $\mathbf{\Gamma}_{2}$& $\textbf{Y}_{1} \oplus \textbf{Y}_2$ & $\textbf{A}_{1} \oplus \textbf{A}_2$ & $\textbf{B}_{1} \oplus \textbf{B}_2$ \\ \hline
  $\blacktriangle \mathbf{\Gamma}_{5}(2)$ & $2\mathbf{\Gamma}_{2}$ & $\textbf{Y}_{1} \oplus \textbf{Y}_2$ & $\textbf{A}_{1} \oplus \textbf{A}_2$ & $\textbf{B}_{1} \oplus \textbf{B}_2$  \\
\hline
  $ \bar{\mathbf{\Gamma}}_{6}(2)$ & $\bar{\mathbf{\Gamma}}_{3}\bar{\mathbf{\Gamma}}_{4}(2)$ & $\bar{\mathbf{Y}}_{3}\bar{\mathbf{Y}}_{4}(2)$& $\bar{\mathbf{A}}_{3}\bar{\mathbf{A}}_{4}(2)$& $\bar{\mathbf{B}}_{3}\bar{\mathbf{B}}_{4}(2)$ \\
\hline
  $ \bar{\mathbf{\Gamma}}_{7}(2)$ & $\bar{\mathbf{\Gamma}}_{3}\bar{\mathbf{\Gamma}}_{4}(2)$ & $\bar{\mathbf{Y}}_{3}\bar{\mathbf{Y}}_{4}(2)$& $\bar{\mathbf{A}}_{3}\bar{\mathbf{A}}_{4}(2)$& $\bar{\mathbf{B}}_{3}\bar{\mathbf{B}}_{4}(2)$ \\
\hline
\end{tabular}
\caption{$G_0^A = C_{4v}$ and $G_0 = P2$}
\end{subtable}
\caption{\justifying With TR and $G^A_0 = C_{4v}$, for the symmorphic space groups $G_0$ listed above, there exists at least one irrep for which the  low-energy $n_d$ moiré bands are necessarily symmetry-enforced semimetallic.
 \label{tab:C4v TR S0} }
\end{table}
\begin{table}[H]
\begin{subtable}{0.49\textwidth}
\centering
\renewcommand{\arraystretch}{1.2}
\begin{tabular}{c|c|c|c}
\hline
$G_0^A = C_{4v}$ & \multicolumn{3}{c}{$G_0 = P4bm$} \\
\hline
$\mathbf{\Gamma}$ & $\mathbf{\Gamma}_M$ & $\textbf{M}_M$ & $\textbf{X}_M$ \\
\hline
  $ \blacktriangle \mathbf{\Gamma}_{1}$ & $\mathbf{\Gamma}_{1}$ & $\textbf{M}_{1} \oplus \textbf{M}_{3} \oplus \textbf{M}_{5}(2)$ & $\textbf{X}_{1}(2)$ \\
  \hline
  $ \blacktriangle \mathbf{\Gamma}_{2}$ & $\mathbf{\Gamma}_{2}$ & $\textbf{M}_{2} \oplus \textbf{M}_{4} \oplus \textbf{M}_{5}(2)$ & $\textbf{X}_{1}(2)$ \\
  \hline
  $\blacktriangle \mathbf{\Gamma}_{3}$ & $\mathbf{\Gamma}_{3}$ & $\textbf{M}_{1} \oplus \textbf{M}_{3} \oplus \textbf{M}_{5}(2)$ & $\textbf{X}_{1}(2)$ \\
  \hline
  $\blacktriangle \mathbf{\Gamma}_{4}$ & $\mathbf{\Gamma}_{4}$ & $\textbf{M}_{2} \oplus \textbf{M}_{4} \oplus \textbf{M}_{5}(2)$ & $\textbf{X}_{1}(2)$ \\
\hline
  $\mathbf{\Gamma}_{5}(2)$ & $\mathbf{\Gamma}_{5}(2)$ & $\textbf{M}_{1} \oplus \textbf{M}_{3} \oplus \textbf{M}_{5}(2)$ & $\textbf{X}_{1}(2)$ \\
  & & $\textbf{M}_{2} \oplus \textbf{M}_{4} \oplus \textbf{M}_{5}(2)$ & \\
\hline
  $\blacktriangle \Bar{\mathbf{\Gamma}}_{6}(2)$ & $\Bar{\mathbf{\Gamma}}_{6}(2)$ & $\bar{\textbf{M}}_{6}(2) \oplus \bar{\textbf{M}}_{7}(2)$ & $\bar{\textbf{X}}_{2}  \bar{\textbf{X}}_{3}(2)$ \\
  & & & $\bar{\textbf{X}}_{4} \bar{\textbf{X}}_{5}(2)$ \\
\hline
  $\blacktriangle \Bar{\mathbf{\Gamma}}_{7}(2)$ & $\Bar{\mathbf{\Gamma}}_{7}(2)$ & $\bar{\textbf{M}}_{6}(2) \oplus \bar{\textbf{M}}_{7}(2)$ & $\bar{\textbf{X}}_{2}  \bar{\textbf{X}}_{3}(2)$ \\
  & & & $\bar{\textbf{X}}_{4}  \bar{\textbf{X}}_{5}(2)$ \\
\hline
\end{tabular}
\caption{$G_0^A = C_{4v}$ and $G_0 = P4bm$}
\end{subtable}\hfill
\begin{subtable}{0.49\textwidth}
\centering
\renewcommand{\arraystretch}{1.2}
\begin{tabular}{c|c|c|c|c}
\hline
$G_0^A = C_{4v}$ & \multicolumn{4}{c}{$G_0 = Pba2$} \\
\hline
$\mathbf{\Gamma}$ & $\mathbf{\Gamma}$ & \textbf{Y} &\textbf{X} & \textbf{S} \\
\hline
  $ \blacktriangle \mathbf{\Gamma}_{1}$ & $ \mathbf{\Gamma}_{1}$ & $\textbf{Y}_{1}(2)$ & $\textbf{X}_{1}(2)$ & $\textbf{S}_{1} \textbf{S}_{2}(2) \oplus \textbf{S}_{3} \textbf{S}_{4}(2)$ \\
  \hline
  $\blacktriangle \mathbf{\Gamma}_{2}$ & $\mathbf{\Gamma}_{1}$  & $\textbf{Y}_{1}(2)$ & $\textbf{X}_{1}(2)$ & $\textbf{S}_{1} \textbf{S}_{2}(2) \oplus \textbf{S}_{3} \textbf{S}_{4}(2)$\\
  \hline
  $\blacktriangle\mathbf{\Gamma}_{3}$ & $\mathbf{\Gamma}_{2}$ & $\textbf{Y}_{1}(2)$ & $\textbf{X}_{1}(2)$  & $\textbf{S}_{1} \textbf{S}_{2}(2) \oplus \textbf{S}_{3} \textbf{S}_{4}(2)$\\
  \hline
  $\blacktriangle\mathbf{\Gamma}_{4}$ & $\mathbf{\Gamma}_{2}$  & $\textbf{Y}_{1}(2)$ & $\textbf{X}_{1}(2)$  & $\textbf{S}_{1} \textbf{S}_{2}(2) \oplus \textbf{S}_{3} \textbf{S}_{4}(2)$\\
\hline
  $\mathbf{\Gamma}_{5}(2)$ & $\mathbf{\Gamma}_{3} \oplus \mathbf{\Gamma}_{4}$ & $\textbf{Y}_{1}(2)$ &$\textbf{X}_{1}(2)$ & $\textbf{S}_{1} \textbf{S}_{2}(2) \oplus \textbf{S}_{3} \textbf{S}_{4}(2)$ \\
\hline
  $\blacktriangle \bar{\mathbf{\Gamma}}_{6}(2)$ & $\bar{\mathbf{\Gamma}}_{5}(2)$ & $\bar{\textbf{Y}}_{2} \bar{\textbf{Y}}_{3}(2)$ &$\bar{\textbf{X}}_{2} \bar{\textbf{X}}_{5}(2)$ & $\bar{\textbf{S}}_{5}  \bar{\textbf{S}}_{5}(4)$ \\
  & & $\bar{\textbf{Y}}_{4} \bar{\textbf{Y}}_{5}(2)$ &$\bar{\textbf{X}}_{3} \bar{\textbf{X}}_{4}(2)$ & \\
\hline
  $\blacktriangle \bar{\mathbf{\Gamma}}_{7}(2)$ & $\bar{\mathbf{\Gamma}}_{5}(2)$ & $\bar{\textbf{Y}}_{2} \bar{\textbf{Y}}_{3}(2)$ &$\bar{\textbf{X}}_{2} \bar{\textbf{X}}_{5}(2)$ & $\bar{\textbf{S}}_{5}  \bar{\textbf{S}}_{5}(4)$ \\
  & & $\bar{\textbf{Y}}_{4} \bar{\textbf{Y}}_{5}(2)$ &$\bar{\textbf{X}}_{3} \bar{\textbf{X}}_{4}(2)$ & \\
\hline
\end{tabular}
\caption{$G_0^A = C_{4v}$ and $G_0 = Pba2$}
\end{subtable}
\begin{subtable}{0.49\textwidth}
\centering
\renewcommand{\arraystretch}{1.2}
\begin{tabular}{c|c|c|c|c}
\hline
$G_0^A = C_{4v}$ & \multicolumn{4}{c}{$G_0 = Pma2$} \\
\hline
$\mathbf{\Gamma}$ & $\mathbf{\Gamma}$ & \textbf{Y} &\textbf{X} & \textbf{S} \\
\hline
  $\blacktriangle \mathbf{\Gamma}_{1}$ & $\mathbf{\Gamma}_{1}$& $\textbf{Y}_{1} \oplus \textbf{Y}_{3}$ & $\textbf{X}_{1}(2)$ & $2\textbf{S}_{1}(2)$ \\
  \hline
  $\blacktriangle \mathbf{\Gamma}_{2}$& $\mathbf{\Gamma}_{1}$& $\textbf{Y}_{1} \oplus \textbf{Y}_{3}$ & $\textbf{X}_{1}(2)$ & $2\textbf{S}_{1}(2)$\\
  \hline
  $\blacktriangle \mathbf{\Gamma}_{3}$ & $\mathbf{\Gamma}_{2}$& $\textbf{Y}_{2} \oplus \textbf{Y}_{4}$ & $\textbf{X}_{1}(2)$ & $2\textbf{S}_{1}(2)$\\
  \hline
  $\blacktriangle \mathbf{\Gamma}_{4}$ & $\mathbf{\Gamma}_{2}$& $\textbf{Y}_{2} \oplus \textbf{Y}_{4}$ & $\textbf{X}_{1}(2)$ & $2\textbf{S}_{1}(2)$\\
\hline
  $\blacktriangle \mathbf{\Gamma}_{5}(2)$ & $\mathbf{\Gamma}_{3} \oplus \mathbf{\Gamma}_{4}$ & $\textbf{Y}_{1} \oplus \textbf{Y}_{3}$ &$\textbf{X}_{1}(2)$ & $2\textbf{S}_{1}(2)$ \\
  & & $\textbf{Y}_{2} \oplus \textbf{Y}_{4}$ & \\
\hline
  $\blacktriangle \bar{\mathbf{\Gamma}}_{6}(2)$ & $\bar{\mathbf{\Gamma}}_{5}(2)$ & $\bar{\textbf{Y}}_{5}(2)$ &$\bar{\textbf{X}}_{2} \bar{\textbf{X}}_{5}(2)$ & $\bar{\textbf{S}}_{2} \bar{\textbf{S}}_{5}(2) \oplus \bar{\textbf{S}}_{3} \bar{\textbf{S}}_{4}(2)$ \\
  & & $\bar{\textbf{Y}}_{5}(2)$ &$\bar{\textbf{X}}_{3} \bar{\textbf{X}}_{4}(2)$ & \\
\hline
  $\blacktriangle \bar{\mathbf{\Gamma}}_{7}(2)$ & $\bar{\mathbf{\Gamma}}_{5}(2)$ & $\bar{\textbf{Y}}_{5}(2)$ &$\bar{\textbf{X}}_{2} \bar{\textbf{X}}_{5}(2)$ & $\bar{\textbf{S}}_{2} \bar{\textbf{S}}_{5}(2) \oplus \bar{\textbf{S}}_{3} \bar{\textbf{S}}_{4}(2)$ \\
  & & $\bar{\textbf{Y}}_{5}(2)$ &$\bar{\textbf{X}}_{3} \bar{\textbf{X}}_{4}(2)$ & \\
\hline
\end{tabular}
\caption{$G_0^A = C_{4v}$ and $G_0 = Pma2$}
\end{subtable}
\hfill
\begin{subtable}{0.49\textwidth}
\centering
\renewcommand{\arraystretch}{1.2}
\begin{tabular}{c|c|c|c|c}
\hline
$G_0^A = C_{4v}$ & \multicolumn{4}{c}{$G_0 = Pb11$} \\
\hline
$\mathbf{\Gamma}$ & $\mathbf{\Gamma}_M$ & $\textbf{Y}_M$ & $\textbf{X}_M$ & $\textbf{S}_M$ \\
\hline
 $ \blacktriangle \mathbf{\Gamma}_{1}$ & $\mathbf{\Gamma}_{1}$ & $\textbf{Y}_{1} \oplus \textbf{Y}_{2}$ & $\textbf{X}_{1}\textbf{X}_{2}(2)$ & $ 2\textbf{S}_{1}\textbf{S}_{2}(2)$  \\
\hline
 $ \blacktriangle \mathbf{\Gamma}_{2}$& $\mathbf{\Gamma}_{1}$ & $\textbf{Y}_{1} \oplus \textbf{Y}_{2}$ & $\textbf{X}_{1}\textbf{X}_{2}(2)$ & $ 2\textbf{S}_{1}\textbf{S}_{2}(2)$  \\
\hline
 $ \blacktriangle \mathbf{\Gamma}_{3}$& $\mathbf{\Gamma}_{2}$ & $\textbf{Y}_{1} \oplus \textbf{Y}_{2}$ & $\textbf{X}_{1}\textbf{X}_{2}(2)$ & $ 2\textbf{S}_{1}\textbf{S}_{2}(2)$  \\
\hline
 $\blacktriangle \mathbf{\Gamma}_{4}$& $\mathbf{\Gamma}_{2}$ & $\textbf{Y}_{1} \oplus \textbf{Y}_{2}$ & $\textbf{X}_{1}\textbf{X}_{2}(2)$ & $ 2\textbf{S}_{1}\textbf{S}_{2}(2)$  \\
\hline
  $\mathbf{\Gamma}_{5}(2)$ & $\mathbf{\Gamma}_{1} \oplus \mathbf{\Gamma}_{2}$ & $\textbf{Y}_{1} \oplus \textbf{Y}_{2}$ & $\textbf{X}_{1}  \textbf{X}_{2}(2)$ & $2\textbf{S}_{1}  \textbf{S}_{2}(2) $ \\
\hline
  $\blacktriangle \bar{\mathbf{\Gamma}}_{6}(2)$ & $\bar{\mathbf{\Gamma}}_{3} \bar{\mathbf{\Gamma}}_{4}(2)$ & $\bar{\textbf{Y}}_{3} \bar{\textbf{Y}}_{4}(2)$ & $\bar{\textbf{X}}_{3} \bar{\textbf{X}}_{3}(2)$ & $\bar{\textbf{S}}_{3} \bar{\textbf{S}}_{3} (2)\oplus \bar{\textbf{S}}_{4} \bar{\textbf{S}}_{4}(2)$ \\
\hline
\end{tabular}
\caption{$G_0^A = C_{4v}$ and $G_0 = Pb11$}
\end{subtable}
\caption{\justifying With TR and $G^A_0 = C_{4v}$, for the nonsymmorphic space groups $G_0$ listed above, there exists at least one irrep for which the  low-energy $n_d$ moiré bands are necessarily symmetry-enforced semimetallic.
 \label{tab:C4v TR S} }
\end{table}
\begin{table}[H]
\centering
\begin{subtable}{0.49\textwidth}
\centering
\renewcommand{\arraystretch}{1.2}
\begin{tabular}{c|c|c|c|c}
\hline
$G_0^A = C_{2v}$ & \multicolumn{4}{c}{$G_0 = Pba2$} \\
\hline
$\mathbf{\Gamma}$ & $\mathbf{\Gamma}_M$ & $\textbf{Y}_M$ & $\textbf{X}_M$ & $\textbf{S}_M$ \\
\hline
  $ \blacktriangle \mathbf{\Gamma}_{1}$ & $ \mathbf{\Gamma}_{1}$ & $\textbf{Y}_{1}(2)$ & $\textbf{X}_{1}(2)$ & $\textbf{S}_{1} \textbf{S}_{2}(2) \oplus \textbf{S}_{3} \textbf{S}_{4}(2)$ \\
  \hline
  $\blacktriangle \mathbf{\Gamma}_{2}$ & $\mathbf{\Gamma}_{2}$  & $\textbf{Y}_{1}(2)$ & $\textbf{X}_{1}(2)$ & $\textbf{S}_{1} \textbf{S}_{2}(2) \oplus \textbf{S}_{3} \textbf{S}_{4}(2)$\\
  \hline
  $\blacktriangle\mathbf{\Gamma}_{3}$ & $\mathbf{\Gamma}_{3}$ & $\textbf{Y}_{1}(2)$ & $\textbf{X}_{1}(2)$  & $\textbf{S}_{1} \textbf{S}_{2}(2) \oplus \textbf{S}_{3} \textbf{S}_{4}(2)$\\
  \hline
  $\blacktriangle\mathbf{\Gamma}_{4}$ & $\mathbf{\Gamma}_{4}$  & $\textbf{Y}_{1}(2)$ & $\textbf{X}_{1}(2)$  & $\textbf{S}_{1} \textbf{S}_{2}(2) \oplus \textbf{S}_{3} \textbf{S}_{4}(2)$\\
\hline
  $\blacktriangle \bar{\mathbf{\Gamma}}_{5}(2)$ & $\bar{\mathbf{\Gamma}}_{5}(2)$ & $\bar{\textbf{Y}}_{2}\bar{\textbf{Y}}_{3}(2)$ & $\bar{\textbf{X}}_{2}\bar{\textbf{X}}_{5}(2)$ & $\bar{\textbf{S}}_{5}\bar{\textbf{S}}_{5}(4)$ \\
  & & $\bar{\textbf{Y}}_{4}\bar{\textbf{Y}}_{5}(2)$ & $\bar{\textbf{X}}_{3}\bar{\textbf{X}}_{4}(2)$ & \\
\hline
\end{tabular}
\caption{$G_0^A = C_{2v}$ and $G_0 = Pba2$}
\end{subtable}
\begin{subtable}{0.49\textwidth}
\centering
\renewcommand{\arraystretch}{1.2}
\begin{tabular}{c|c|c|c|c}
\hline
$G_0^A = C_{2v}$ & \multicolumn{4}{c}{$G_0 = Pma2$} \\
\hline
$\mathbf{\Gamma}$ & $\mathbf{\Gamma}_M$ & $\textbf{Y}_M$ & $\textbf{X}_M$ & $\textbf{S}_M$ \\
\hline
  $\blacktriangle \mathbf{\Gamma}_{1}$ & $\mathbf{\Gamma}_{1}$& $\textbf{Y}_{1} \oplus \textbf{Y}_{3}$ & $\textbf{X}_{1}(2)$ & $2\textbf{S}_{1}(2)$ \\
  \hline
  $\blacktriangle \mathbf{\Gamma}_{2}$& $\mathbf{\Gamma}_{2}$& $\textbf{Y}_{2} \oplus \textbf{Y}_{4}$ & $\textbf{X}_{1}(2)$ & $2\textbf{S}_{1}(2)$\\
  \hline
  $\blacktriangle \mathbf{\Gamma}_{3}$ & $\mathbf{\Gamma}_{3}$& $\textbf{Y}_{1} \oplus \textbf{Y}_{3}$ & $\textbf{X}_{1}(2)$ & $2\textbf{S}_{1}(2)$\\
  \hline
  $\blacktriangle \mathbf{\Gamma}_{4}$ & $\mathbf{\Gamma}_{4}$& $\textbf{Y}_{2} \oplus \textbf{Y}_{4}$ & $\textbf{X}_{1}(2)$ & $2\textbf{S}_{1}(2)$\\
\hline
  $\blacktriangle \bar{\mathbf{\Gamma}}_{5}(2)$ & $\bar{\mathbf{\Gamma}}_{5}(2)$ & $\bar{\textbf{Y}}_{5}(2)$ & $\bar{\textbf{X}}_{2}\bar{\textbf{X}}_{5}(2)$ & $\bar{\textbf{S}}_{2}\bar{\textbf{S}}_{5} (2)\oplus \bar{\textbf{S}}_{3}\bar{\textbf{S}}_{4}(2)$ \\
  & & & $\bar{\textbf{X}}_{3}\bar{\textbf{X}}_{4}(2)$ & \\
\hline
\end{tabular}
\caption{$G_0^A = C_{2v}$ and $G_0 = Pma2$}
\end{subtable}
\RaggedRight
\begin{subtable}{0.49\textwidth}
\centering
\renewcommand{\arraystretch}{1.2}
\begin{tabular}{c|c|c|c|c}
\hline
$G_0^A = C_{2v}$ & \multicolumn{4}{c}{$G_0 = Pb11$} \\
\hline
$\mathbf{\Gamma}$ & $\mathbf{\Gamma}_M$ & $\textbf{Y}_M$ & $\textbf{X}_M$ & $\textbf{S}_M$ \\
\hline
 $ \blacktriangle \mathbf{\Gamma}_{1}$ & $\mathbf{\Gamma}_{1}$ & $\textbf{Y}_{1} \oplus \textbf{Y}_{2}$ & $\textbf{X}_{1}\textbf{X}_{2}(2)$ & $ 2\textbf{S}_{1}\textbf{S}_{2}(2)$  \\
\hline
 $ \blacktriangle \mathbf{\Gamma}_{2}$& $\mathbf{\Gamma}_{1}$ & $\textbf{Y}_{1} \oplus \textbf{Y}_{2}$ & $\textbf{X}_{1}\textbf{X}_{2}(2)$ & $ 2\textbf{S}_{1}\textbf{S}_{2}(2)$  \\
\hline
 $ \blacktriangle \mathbf{\Gamma}_{3}$& $\mathbf{\Gamma}_{2}$ & $\textbf{Y}_{1} \oplus \textbf{Y}_{2}$ & $\textbf{X}_{1}\textbf{X}_{2}(2)$ & $ 2\textbf{S}_{1}\textbf{S}_{2}(2)$  \\
\hline
 $\blacktriangle \mathbf{\Gamma}_{4}$& $\mathbf{\Gamma}_{2}$ & $\textbf{Y}_{1} \oplus \textbf{Y}_{2}$ & $\textbf{X}_{1}\textbf{X}_{2}(2)$ & $ 2\textbf{S}_{1}\textbf{S}_{2}(2)$  \\
\hline
  $\blacktriangle \bar{\mathbf{\Gamma}}_{5}(2)$ & $\bar{\mathbf{\Gamma}}_{3}\bar{\mathbf{\Gamma}}_{4}(2)$ & $\bar{\textbf{Y}}_{3}\bar{\textbf{Y}}_{4}(2)$ & $\bar{\textbf{X}}_{3}\bar{\textbf{X}}_{3}(2)$ & $\bar{\textbf{S}}_{3}\bar{\textbf{S}}_{3} (2)\oplus \bar{\textbf{S}}_{4}\bar{\textbf{S}}_{4}(2)$ \\
  & & & $\bar{\textbf{X}}_{4}\bar{\textbf{X}}_{4}(2)$ & \\
\hline
\end{tabular}
\caption{$G_0^A = C_{2v}$ and $G_0 = Pb11$}
\end{subtable}
\caption{\justifying With TR and $G^A_0 = C_{2v}$, for the  nonsymmorphic space groups $G_0$ listed above, there exists at least one irrep for which the low-energy  moiré bands are necessarily symmetry-enforced semimetallic.
 \label{tab:C2v TR S} }
\end{table}

\begin{table}[H]
\centering
\renewcommand{\arraystretch}{1.2}
\begin{tabular}{c|c|c|c|c}
\hline
$G_0^A = C_{s}$ & \multicolumn{4}{c}{$G_0 = Pb11$} \\
\hline
$\mathbf{\Gamma}$ & $\mathbf{\Gamma}_M$ & $\textbf{Y}_M$ & $\textbf{X}_M$ & $\textbf{S}_M$ \\
\hline
$ \blacktriangle \mathbf{\Gamma}_{1}$ & $\mathbf{\Gamma}_{1}$ & $\textbf{Y}_{1} \oplus \textbf{Y}_{2}$ & $\textbf{X}_{1}\textbf{X}_{2}(2)$ & $ 2\textbf{S}_{1}\textbf{S}_{2}(2)$  \\
\hline
 $ \blacktriangle \mathbf{\Gamma}_{2}$& $\mathbf{\Gamma}_{2}$ & $\textbf{Y}_{1} \oplus \textbf{Y}_{2}$ & $\textbf{X}_{1}\textbf{X}_{2}(2)$ & $ 2\textbf{S}_{1}\textbf{S}_{2}(2)$ \\
\hline
  $\blacktriangle \bar{\mathbf{\Gamma}}_{3}\bar{\mathbf{\Gamma}}_{4}(2)$ & $\bar{\mathbf{\Gamma}}_{3}\bar{\mathbf{\Gamma}}_{4}(2)$ & $\bar{\textbf{Y}}_{3}\bar{\textbf{Y}}_{4}(2)$ & $\bar{\textbf{X}}_{3}\bar{\textbf{X}}_{3}(2)$ & $\bar{\textbf{S}}_{3}\bar{\textbf{S}}_{3} (2)\oplus \bar{\textbf{S}}_{4}\bar{\textbf{S}}_{4}(2)$ \\
  & & & $\bar{\textbf{X}}_{4}\bar{\textbf{X}}_{4}(2)$ & \\
\hline
\end{tabular}
\caption{\justifying With TR and $G^A_0 = C_{s}$, for the  space groups $Pb11$, there exists at least one irrep for which the low-energy  moiré bands are necessarily symmetry-enforced semimetallic.
 \label{tab:Cs TR S} }
\end{table}

\begin{table}[H]

\begin{subtable}{0.49\textwidth}
\centering
\renewcommand{\arraystretch}{1.2}
\begin{tabular}{c|c|c|c}
\hline
$G_0^A = C_{6}$ & \multicolumn{3}{c}{$G_0 = P6$} \\
\hline
$\mathbf{\Gamma}$ & $\mathbf{\Gamma}_M$ & $\textbf{M}_M$ & $\textbf{K}_M$ \\
\hline
$\mathbf{\Gamma}_{1}$ & $\mathbf{\Gamma}_{1}$ & $\textbf{M}_{1} \oplus \textbf{M}_{2}$ & $\textbf{K}_{1} \oplus \textbf{K}_{2}\oplus \textbf{K}_{3}$ \\
\hline
$\mathbf{\Gamma}_{2}$ & $\mathbf{\Gamma}_{2}$ & $\textbf{M}_{1} \oplus \textbf{M}_{2}$ & $\textbf{K}_{1} \oplus \textbf{K}_{2}\oplus \textbf{K}_{3}$ \\
\hline
  $\blacktriangle\mathbf{\Gamma}_{3}\mathbf{\Gamma}_{5}(2)$ & $\mathbf{\Gamma}_{3}\mathbf{\Gamma}_{5}(2)$ & $\textbf{M}_{1}\oplus \textbf{M}_{2}$ & $\textbf{K}_{1}\oplus \textbf{K}_{2}\oplus \textbf{K}_{3}$ \\
\hline
  $\blacktriangle\mathbf{\Gamma}_{4}\mathbf{\Gamma}_{6}(2)$ & $\mathbf{\Gamma}_{4}\mathbf{\Gamma}_{6}(2)$ & $\textbf{M}_{1}\oplus \textbf{M}_{2}$ & $\textbf{K}_{1}\oplus \textbf{K}_{2}\oplus \textbf{K}_{3}$ \\
\hline
  $\bigstar \mathbf{\Bar{\mathbf{\Gamma}}}_{7}\mathbf{\Bar{\mathbf{\Gamma}}}_{8}(2)$ & $\mathbf{\Bar{\mathbf{\Gamma}}}_{7}\mathbf{\Bar{\mathbf{\Gamma}}}_{8}(2)$ & $\textbf{\=M}_{3}\textbf{\=M}_{4}(2)$ & $\textbf{\={K}}_{4}\oplus \textbf{\={K}}_{5}\oplus \textbf{\={K}}_{6}$ \\
\hline
  $\mathbf{\Bar{\mathbf{\Gamma}}}_{10}\mathbf{\Bar{\mathbf{\Gamma}}}_{11}(2)$ & $\mathbf{\Bar{\mathbf{\Gamma}}}_{10}\mathbf{\Bar{\mathbf{\Gamma}}}_{11}(2)$ & $\textbf{\=M}_{3}\textbf{\=M}_{4}(2)$ & $\textbf{\={K}}_{4}\oplus \textbf{\={K}}_{5}\oplus \textbf{\={K}}_{6}$ \\
\hline
  $\mathbf{\Bar{\mathbf{\Gamma}}}_{9}\mathbf{\Bar{\mathbf{\Gamma}}}_{12}(2)$ & $\mathbf{\Bar{\mathbf{\Gamma}}}_{9}\mathbf{\Bar{\mathbf{\Gamma}}}_{12}(2)$ & $\textbf{\=M}_{3}\textbf{\=M}_{4}(2)$ & $\textbf{\={K}}_{4}\oplus \textbf{\={K}}_{5}\oplus \textbf{\={K}}_{6}$ \\
\hline
\end{tabular}
\caption{$G_0^A = C_{6}$ and $G_0 = P6$}
\end{subtable}
\begin{subtable}{0.49\textwidth}
\centering
\renewcommand{\arraystretch}{1.2}
\begin{tabular}{c|c|c|c|c}
\hline
$G_0^A = C_{6}$ & \multicolumn{4}{c}{$G_0 = P2$} \\
\hline
$\mathbf{\Gamma}$ & $\mathbf{\Gamma}_M$ & $\textbf{Y}_M$ & $\textbf{A}_M$& $\textbf{B}_M$ \\
\hline
$\mathbf{\Gamma}_{1}$ & $\mathbf{\Gamma}_{1}$  & $\textbf{Y}_{1} \oplus \textbf{Y}_2$ & $\textbf{A}_{1} \oplus \textbf{A}_2$ & $\textbf{B}_{1} \oplus \textbf{B}_2$ \\
\hline
$\mathbf{\Gamma}_{2}$ & $\mathbf{\Gamma}_{1}$ & $\textbf{Y}_{1} \oplus \textbf{Y}_2$ & $\textbf{A}_{1} \oplus \textbf{A}_2$ & $\textbf{B}_{1} \oplus \textbf{B}_2$ \\
\hline
  $\blacktriangle  \mathbf{\Gamma}_{3}\mathbf{\Gamma}_{5}(2)$ & $2\mathbf{\Gamma}_{1}$ & $\textbf{Y}_{1} \oplus \textbf{Y}_2$ & $\textbf{A}_{1} \oplus \textbf{A}_2$ & $\textbf{B}_{1} \oplus \textbf{B}_2$  \\
\hline
  $\blacktriangle \mathbf{\Gamma}_{4}\mathbf{\Gamma}_{6}(2)$  & $2\mathbf{\Gamma}_{2}$ & $\textbf{Y}_{1} \oplus \textbf{Y}_2$ & $\textbf{A}_{1} \oplus \textbf{A}_2$ & $\textbf{B}_{1} \oplus \textbf{B}_2$  \\
\hline
  $ \bar{\mathbf{\Gamma}}_{7}\bar{\mathbf{\Gamma}}_{8}(2)$ & $\bar{\mathbf{\Gamma}}_{3}\bar{\mathbf{\Gamma}}_{4}(2)$ & $\bar{\mathbf{Y}}_{3}\bar{\mathbf{Y}}_{4}(2)$& $\bar{\mathbf{A}}_{3}\bar{\mathbf{A}}_{4}(2)$& $\bar{\mathbf{B}}_{3}\bar{\mathbf{B}}_{4}(2)$ \\
\hline
  $ \bar{\mathbf{\Gamma}}_{10}\bar{\mathbf{\Gamma}}_{11}(2)$ & $\bar{\mathbf{\Gamma}}_{3}\bar{\mathbf{\Gamma}}_{4}(2)$ & $\bar{\mathbf{Y}}_{3}\bar{\mathbf{Y}}_{4}(2)$& $\bar{\mathbf{A}}_{3}\bar{\mathbf{A}}_{4}(2)$& $\bar{\mathbf{B}}_{3}\bar{\mathbf{B}}_{4}(2)$ \\
\hline
  $\bar{\mathbf{\Gamma}}_{9}\bar{\mathbf{\Gamma}}_{12}(2)$ & $\bar{\mathbf{\Gamma}}_{3}\bar{\mathbf{\Gamma}}_{4}(2)$ & $\bar{\mathbf{Y}}_{3}\bar{\mathbf{Y}}_{4}(2)$& $\bar{\mathbf{A}}_{3}\bar{\mathbf{A}}_{4}(2)$& $\bar{\mathbf{B}}_{3}\bar{\mathbf{B}}_{4}(2)$ \\
\hline
\end{tabular}
\caption{$G_0^A = C_{6}$ and $G_0 = P2$}
\end{subtable}
\begin{subtable}{0.49\textwidth}
\centering
\renewcommand{\arraystretch}{1.2}
\begin{tabular}{c|c|c|c}
\hline$G_0^A = C_{6}$ & \multicolumn{3}{c}{$G_0 = P3$} \\
\hline
$\mathbf{\Gamma}$& $\mathbf{\Gamma}_M$ & $\textbf{M}_M$ & $\textbf{K}_M(\textbf{KA})_M$ \\
\hline
$\mathbf{\Gamma}_{1}$ & $\mathbf{\Gamma}_{1}$ & $2\textbf{M}_{1} $ & $\textbf{K}_{1}\oplus \textbf{K}_{2}\oplus \textbf{K}_{3}$ $(\textbf{KA}_{1}\oplus \textbf{KA}_{2}\oplus \textbf{KA}_{3})$  \\
\hline
$\mathbf{\Gamma}_{2}$ & $\mathbf{\Gamma}_{1}$ & $2\textbf{M}_{1} $  & $\textbf{K}_{1}\oplus \textbf{K}_{2}\oplus \textbf{K}_{3}$ $(\textbf{KA}_{1}\oplus \textbf{KA}_{2}\oplus \textbf{KA}_{3})$  \\
\hline
  $\mathbf{\Gamma}_{3}\mathbf{\Gamma}_{5}(2)$ & $\mathbf{\Gamma}_{2}\mathbf{\Gamma}_{3}(2)$ & $2\textbf{M}_{1}$ & $\textbf{K}_{1}\oplus \textbf{K}_{2}\oplus \textbf{K}_{3}$ $(\textbf{KA}_{1}\oplus \textbf{KA}_{2}\oplus \textbf{KA}_{3})$ \\
\hline
  $\mathbf{\Gamma}_{4}\mathbf{\Gamma}_{6}(2)$ & $\mathbf{\Gamma}_{2}\mathbf{\Gamma}_{3}(2)$ & $2\textbf{M}_{1}$ & $\textbf{K}_{1}\oplus \textbf{K}_{2}\oplus \textbf{K}_{3}$ $(\textbf{KA}_{1}\oplus \textbf{KA}_{2}\oplus \textbf{KA}_{3})$ \\
\hline
  $\bigstar \mathbf{\Bar{\mathbf{\Gamma}}}_{7}\mathbf{\Bar{\mathbf{\Gamma}}}_{8}(2)$ & $\Bar{\mathbf{\Gamma}}_{4}\Bar{\mathbf{\Gamma}}_{4}(2)$ & $\textbf{\=M}_{2}\textbf{\=M}_{2}(2)$ & $\textbf{\={K}}_{4}\oplus \textbf{\={K}}_{5}\oplus \bar{\textbf{K}}_{6}$ $(\bar{\textbf{{KA}}}_{4}\oplus \bar{\textbf{{KA}}}_{5}\oplus \bar{\textbf{{KA}}}_{6})$  \\
\hline
  $\mathbf{\Bar{\mathbf{\Gamma}}}_{10}\mathbf{\Bar{\mathbf{\Gamma}}}_{11}(2)$ & $\mathbf{\Bar{\mathbf{\Gamma}}}_{5}\mathbf{\Bar{\mathbf{\Gamma}}}_{6}(2)$ & $\textbf{\=M}_{2}\textbf{\=M}_{2}(2)$ & $\textbf{\={K}}_{4}\oplus \textbf{\={K}}_{5}\oplus \bar{\textbf{K}}_{6}$ $(\bar{\textbf{{KA}}}_{4}\oplus \bar{\textbf{{KA}}}_{5}\oplus \bar{\textbf{{KA}}}_{6})$ \\
\hline
\end{tabular}
\caption{$G_0^A = C_6$ and $G_0 = P3$}
\end{subtable}
\caption{With TR and $G^A_0 = C_{6}$, for the space groups $G_0$ listed above, there exists at least one irrep for symmetry-enforced topologically insulating or semimetallic moir\'e bands.\label{tab: C6 TR T}}
\end{table}

\begin{table}[H]

\centering
\renewcommand{\arraystretch}{1.2}
\begin{tabular}{c|c|c|c|c}
\hline
$G_0^A = C_{4}$ & \multicolumn{4}{c}{$G_0 = P2$} \\
\hline
$\mathbf{\Gamma}$ & $\mathbf{\Gamma}_M$ & $\textbf{Y}_M$ & $\textbf{A}_M$& $\textbf{B}_M$ \\
\hline
$\mathbf{\Gamma}_{1}$ & $\mathbf{\Gamma}_{1}$  & $\textbf{Y}_{1} \oplus \textbf{Y}_2$ & $\textbf{A}_{1} \oplus \textbf{A}_2$ & $\textbf{B}_{1} \oplus \textbf{B}_2$ \\
\hline
$\mathbf{\Gamma}_{2}$ & $\mathbf{\Gamma}_{1}$ & $\textbf{Y}_{1} \oplus \textbf{Y}_2$ & $\textbf{A}_{1} \oplus \textbf{A}_2$ & $\textbf{B}_{1} \oplus \textbf{B}_2$ \\
\hline
  $\blacktriangle \mathbf{\Gamma}_{3}\mathbf{\Gamma}_{4}(2)$ & $2\mathbf{\Gamma}_{2}$ & $\textbf{Y}_{1} \oplus \textbf{Y}_2$ & $\textbf{A}_{1} \oplus \textbf{A}_2$ & $\textbf{B}_{1} \oplus \textbf{B}_2$  \\
\hline
  $ \bar{\mathbf{\Gamma}}_{5}\bar{\mathbf{\Gamma}}_{7}(2)$ & $\bar{\mathbf{\Gamma}}_{3}\bar{\mathbf{\Gamma}}_{4}(2)$ & $\bar{\mathbf{Y}}_{3}\bar{\mathbf{Y}}_{4}(2)$& $\bar{\mathbf{A}}_{3}\bar{\mathbf{A}}_{4}(2)$& $\bar{\mathbf{B}}_{3}\bar{\mathbf{B}}_{4}(2)$ \\
\hline
  $\bar{\mathbf{\Gamma}}_{6}\bar{\mathbf{\Gamma}}_{8}(2)$ & $\bar{\mathbf{\Gamma}}_{3}\bar{\mathbf{\Gamma}}_{4}(2)$ & $\bar{\mathbf{Y}}_{3}\bar{\mathbf{Y}}_{4}(2)$& $\bar{\mathbf{A}}_{3}\bar{\mathbf{A}}_{4}(2)$& $\bar{\mathbf{B}}_{3}\bar{\mathbf{B}}_{4}(2)$ \\
\hline
\end{tabular}

 \caption{\justifying With TR and $G^A_0 = C_{4}$, for the space groups $P2$, there exists at least one irrep for which the low-energy  moiré bands are necessarily symmetry-enforced semimetallic.
 \label{tab:C4 TR T} }
\end{table}
\begin{table}[H]

\centering
\renewcommand{\arraystretch}{1.2}
\begin{tabular}{c|c|c|c}
\hline
$G_0^A = C_{3}$ & \multicolumn{3}{c}{$G_0 = P3$} \\
\hline
$\mathbf{\Gamma}$ & $\mathbf{\Gamma}_M$ & $\textbf{M}_M$ & $\textbf{K}_M(\textbf{KA})_M$ \\
\hline
$\mathbf{\Gamma}_{1}$ & $\mathbf{\Gamma}_{1}$ & $2\textbf{M}_{1} $ & $\textbf{K}_{1}\oplus \textbf{K}_{2}\oplus \textbf{K}_{3}$ $(\textbf{KA}_{1}\oplus \textbf{KA}_{2}\oplus \textbf{KA}_{3})$  \\
\hline
  $\mathbf{\Gamma}_{2}\mathbf{\Gamma}_{3}(2)$ & $\mathbf{\Gamma}_{2}\mathbf{\Gamma}_{3}(2)$ & $2\textbf{M}_{1}$ & $\textbf{K}_{1}\oplus \textbf{K}_{2}\oplus \textbf{K}_{3}$ $(\textbf{KA}_{1}\oplus \textbf{KA}_{2}\oplus \textbf{KA}_{3})$ \\
\hline
  $\bigstar \Bar{\mathbf{\Gamma}}_{4}\Bar{\mathbf{\Gamma}}_{4}(2)$ & $\Bar{\mathbf{\Gamma}}_{4}\Bar{\mathbf{\Gamma}}_{4}(2)$ & $\bar{\textbf{M}}_{2}\bar{\textbf{M}}_{2}(2)$ & $\bar{\textbf{K}}_{4}\oplus \bar{\textbf{K}}_{5}\oplus \bar{\textbf{K}}_{6}$ $(\bar{\textbf{KA}}_{4}\oplus \bar{\textbf{KA}}_{5}\oplus \bar{\textbf{KA}}_{6})$ \\
\hline
  $\mathbf{\Bar{\mathbf{\Gamma}}}_{5}\mathbf{\Bar{\mathbf{\Gamma}}}_{6}(2)$ & $\mathbf{\Bar{\mathbf{\Gamma}}}_{5}\mathbf{\Bar{\mathbf{\Gamma}}}_{6}(2)$ & $\bar{\textbf{M}}_{2}\bar{\textbf{M}}_{2}(2)$ & $\bar{\textbf{K}}_{4}\oplus \bar{\textbf{K}}_{5}\oplus \bar{\textbf{K}}_{6}$ $(\bar{\textbf{KA}}_{4}\oplus \bar{\textbf{KA}}_{5}\oplus \bar{\textbf{KA}}_{6})$ \\
\hline
\end{tabular}

\caption{\justifying With TR and $G^A_0 = C_{3}$, for the space group $P3$, there exists at least one irrep for symmetry-enforced topologically insulating moir\'e bands. \label{tab:C3 TR T}}
\end{table}

\begin{table}[h!]
\centering
\renewcommand{\arraystretch}{1.2}
\centering
\begin{subtable}{\textwidth}
 \begin{ruledtabular}
\begin{tabular}{ c c c c c c c}
$C_{6v}$ & $\mathcal{E}$ & $C_{3z}, C_{3z}^{-1}$ & $C_{2z}$ & $C_{6z}, C_{6z}^{-1}$   & ${C}_{3z}{M}_{x}{C}^{-1}_{3z}, {M}_{x}, {C}^{-1}_{3z}{M}_{x}{C}_{3z}$ & ${C}_{3z}{M}_{y}{C}^{-1}_{3z}, {M}_{y}, {C}^{-1}_{3z}{M}_{y}{C}_{3z}$ \\ 
\hline
$ \mathbf{\Gamma}_1$ & 1 & 1 & 1 & 1 & 1 & 1 \\ 
$ \mathbf{\Gamma}_2$& 1 & 1 & 1 & 1 & -1 & -1 \\ 
$ \mathbf{\Gamma}_3$ & 1 & 1 & -1 & -1 & -1 & 1  \\ 
$\mathbf{\Gamma}_4$ & 1 & 1 & -1 & -1 & 1 & -1  \\ 
$ \mathbf{\Gamma}_5$ & 2 & -1 & 2 & -1 & 0 & 0 \\ 
$ \mathbf{\Gamma}_6$ & 2 & -1 & -2 & 1 & 0 & 0 \\ 
$\mathbf{\bar{\Gamma}}_7$ & 2 & -2 & 0 & 0 & 0 & 0 \\ 
$\mathbf{\bar{\Gamma}}_8$ & 2 & 1 &  0 &$-\sqrt{3}$   & 0 & 0 \\ 
$\mathbf{\bar{\Gamma}}_9$ & 2 & 1 & 0 & $\sqrt{3}$   & 0 & 0 \\ 
\end{tabular}
\caption{
The character table for point group $C_{6v}$.}
  \end{ruledtabular}
\end{subtable}
\begin{subtable}{\textwidth}
 \begin{ruledtabular}
\begin{tabular}{ c c c  c c c }
$C_{4v}$ & $\mathcal{E}$  & $C_{2z}$ & $C_{4z}, C_{4z}^{-1}$   & $ {M}_{y}, {M}_{x}$ & $C_{4z}{M}_{x},{M}_{x}C_{4z}$ \\ 
\hline
$ \mathbf{\Gamma}_1$ & 1 & 1 & 1 & 1 & 1  \\ 
$ \mathbf{\Gamma}_2$& 1 & 1 & -1 & 1 & -1  \\ 
$ \mathbf{\Gamma}_3$ & 1 & 1 & -1 & -1 & 1  \\ 
$\mathbf{\Gamma}_4$ & 1 & 1 & 1 & -1 & -1   \\ 
$ \mathbf{\Gamma}_5$ & 2 & -2 & 0 & 0 & 0  \\ 
$\mathbf{\bar{\Gamma}}_6$ & 2 & 0 & $-\sqrt{2}$ & 0 & 0  \\ 
$\mathbf{\bar{\Gamma}}_7$ & 2 & 0 &  $\sqrt{2}$ &0  & 0  \\ 
\end{tabular}
\caption{
The character table for point group $C_{4v}$.}
  \end{ruledtabular}
\end{subtable}
\begin{subtable}{\textwidth}
 \begin{ruledtabular}
        \begin{tabular}{c c c c}
            $C_{3v}$ & $\mathcal{E}$ & ${C}_{3z}, {C}_{3z}^{-1}$ & ${C}_{3z}{M}_{y}{C}^{-1}_{3z}, {M}_{y}, {C}^{-1}_{3z}{M}_{y}{C}_{3z}$ \\ 
            \hline
            $\mathbf{\Gamma}_1$ & 1 & 1 & 1 \\ 
            $\mathbf{\Gamma}_2$ & 1 & 1 & -1 \\ 
            $\mathbf{\Gamma}_3$ & 2 & -1 & 0 \\ 
            $\mathbf{\Bar{\Gamma}}_4$ & 1 & -1 & $-i$ \\ 
            $\mathbf{\Bar{\Gamma}}_5$ & 1 & -1 & $+i$ \\ 
            $\mathbf{\Bar{\Gamma}}_6$ & 2 & 1 & 0 \\ 
        \end{tabular}
    \end{ruledtabular}
     \caption{
The character table for point group $C_{3v}$.}
    \end{subtable}

   \begin{subtable}{0.49\textwidth}
 \begin{ruledtabular}
        \begin{tabular}{c c c c c}
            $C_{2v}$ & $\mathcal{E}$ & ${C}_{2z}$ & $ {M}_{x}$  & $ {M}_{y}$ \\ 
            \hline
            $\mathbf{\Gamma}_1$ & 1 & 1 & 1 & 1 \\ 
            $\mathbf{\Gamma}_2$ & 1 & 1 & -1 & -1 \\ 
            $\mathbf{\Gamma}_3$ & 1 & -1 & -1 & 1 \\ 
            $\mathbf{{\Gamma}}_4$ & 1 & -1 & 1 & -1 \\ 
            $\mathbf{\Bar{\Gamma}}_5$ & 2 & 0 & 0 & 0\\ 
        \end{tabular}
    \end{ruledtabular}
     \caption{
The character table for point group $C_{2v}$.}
    \end{subtable}
     \begin{subtable}{0.49\textwidth}
 \begin{ruledtabular}
        \begin{tabular}{c c c}
            $C_{s}$ & $\mathcal{E}$ & $M_x$   \\ 
            \hline
            $\mathbf{\Gamma}_1$ & 1 & 1 \\ 
            $\mathbf{\Gamma}_2$ & 1  & -1\\ 
            $\mathbf{\Bar{\Gamma}}_3$ & 1 & $-i$ \\ 
            $\mathbf{\Bar{\Gamma}}_4$ & 1 & $i$ \\ 
        \end{tabular}
    \end{ruledtabular}
     \caption{
The character table for point group $C_{s}$, where $C_{s}=C_{1v}$}
    \end{subtable}
   
\caption{The character table for the point group $C_{nv}$ with $n=1,2,3,4,6$.}
\label{tab: character table point group}
\end{table}

\begin{table}[h!]
\centering
\renewcommand{\arraystretch}{1.2}
\centering
\begin{subtable}{0.49\textwidth}
\RaggedRight
 \begin{ruledtabular}
        \begin{tabular}{c c c c}
            $C_{3}$ & $\mathcal{E}$ & ${C}_{3z}$ & ${C}_{3z}^{-1}$ \\ 
            \hline
            $\mathbf{\Gamma}_1$ & 1 & 1 & 1 \\ 
            $\mathbf{\Gamma}_2$ & 1 &  $e^{\frac{i2 \pi}{3}}$ & $e^{-\frac{i2 \pi}{3}}$ \\ 
            $\mathbf{\Gamma}_3$ & 1 & $e^{-\frac{i2 \pi}{3}}$ & $e^{\frac{i2 \pi}{3}}$\\ 
           
            $\mathbf{\Bar{\Gamma}}_4$ & 1 & -1 & -1 \\ 
            $\mathbf{\Bar{\Gamma}}_5$ & 1 & $e^{-\frac{i \pi}{3}}$& $e^{\frac{i \pi}{3}}$ \\ 
            $\mathbf{\Bar{\Gamma}}_6$ & 1 &  $e^{\frac{i \pi}{3}}$ &  $e^{-\frac{i \pi}{3}}$\\ 
        \end{tabular}
    \end{ruledtabular}
     \caption{
The character table for point group $C_{3}$.}
    \end{subtable}
\begin{subtable}{0.49\textwidth}
 \begin{ruledtabular}
\begin{tabular}{ c c c  c  c}
$C_4$ & $\mathcal{E}$  & $C_{2z}$ & $C_{4z}$ & $C^{-1}_{4z}$ \\ 
\hline
$ \mathbf{\Gamma}_1$ & 1 & 1 & 1 & 1   \\ 
$ \mathbf{\Gamma}_2$& 1 & 1 & -1 & -1  \\ 
$ \mathbf{\Gamma}_3$ & 1 & 1 & $i$ & $-i$   \\ 
$\mathbf{\Gamma}_4$ & 1 & 1 & $-i$ & $i$  \\ 
$\mathbf{\bar{\Gamma}}_5$ & 1 & $-i$ & $e^{\frac{i3 \pi}{4}}$ & $e^{-\frac{i3 \pi}{4}}$  \\ 
$\mathbf{\bar{\Gamma}}_6$ & 1 & $-i$& $e^{-\frac{i \pi}{4}}$ & $e^{\frac{i \pi}{4}}$  \\ 
$\mathbf{\bar{\Gamma}}_7$ & 1 & $i$ & $e^{-\frac{i3 \pi}{4}}$ & $e^{\frac{i3 \pi}{4}}$  \\
$\mathbf{\bar{\Gamma}}_8$ & 1 & $i$ & $e^{\frac{i \pi}{4}}$ & $e^{-\frac{i \pi}{4}}$  \\ 
\end{tabular}
\caption{
The character table for point group $C_{4}$.}
  \end{ruledtabular}
\end{subtable}
\RaggedRight

\begin{subtable}{0.49\textwidth}
 \begin{ruledtabular}
\begin{tabular}{ c c c c c c c}
$C_{6}$ & $\mathcal{E}$ & $C_{3z}$ & $C_{3z}^{-1}$ &$C_{2z}$ & $C^{-1}_{6z}$ &$C_{6z}$   \\ 
\hline
$ \mathbf{\Gamma}_1$ & 1 & 1 & 1 & 1 & 1 & 1 \\ 
$ \mathbf{\Gamma}_2$ & 1 & 1 & 1 & -1 & -1 & -1 \\ 
$ \mathbf{\Gamma}_3$& 1 & $e^{\frac{i 2\pi}{3}}$ & $e^{-\frac{i 2\pi}{3}}$ & 1 & $e^{\frac{i 2\pi}{3}}$ & $e^{-\frac{i 2\pi}{3}}$ \\ 
$ \mathbf{\Gamma}_4$& 1 & $e^{\frac{i 2\pi}{3}}$ & $e^{-\frac{i 2\pi}{3}}$ & -1 & $e^{-\frac{i \pi}{3}}$ & $e^{\frac{i \pi}{3}}$ \\ 
$\mathbf{\Gamma}_5$ & 1 & $e^{-\frac{i 2\pi}{3}}$ & $e^{\frac{i 2\pi}{3}}$ & 1 & $e^{-\frac{i 2\pi}{3}}$ &  $e^{\frac{i 2\pi}{3}}$ \\ 
$ \mathbf{\Gamma}_6$ & 1 & $e^{-\frac{i 2\pi}{3}}$ & $e^{\frac{i 2\pi}{3}}$ & -1 & $e^{\frac{i \pi}{3}}$ & $e^{-\frac{i \pi}{3}}$ \\
$\mathbf{\bar{\Gamma}}_7$  & 1 & -1 & -1 & $-i$ & $i$ & $-i$ \\
$\mathbf{\bar{\Gamma}}_8$  & 1 & -1 & -1 & $i$ & $-i$ & $i$ \\
$\mathbf{\bar{\Gamma}}_9$  & 1 & $e^{-\frac{i \pi}{3}}$ & $e^{\frac{i \pi}{3}}$ & $-i$  & $e^{-\frac{i5 \pi}{6}}$ & $e^{\frac{i5 \pi}{6}}$ \\
$\mathbf{\bar{\Gamma}}_{10}$  & 1 & $e^{-\frac{i \pi}{3}}$  & $e^{\frac{i \pi}{3}}$ & $i$  & $e^{\frac{i \pi}{6}}$ & $e^{-\frac{i \pi}{6}}$\\
$\mathbf{\bar{\Gamma}}_{11}$  & 1 & $e^{\frac{i \pi}{3}}$  & $e^{-\frac{i \pi}{3}}$ & $-i$  & $e^{-\frac{i \pi}{6}}$ & $e^{\frac{i \pi}{6}}$ \\
$\mathbf{\bar{\Gamma}}_{12}$  & 1 & $e^{\frac{i \pi}{3}}$  & $e^{-\frac{i \pi}{3}}$ & $i$  & $e^{\frac{i5 \pi}{6}}$ & $e^{-\frac{i5 \pi}{6}}$ \\
\end{tabular}
\caption{
The character table for point group $C_{6}$.}
  \end{ruledtabular}
\end{subtable}

\caption{The character table for the point group $C_{n}$ with $n=3,4,6$.}
\label{tab: character table point group rotation}
\end{table}

\begin{table}[H]
\centering
\renewcommand{\arraystretch}{1.2}
\centering
\begin{subtable}{\textwidth}
 \begin{ruledtabular}
%
\caption{
The character table for the wave vector group $G_{\mathbf{\mathbf{\Gamma}}_M} = C_{6v}$.}
  \end{ruledtabular}
\end{subtable}
\begin{subtable}{\textwidth}
 \begin{ruledtabular}
        %
    \end{ruledtabular}
     \caption{
The character table for the wave vector group $G_{\mathbf{K}_M} = C_{3v}$.}
    \end{subtable}

   \begin{subtable}{\textwidth}
 \begin{ruledtabular}
        \begin{tabular}{c c c c c}
            $C_{2v}$ & $\mathcal{E}$ & ${C}_{2z}$ & $ {M}_{x}$  & $ {M}_{y}$ \\ 
            \hline
            $\mathbf{M}_1$ & 1 & 1 & 1 & 1 \\ 
            $\mathbf{M}_2$ & 1 & 1 & -1 & -1 \\ 
            $\mathbf{M}_3$ & 1 & -1 & -1 & 1 \\ 
            $\mathbf{{M}}_4$ & 1 & -1 & 1 & -1 \\ 
            $\mathbf{\Bar{M}}_5$ & 2 & 0 & 0 & 0\\ 
        \end{tabular}
    \end{ruledtabular}
     \caption{
The character table for the wave vector group $G_{\mathbf{M}_M} = C_{2v}$.}
    \end{subtable}
   
\caption{The character table for the little groups of  $P6mm$}
\label{tab: character table P6mm}
\end{table}

\begin{table}[h!]
\centering
\renewcommand{\arraystretch}{1.2}
\centering
 \begin{ruledtabular}
%
  \end{ruledtabular}
\caption{The character table for the group $G^A_{\mathbf{\Gamma}_M\mathbf{M}_M}=G^A_{\mathbf{\Gamma}_M\mathbf{K}_M}=C_s$}
\label{tab: character table Cs}
\end{table}

\begin{table}[h!]
\renewcommand{\arraystretch}{1.5}
%
\centering
\caption{\centering  2D spinful EBR table for the $P6mm$ group with TR.}

\label{TAB:EBR}
\end{table}

\subsection{Cases without TR symmetry} \label{sec: cases withour TR}

Table II in the main text summarizes all combinations of $G^A_0$, $\Lambda^A_{\Gamma}$, and $G_0$ that lead to symmetry-enforced moir\'e topology in a spinless TR-breaking system. Table \ref{tab:summary of No TR} summarizes all the combinations of $G^A_0$, $\Lambda^A_{\Gamma}$ and $G_0$ that lead to symmetry-enforced moir\'e topology in a spinful TR-breaking system. Without TR, we find 24 cases for spinful systems, and 66 cases for the spinless systems. Among the 66 spinless cases, we find that there are 5 cases that allow for isolated symmetry-enforced low-energy topological moir\'e bands, while all other cases only have topological semi-metal phases for the low-energy moir\'e bands. For an atomic 1D irrep $\Lambda^{A}_{\Gamma}$, symmetry-enforced moiré semi-metal phases only appear when $G_0$ is a non-symmorphic plane group, which is similar to the cases with TR. Tables \ref{tab:C6v NTR T}-- \ref{tab:Cs NTR S} show the possible representations at high-symmetry momenta in the mBZ for all the non-trivial cases in 
Table \ref{tab:summary of No TR} and Table II in the main text. We organize these tables according to $G^A_0$, namely, Tables \ref{tab:C6v NTR T}-\ref{tab:C6v NTR S} are for $G^A_0=C_{6v}$, Tables \ref{tab:C4v NTR T}-\ref{tab:C4v NTR S} are for $G^A_0=C_{4v}$, Tables \ref{tab:C3v NTR S} is for $G^A_0=C_{3v}$, Table \ref{tab:C2v NTR S} is for $G^A_0=C_{2v}$, and Table \ref{tab:Cs NTR S} is for $G^A_0=C_{s}$. 

To illustrate the symmetry-enforced moiré topology with a spinless atomic irrep when TR is broken, we consider the case with $\Lambda^{A}_{\Gamma}=\mathbf{\Gamma}_5$, $G^A_0 = C_{4v}$ and $G_0=P2$ in Table. \ref{tab:C4v NTR T}. The mBZ is shown in Fig.\ref{fig:BZ Figure}(e), and the band representations in the mBZ of low-energy moir\'e bands for this case are 
\begin{eqnarray} \label{eq_SM:bandirreps_C4vP2_gamma5}
    (2\mathbf{\Gamma}_{2}, \textbf{Y}_{1} \oplus \textbf{Y}_2, \textbf{A}_{1} \oplus \textbf{A}_2, \textbf{B}_{1} \oplus \textbf{B}_2),
\end{eqnarray}
as listed in Table. \ref{tab:C4v NTR T}.  With a nonzero moiré potential respecting $G_0 = P2$, the little groups at the high-symmetry momenta $\mathbf{\Gamma}_M$, $\textbf{A}_M$, $\textbf{B}_M$, and $\textbf{Y}_M$ are all $C_2$. Therefore, the irrep at any of these high-symmetry points is characterized by the $C_{2z}$ eigenvalue, denoted as $\zeta_i(\kk_M)$, where $i$ labels the moiré band. According to Ref.\cite{fang2012bulk},  the Chern number $C$ of the low-energy two bands is associated with the $\zeta_i$ by 
\begin{eqnarray}\label{eq;chern numbaer c2}
(-1)^C  = \prod^2_{i=1} \zeta_i(\mathbf{\Gamma}_M) \zeta_i(\mathbf{A}_M) \zeta_i(\mathbf{B}_M) \zeta_i(\mathbf{Y}_M).
\end{eqnarray}
According to Eq.(\ref{eq;chern numbaer c2}) and the band irreps in Eq.(\ref{eq_SM:bandirreps_C4vP2_gamma5}), we get 
\begin{eqnarray}
    (-1)^C=-1,
\end{eqnarray}
which implies that the total Chern number of the two low-energy bands must be odd.  We can use the same method to demonstrate the symmetry-enforced topological moiré Chern bands in the combination of $G^A_0=C_{6v}$ and $G_0=P2$ as well as the combination of  $G^A_0=C_{6v}$ and $G_0=P6$.
Without TR, all symmetry-enforced moir\'e topologically insulating bands have non-zero total Chern number.

\begin{table}[H]
\centering
\begin{tabular}{|c|c|c|c|c|c|}
\hline
 \diagbox{$G_0$}{$G^A_0$} & $C_{6v}$ & $C_{4v}$ & $C_{3v}$ & $C_{2v}$ & $C_{s}$
\\
\hline
 $P4bm$ &  & $\Bar{\mathbf{\Gamma}}_6$, $\Bar{\mathbf{\Gamma}}_7$  &  &  &
 \\
 $Pba2$ & $\Bar{\mathbf{\Gamma}}_7$, $\Bar{\mathbf{\Gamma}}_8$, $\Bar{\mathbf{\Gamma}}_9$ & $\Bar{\mathbf{\Gamma}}_6$, $\Bar{\mathbf{\Gamma}}_7$  &  &  $\Bar{\mathbf{\Gamma}}_5$&
 \\
 $Pma2$ & $\Bar{\mathbf{\Gamma}}_7$, $\Bar{\mathbf{\Gamma}}_8$, $\Bar{\mathbf{\Gamma}}_9$ & $\Bar{\mathbf{\Gamma}}_6$, $\Bar{\mathbf{\Gamma}}_7$ &  & $\Bar{\mathbf{\Gamma}}_5$&
 \\
 $Pb11$ & $\Bar{\mathbf{\Gamma}}_7$, $\Bar{\mathbf{\Gamma}}_8$, $\Bar{\mathbf{\Gamma}}_9$ & $\Bar{\mathbf{\Gamma}}_6$, $\Bar{\mathbf{\Gamma}}_7$ &  $\Bar{\mathbf{\Gamma}}_4$, $\Bar{\mathbf{\Gamma}}_5$ & $\Bar{\mathbf{\Gamma}}_5$ &$\Bar{\mathbf{\Gamma}}_3$, $\Bar{\mathbf{\Gamma}}_4$
 \\
\hline
\end{tabular}
  \caption{\justifying \textbf{Spinful TR-breaking cases for low-energy moir\'e topology.}
List of combinations of atomic point group $G_0^A$, atomic $\Gamma$ irrep, and moiré plane group $G_0$ that lead to symmetry-enforced nontrivial moiré topology, in the presence of SOC and without TR symmetry. For a given $G_0^A$ and $G_0$, the table entry indicates the required atomic $\Gamma$ irrep to give symmetry-enforced topology.  The low-energy bands are semimetallic.  \label{tab:summary of No TR}
}
\end{table}

\begin{table}[H]
\begin{subtable}{0.49\textwidth}
\centering
\renewcommand{\arraystretch}{1.2}

\caption{$G_0^A = C_{6v}$ and $G_0 = P6$}
\end{subtable}
\begin{subtable}{0.49\textwidth}
\centering
\renewcommand{\arraystretch}{1.2}
%
\caption{$G_0^A = C_{6v}$ and $G_0 = P2$}
\end{subtable}
 \caption{\justifying Without TR and $G^A_0 = C_{6v}$, for the space groups $G_0$ listed above, there exists at least one irrep for symmetry-enforced topologically insulating moir\'e bands.
 \label{tab:C6v NTR T} }
\end{table}
\begin{table}[H]
\begin{subtable}{0.49\textwidth}
\centering
\renewcommand{\arraystretch}{1.2}
%
\caption{$G_0^A = C_{6v}$ and $G_0 = P6mm$}
\end{subtable}
\begin{subtable}{0.49\textwidth}
\renewcommand{\arraystretch}{1.2}
\centering
%
\caption{$G_0^A = C_{6v}$ and $G_0 = P3m1$}
\end{subtable}
\hfill
\begin{subtable}{0.49\textwidth}
\centering
\renewcommand{\arraystretch}{1.2}
%
\caption{$G_0^A = C_{6v}$ and $G_0 = Pmm2$}
\end{subtable}
\begin{subtable}{0.49\textwidth}
\centering
\renewcommand{\arraystretch}{1.2}
%
\caption{$G_0^A = C_{6v}$ and $G_0 = Pm11$}
\end{subtable}
 \begin{subtable}{0.49\textwidth}
\centering
\renewcommand{\arraystretch}{1.2}
%
\caption{$G_0^A = C_{6v}$ and $G_0 = C2mm$}
\end{subtable}
\caption{\justifying Without TR and $G^A_0=C_{6v}$,  for the symmorphic space groups $G_0$ listed above, there exists at least one irrep for which the low-energy  moiré bands are necessarily symmetry-enforced semimetallic.
 \label{tab:C6v NTR S0} }
\end{table}
\begin{table}[H]
\begin{subtable}{0.49\textwidth}
\centering
\renewcommand{\arraystretch}{1.2}
%

\caption{$G_0^A = C_{6v}$ and $G_0 = Pba2$}
\end{subtable}
\begin{subtable}{0.49\textwidth}
\centering
\renewcommand{\arraystretch}{1.2}
%
\caption{$G_0^A = C_{6v}$ and $G_0 = Pma2$}
\end{subtable}

\begin{subtable}{0.49\textwidth}
\centering
\renewcommand{\arraystretch}{1.2}
%
\caption{$G_0^A = C_{6v}$ and $G_0 = Pb11$}
\end{subtable}
\caption{\justifying Without TR and $G^A_0=C_{6v}$,  for the nonsymmorphic space groups $G_0$ listed above, there exists at least one irrep for which the low-energy  moiré bands are necessarily symmetry-enforced semimetallic.
 \label{tab:C6v NTR S} }
\end{table}

\begin{table}[H]
\centering
\renewcommand{\arraystretch}{1.2}
%
 \caption{\justifying Without TR and $G^A_0 = C_{4v}$, for the space group $P2$, there exists at least one irrep for symmetry-enforced topologically insulating moir\'e bands.
 \label{tab:C4v NTR T} }
\end{table}
\begin{table}[H]
\centering
\begin{subtable}{0.49\textwidth}
\centering
\renewcommand{\arraystretch}{1.2}
%
\caption{$G_0^A = C_{4v}$ and $G_0 = P4mm$}
\end{subtable}\hfill
\begin{subtable}{0.49\textwidth}
\centering
\renewcommand{\arraystretch}{1.2}
%
\caption{$G_0^A = C_{4v}$ and $G_0 = Pmm2$}
\end{subtable}

\begin{subtable}{0.49\textwidth}
\centering
\renewcommand{\arraystretch}{1.2}
%
\caption{$G_0^A = C_{4v}$ and $G_0 = C2mm$}
\end{subtable}
\begin{subtable}{0.49\textwidth}
\centering
\renewcommand{\arraystretch}{1.2}
%
\caption{$G_0^A = C_{4v}$ and $G_0 = Pm11$}
\end{subtable}
\caption{\justifying Without TR and $G^A_0 = C_{4v}$, for the symmorphic space groups $G_0$ listed above, there exists at least one irrep for which the low-energy  moiré bands are necessarily symmetry-enforced semimetallic.
 \label{tab:C4v NTR S0} }
\end{table}
\begin{table}[H]
\begin{subtable}{0.49\textwidth}
\centering
\renewcommand{\arraystretch}{1.2}
%
\caption{$G_0^A = C_{4v}$ and $G_0 = P4bm$}
\end{subtable}
\begin{subtable}{0.49\textwidth}
\centering
\renewcommand{\arraystretch}{1.2}
%
\caption{$G_0^A = C_{4v}$ and $G_0 = Pba2$}
\end{subtable}
\begin{subtable}{0.49\textwidth}
\centering
\renewcommand{\arraystretch}{1.2}
%
\caption{$G_0^A = C_{4v}$ and $G_0 = Pma2$}
\end{subtable}
\begin{subtable}{0.49\textwidth}
\centering
\renewcommand{\arraystretch}{1.2}
%
\caption{$G_0^A = C_{4v}$ and $G_0 = Pb11$}
\end{subtable}
\caption{\justifying Without TR and $G^A_0 = C_{4v}$, for the nonsymmorphic space groups $G_0$ listed above, there exists at least one irrep for which the low-energy  moiré bands are necessarily symmetry-enforced semimetallic.
 \label{tab:C4v NTR S} }
\end{table}

\begin{table}[H]
\centering
\centering
\renewcommand{\arraystretch}{1.2}
%

\caption{\justifying Without TR and $G^A_0 = C_{3v}$, for the space group $Pb11$, there exists at least one irrep for which the low-energy  moiré bands are necessarily symmetry-enforced semimetallic.
 \label{tab:C3v NTR S} }
\end{table}

\begin{table}[H]
\centering
\begin{subtable}{0.49\textwidth}
\centering
\renewcommand{\arraystretch}{1.2}
%
\caption{$G_0^A = C_{2v}$ and $G_0 = Pba2$}
\end{subtable}
\begin{subtable}{0.49\textwidth}
\centering
\renewcommand{\arraystretch}{1.2}
%
\caption{$G_0^A = C_{2v}$ and $G_0 = Pma2$}
\end{subtable}
\RaggedRight
\begin{subtable}{0.49\textwidth}
\centering
\renewcommand{\arraystretch}{1.2}
%
\caption{$G_0^A = C_{2v}$ and $G_0 = Pb11$}
\end{subtable}
\caption{\justifying Without TR and $G^A_0 = C_{2v}$, for the  space groups $G_0$ listed above, there exists at least one irrep for which the low-energy  moiré bands are necessarily symmetry-enforced semimetallic.
 \label{tab:C2v NTR S} }
\end{table}

\begin{table}[H]
\centering

\centering
\renewcommand{\arraystretch}{1.2}
%
\caption{\justifying Without TR and $G^A_0 = C_{s}$, for the space group $Pb11$, there exists at least one irrep for which the low-energy  moiré bands are necessarily symmetry-enforced semimetallic.
 \label{tab:Cs NTR S} }
\end{table}

\newpage

\section{Two models for Symmetry-enforced Moiré Topology}
In this section, we will present two toy models for the symmetry-enforced moiré topology, namely the moir\'e cubic Rashba model that preserves TR symmetry and a spinless two-band model that breaks TR symmetry.  

\subsection{Moir\'e cubic Rashba model}\label{sec:cubic Rashba}
We consider a 2D electron model with the cubic Rashba spin-orbit coupling (SOC) \cite{nakamura2012experimental, winkler2003spin} and a moir\'e superlattice potential that belongs to the 2D plane group $P6mm$.
The model Hamiltonian is written as
\begin{eqnarray} \label{eq: cubic rashba}
&&\hat{H}=\hat{H}^{A}+\hat{H}_M,\nonumber \\
&& \hat{H}^{A} = \sum_{\kk}\sum_{n, m= 1, 2}c_{n \kk }^\dagger 
 [H^{A}(\kk)]_{n m} c_{m \kk}, \nonumber \\
  &&H^{A}(\kk)=\begin{pmatrix}
 \alpha k^2+\beta  k^4  &  i R_3 k^3_+ \\
 -i R_3 k^3_-& \alpha  k^2+\beta k^4
  \end{pmatrix},\nonumber \\
 && \hat{H} _ M  =  \sum_{n= 1, 2 }\int d^2r  c_{n \rr }^\dagger H _ M (\rr) c_{n \rr },
\end{eqnarray}
where $k_\pm = k_x \pm \mathrm{i} k_y$, $c_{n \rr }=\frac{1}{\sqrt{V}}\int d^2k c_{n \kk } e ^{i \kk \cdot \rr}$ with the integral over the momentum $\kk$ in the ABZ and the system area $V$. Here we include the $k^4$ terms in the diagonal part of $H^A$ to ensure that the Hamiltonian is bounded.

We consider the second harmonics approximation for the moir\'e potential to show that the symmetry-enforced moir\'e topology does not rely on the form of the moir\'e potential. 
$H_M(\rr)$ can be expressed as $H_M(\rr)=\sum_{\gs \in \GG^{\mathds{1}}_M} \Delta_1 e^{i \gs \cdot \rr}+\sum_{\gs \in \GG^{\mathbb{2}}_M} \Delta_2 e^{i \gs \cdot \rr}$, where $\GG^{\mathds{1}}_M$ consists of $\bb^M_1$ and its partners under six-fold rotation symmetries, while $\GG^{\mathbb{2}}_M$ consists of $\bb^M_1 + \bb^M_2$ and its partners under six-fold rotation symmetries. $\bb^M_1= 2\pi(1,-1/\sqrt{3})/a_{M}$ and $\bb^M_2 =  2\pi(1,1/\sqrt{3})/a_{M}$ are primitive reciprocal lattice vectors, and $a_{M}$ is the moir\'e lattice constant.  
We rewrite Eq.(\ref{eq: cubic rashba}) into the momentum space as 
\begin{eqnarray}\label{eq:moire cubic rashba K space}
&& \hat{H}^A = \sum_{\kk \gs}\sum_{ n,m=1,2}c_{n \kk \gs}^\dagger 
[H^{A}(\kk+\gs)]_{n m} c_{m \kk \gs} \label{eq: cubic rashba_kspace} \\
&& \hat{H}_{M} = \sum_{\kk \gs \gs'} \sum_{n=1,2}\left(\Delta_1\sum_{\qq \in \GG^{\mathds{1}}_M } \delta(\gs' -\gs -\qq)+\Delta_2\sum_{\qq \in \GG^{\mathbb{2}}_M } \delta(\gs' -\gs -\qq) \right)c_{n \kk \gs}^\dagger c_{n \kk \gs'}\label{eq:HM_kspace},
\end{eqnarray}
where $c_{n \kk \gs} = c_{n, \kk+\gs}$ with 
$\kk$ to be the momentum in the first mBZ and $\gs, \gs'\in \GG_M$ ($\GG_M$ is the group of moir\'e reciprocal lattice vectors).
For  the calculation in the main text, $\alpha=-2$  $nm^2\cdot$ eV, $\beta=-6.1$  $nm^4\cdot$eV, $R_3=5$  $nm^3\cdot$eV and $a_{M}$=$11.46$ $nm$. Both   $\Delta_1$ and $\Delta_2$ are treated as tunable parameters. 

\subsubsection{Perturbation theory at \textbf{K} point}
In the main text, we have demonstrated that the topmost two bands of the moir\'e cubic Rashba model must be topologically insulating when $\Delta_1$ is negative. In this section, we will implement perturbation theory to support this claim.  

We first consider the eigen-energy and eigen-wavefunctions of the Hamiltonian $H^A(\kk)$ in Eq.(\ref{eq: cubic rashba}) for zero $\Delta_1$ and $\Delta_2$, which are given by 
\begin{eqnarray} \label{eq: eigen energy}
E_\pm(k)=\alpha k^2+\beta k^4\pm R_3 k^3.
\end{eqnarray}
and 
\begin{eqnarray} \label{eq: eigen state}
\ket{\kk,\pm}=\frac{1}{\sqrt{2}}(\pm i e^{-3i\phi},1),
\end{eqnarray}
where $\phi = \arctan\left(\frac{k_y}{k_x}\right)$ is the momentum angle. 

\begin{figure}[h!]
\centering
\includegraphics[width=0.4\textwidth]{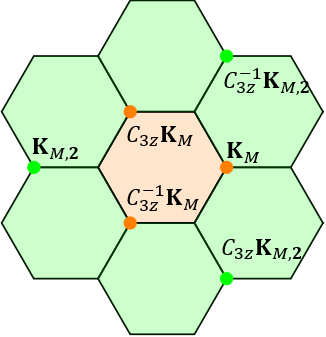}
\caption{\justifying\label{fig:6K} 
Extended mBZ with high symmetry $\textbf{K}_{M}$ points. Orange and green points  are orbits of $\textbf{K}_{M}$  and $\textbf{K}_{M,2}$  respectively}
\end{figure}

Since $R_3 >0$ in our numerical calculation, we focus on the eigen-state $\ket{\kk,+}$ with a higher energy. Along the $\mathbf{\Gamma}_M\textbf{K}_M$ line, the $M_y$ eigenvalue of $\ket{\kk,+}$ is $-i$ and belongs to irrep $\bar{\mathbf{\Gamma}\mathbf{K}}_2$, according to the character table in Table \ref{tab: character table Cs}. Therefore, as discussed in Sec.\ref{sec: Cases with TR}, $\ket{\textbf{K}_M(\phi=0),+}$, $\ket{ C_{3z}\textbf{K}_M(\phi=\frac{2\pi}{3}),+}$, $\ket{C^{-1}_{3z}\textbf{K}_M(\phi=\frac{4\pi}{3}),+}$ are folded into $\textbf{K}_M$ point in the mBZ and form the representation $\bar{\mathbf{K}}_4\oplus\bar{\mathbf{K}}_6$ at zero moir\'e potential. The $\phi$ value for each $\kk$ point is indicated in parentheses. 
The character for $G_{\mathbf{K}_M}=C_{3v}$ is shown in Table. \ref{tab: character table P6mm}(b). In addition, for perturbation theory, we also consider the momenta in the second momentum shell that are folded into $\textbf{K}_M$, including three momenta $\textbf{K}_{M,2}(\phi=\pi)$=$\textbf{K}_M-\bb^M_1-\bb^M_2$, $C_{3z}\textbf{K}_{M,2}(\phi=\frac{5\pi}{3})$ and $C_{3z}^{-1}\textbf{K}_{M,2}(\phi=\frac{\pi}{3})$, because those momenta can be connected to the orbit of $\mathbf{K}_M$ by moir\'e potential. The momenta that are folded into $\textbf{K}_M$ in both the orbits of $\textbf{K}_{M}$  and $\textbf{K}_{M,2}$ are depicted by the orange and green dots in Fig.\ref{fig:6K}, respectively.

We next treat $\hat{H}_M$ as a perturbation, and project it into the eigen-state basis of $H^A(\kk)$ discussed above to understand the energy splitting between the $\bar{\mathbf{K}}_4$ state and the $\bar{\mathbf{K}}_6$ states. From the general wavefunction form in Eq.(\ref{eq: eigen state}), the eigenstate wavefunctions at these six momenta can be explicitly written as 
\begin{eqnarray}
 &&\ket{\textbf{K}_{M},+}=\ket{ C_{3z}\textbf{K}_{M},+}=\ket{ C^{-1}_{3z}\textbf{K}_{M},+}=
 \ket{\textbf{K}_{M,2},-}=\ket{ C_{3z}\textbf{K}_{M,2},-}=\ket{ C^{-1}_{3z}\textbf{K}_{M,2},-}=\frac{1}{\sqrt{2}}(i,1),\label{eq: 6K states}\\
 &&\ket{\textbf{K}_{M},-}=\ket{ C_{3z}\textbf{K}_{M},-}=\ket{ C^{-1}_{3z}\textbf{K}_{M},-}=
 \ket{\textbf{K}_{M,2},+}=\ket{ C_{3z}\textbf{K}_{M,2},+}=\ket{ C^{-1}_{3z}\textbf{K}_{M,2},+}=\frac{1}{\sqrt{2}}(-i,1). \label{eq: 6K states2}
\end{eqnarray} Next, we evaluate the matrix elements of $\Hat{H}$ in the basis formed by these 12 eigenstate wavefunctions. We find that
\begin{eqnarray}
\bra{\textbf{K}_{M},+} \hat{H}\ket{\textbf{K}_{M,2},+}=\bra{\textbf{K}_{M},+}\hat{H}^A\ket{\textbf{K}_{M,2},+}+\bra{\textbf{K}_{M},+}\hat{H}_M\ket{\textbf{K}_{M,2},+}=0,
\end{eqnarray} which indicates that $\ket{\textbf{K}_{M},+}$ and $\ket{\textbf{K}_{M,2},+}$ are decoupled. Similarly, we find that the matrix elements of $\Hat{H}$ between the six eigenstates in Eq.~(\ref{eq: 6K states}) and those in Eq.~(\ref{eq: 6K states2}) vanish. Therefore, we only need to consider the basis formed by 
$\ket{\textbf{K}_{M},+}$, 
$\ket{C_{3z}\textbf{K}_{M},+}$, 
$\ket{C^{-1}_{3z}\textbf{K}_{M},+}$, 
$\ket{\textbf{K}_{M,2},-}$, 
$\ket{C_{3z}\textbf{K}_{M,2},-}$, and 
$\ket{C^{-1}_{3z}\textbf{K}_{M,2},-}$, 
in which the effective Hamiltonian takes the form:  
\begin{equation} \label{eq:K rashba eff}
\begin{split}
  H_{eff,\textbf{K}_M}  = \begin{pmatrix}
 E_0 & \Delta_1 & \Delta_1 & \Delta_2 & \Delta_1 & \Delta_1 \\
 \Delta_1 & E_0 & \Delta_1 & \Delta_1 & \Delta_2 & \Delta_1 \\ 
 \Delta_1 & \Delta_1 & E_0 & \Delta_1 & \Delta_1 &\Delta_2 \\
 \Delta_2 & \Delta_1 & \Delta_1 & E_1 & 0 & 0 & \\
 \Delta_1 & \Delta_2 & \Delta_1 & 0 & E_1 & 0 &\\ 
 \Delta_1 & \Delta_1 &\Delta_2 & 0 & 0 & E_1 &\\
  \end{pmatrix},
\end{split}
\end{equation}
where $ E_0=E_+(\mathbf{K}_M)$ and $ E_1=E_-(\mathbf{K}_{M,2})$ are given in Eq. (\ref{eq: eigen energy}). We do the second order perturbation theory by projecting out $\ket{\textbf{K}_{M,2},-}$, $\ket{ C_{3z}\textbf{K}_{M,2},-}$, and $\ket{C^{-1}_{3z}\textbf{K}_{M,2},-}$ states, 
\begin{eqnarray}
&& H'_{eff,\textbf{K}_M}=\begin{pmatrix}
  E_0 & \Delta_1 & \Delta_1 &\\
  \Delta_1 & E_0 & \Delta_1\\
 \Delta_1 & \Delta_1 & E_0
\end{pmatrix}+H^2_{eff,\textbf{K}_M},\nonumber\\
&&H^2_{eff,\textbf{K}_M}=\sum_{\alpha,\alpha'=1,2,3}\sum_{\beta,\beta'=4,5,6}\frac{\bra{\alpha'}H_{eff,\textbf{K}_M}\ket{\beta'}\bra{\beta}H_{eff,\textbf{K}_M}\ket{\alpha}}{E_0-E_1}, 
\label{eq: Heff(2)f}
\end{eqnarray}  where, $\alpha=1,2,3$ index the basis $\ket{\textbf{K}_{M},+}$, $\ket{ C_{3z}\textbf{K}_{M},+}$, and $\ket{C^{-1}_{3z}\textbf{K}_{M},+}$,  $\beta=4,5,6$ index the basis  $\ket{\textbf{K}_{M,2},-}$, $\ket{ C_{3z}\textbf{K}_{M,2},-}$, and $\ket{C^{-1}_{3z}\textbf{K}_{M,2},-}$. The energy eigenvalues of $H'_{eff,\textbf{K}_M} $  in Eq. (\ref{eq: Heff(2)f}) are given by
\begin{eqnarray} \label{eq:K rashba eff enegry}
&&E(\Bar{\mathbf{K}}_6)=E_0 -\Delta_1 
+ \frac{\Delta_1^{2}}{E_0 - E_1} 
+ \frac{\Delta_2^{2}}{E_0 - E_1} 
- \frac{2\Delta_1\Delta_2}{E_0 - E_1};\nonumber\\
&&E(\Bar{\mathbf{K}}_4)=E_0 +2\Delta_1
+ \frac{4\Delta_1^{2}}{E_0 - E_1} 
+ \frac{\Delta_2^{2}}{E_0 - E_1} 
+  \frac{4\Delta_1\Delta_2}{E_0 - E_1},
\end{eqnarray}
where $\Bar{\mathbf{K}}_{4}$ and  $\Bar{\mathbf{K}}_{6}$  in the bracket represent the irrep for eigenstates. When $\Delta_1$=0, $E(\Bar{\mathbf{K}}_6)=E(\Bar{\mathbf{K}}_4)$, so that the band dispersion remains gapless at $\mathbf{K}_M$.  For small $\Delta_1$ and $\Delta_2$, the leading orders of $E(\Bar{\mathbf{K}}_6)$ and $E(\Bar{\mathbf{K}}_4)$ are $E_0 -\Delta_1$ and $E_0 +2\Delta_1$ respectively. Therefore, the 2D irrep $\bar{\mathbf{K}}_{6}$ has a higher energy when $\Delta_1<0$ and the low-energy two moir\'e bands are topologically insulating. In contrast, if $\Delta_1>0$, $\bar{\mathbf{K}}_{4}$ has a higher energy, and the low-energy moir\'e bands are semi-metallic.

\subsubsection{Linear Rashba model with TR}
\begin{figure}[h!]
\centering
\includegraphics[width=0.7\textwidth]{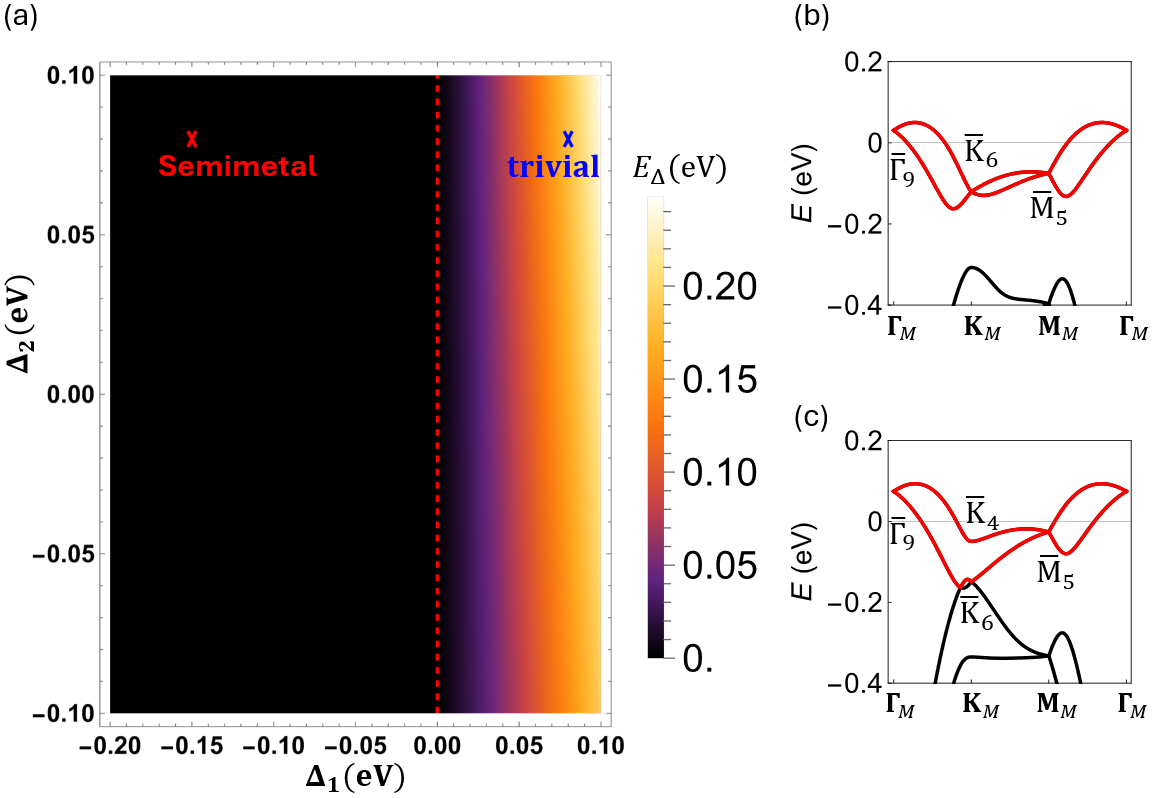}
\caption{\justifying\label{fig:linear_rashba} 
(a) Phase diagram of the low-energy two bands. $E_\Delta$ denotes the direct band gap between the second and third low-energy  bands. The red dashed lines separate the insulator region and semimetal regions. (b) The Band dispersions with parameters labeled by the blue cross in the (a). The topomost bands are trivial. (c) The Band dispersions with parameters labeled by the red cross in the (a). The topomost bands connect to the third low-energy  band. }

\end{figure}

In this section, we consider the moir\'e linear Rashba model for comparison. The moir\'e linear Rashba model is given by
\begin{eqnarray} \label{eq: linear rashba}
&&\hat{H}=\hat{H}^{A}_l+\hat{H}_M,\nonumber \\
&& \hat{H}^{A}_l = \sum_{\kk}\sum_{n, m= 1, 2}c_{n \kk }^\dagger 
 [H^{A}_l(\kk)]_{n m} c_{m \kk}, \nonumber \\
  &&H^{A}(\kk)_l=\begin{pmatrix}
 \alpha k^2+\beta  k^4  &  i R_1  k_+ \\
 -i R_1  k_-& \alpha  k^2+\beta k^4
  \end{pmatrix},\nonumber \\
 && \hat{H} _ M  =  \sum_{n= 1, 2 }\int d^2r  c_{n \rr }^\dagger H _ M (\rr) c_{n \rr }.
\end{eqnarray}
We consider the same form for
$H_M (\rr)$ in Eq.(\ref{eq: linear rashba}) as that in Eq.(\ref{eq: cubic rashba}) for the moir\'e cubic Rashba model. The values of $\alpha$ and $\beta$ take the same value as those in the cubic Rashba model, and $R_1=0.4$  $nm\cdot$eV. The direct band gap between the second top and the third top band as a function of $\Delta_1$ and $\Delta_2$ is shown in Fig. \ref{fig:linear_rashba}(a). When $\Delta_1>0$, the low-energy two bands are isolated trivial, while  $\Delta_1<0$, the low-energy moir\'e bands connect to the third low-energy band. In contrast to the cubic Rashba model with moir\'e potential, the low-energy two bands in the linear Rashba model with moir\'e potential can be isolated trivial. The irreps of the isolated low-energy two bands are ($\mathbf{\bar{\Gamma}}_9$, $\mathbf{\bar{K}}_6$, $\mathbf{\bar{M}}_5$) in the trivial phase, which is the EBR $\bar{E}_1 \uparrow G(2)$ at the Wyckoff position 1a (the center of the honeycomb moiré unit cell). 

The difference between linear and cubic Rashba model lies in the atomic irrep at $\Gamma$, $\Lambda_{\Gamma}^A$. $\Lambda_{\Gamma}^A$ is $\bar{\Gamma}_9$ for the linear Rashba, while it is $\bar{\Gamma}_7$ for the cubic Rashba. This change in atomic $\Lambda_{\Gamma}^A$ leads to the change of irrep for the low-energy moiré bands at $\Gamma_M$, leading to the difference in the symmetry-enforced moiré topology.

\subsection{Spinless two-band model without TR }\label{Sec:spinless two band}

In Sec.\ref{sec: cases withour TR}, we show that the combination of $G^A_0=C_{4v}$ and $G_0=P2$ can yield symmetry-enforced moiré topology for $\Lambda^A_\Gamma=\mathbf{\Gamma}_5$. In this section, we will verify this statement in a spinless two-band model.

The $2\times 2$ effective Hamiltonian has a general form: 
\begin{eqnarray}
  H^A_{2}(\kk)=\sum^{3}_{i=0} F^{i}(\kk)\sigma_i,
\end{eqnarray}
where the expansion coefficients $F^{i}(\kk)$ are the polynomials of $\kk$, $\sigma_0$ and $\sigma_{1,2,3}$ are the $2\times 2$ identity and Pauli matrices. The $C_{4v}$ group is generated by the four-fold rotation  along the out-of-plane direction $C_{4z}$ and  $M_{y}$. The $\sigma$  matrices transform as
\begin{eqnarray}
 D(C_{4z}) \sigma_i D^{-1}(C_{4z})=\sum_j\mathcal{D}^{ji}(C_{4z}) \sigma_j;\
 D(M_{y}) \sigma_i D^{-1}(M_{y})=\sum_j\mathcal{D}^{ji}(M_{y}) \sigma_j,
\end{eqnarray}
where $D(C_{4z})$ and $D(M_{y})$ are representation matrices of $C_{4z}$ and $M_{y}$, respectively, given by 
\begin{eqnarray}
  D(C_{4z})=i \sigma_3 ;\ D(M_y)= \sigma_1
\end{eqnarray}

Using these representation matrices of $C_{4z}$ and $M_y$, we can classify the $\sigma$ matrices into the irrep table of the $C_{4v}$ group, as shown in Table. \ref{TAB: symmetry C4v matrix}.

In the Table. \ref{TAB: symmetry C4v matrix}, the first column shows the irrep, the second and third columns show the construction of representations using $\sigma$ matrices and $k$-polynomials up to the $k^4$ order, respectively. The + (-) sign in parentheses after the representation matrices and $k$-polynomials shows their even (odd) parity under TR. According to Table \ref{TAB: symmetry C4v matrix}, we need to keep the fourth-order term ($k_xk_y(k^2_x-k_y^2)$)$\sigma_3$ to break TR symmetry, and can write down an effective model,
\begin{eqnarray} \label{eq: two band spinless}
  H^A_{2}(\kk)=(- k^2-A k^4)\sigma_0+B(k_xk_y(k^2_x-k_y^2))\sigma_3+(k^2_x-k^2_y)\sigma_1-(2k_xk_y)\sigma_2.
\end{eqnarray}
The $-A k^4$ term in Eq.(\ref{eq: two band spinless}), where $A > 0$, plays a similar role to $\beta k^4$ in Eq.(\ref{eq: cubic rashba}), ensuring boundedness of $H_2(\kk)$. To obtain $G_0=P2$, the moir\'e potential term needs to keep $C_2$ symmetry and break other symmetries. Similar to the previous sections, the moir\'e potential is kept up to the second harmonics and written as
\begin{eqnarray} \label{eq: second harmonics potential P2}
   &&H_{M,2}(\rr)=\frac{\Delta_{1, 1} }{2}\cos(\bb^M_1\cdot\rr)+\frac{\Delta_{1, 2} }{2}\cos(C_{3z}\bb^M_1\cdot\rr)+\frac{\Delta_{1, 3} }{2}\cos(C^{-1}_{3z}\bb^M_1\cdot\rr)\nonumber\\
   &&+\frac{\Delta_{2, 1} }{2} \cos((\bb^M_1+\bb^M_2)\cdot\rr)+\frac{\Delta_{2, 2} }{2} \cos(C_{3z}(\bb^M_1+\bb^M_2)\cdot\rr)+\frac{\Delta_{2, 3} }{2} \cos(C^{-1}_{3z}(\bb^M_1+\bb^M_2)\cdot\rr). 
\end{eqnarray} The first three terms correspond to the first harmonic components, while the remaining three terms are for the second harmonic components. All moir\'e potential parameters $\Delta_{i,j}$  are independent. It should be noted that the form of moir\'e potential is not unique, and the example provided is merely illustrative. The Hamiltonian with moir\'e potential can be rewritten in the second quantization form, 
\begin{eqnarray} \label{eq: spinless moire}
&&\hat{H}_2=\hat{H}^{A}_2+\hat{H}_{M,2}= \sum_{\kk \gs \gs'}\sum_{ n,m=1,2}c_{n \kk \gs'}^\dagger 
[H_2(\kk)]_{n \gs', m \gs} c_{m \kk \gs},\nonumber \\
&& \hat{H}^{A}_2 = \sum_{\kk}\sum_{ n,m= 1, 2}c_{n \kk }^\dagger 
 [H^{A}_2(\kk)]_{n m} c_{m \kk}, \nonumber \\
 && \hat{H} _ {M,2}  = \sum_{n= 1, 2 }\int d^2r  c_{n \rr }^\dagger H  _ {M,2} (\rr) c_{n \rr }.
\end{eqnarray}

\begin{table}[h!]
\centering
\renewcommand{\arraystretch}{1.5}
\begin{tabular}{|c|c|c|c|}
\hline
\(C_{3v}\) & \(\sigma\) & \(k\) \\ 
\hline
\(\mathbf{\Gamma}_1 (A_1)\) & \(\sigma_0(+)\) & \(k_x^2 + k_y^2(+), (k_x^4 + k_y^4)(+), k_x^2 k_y^2(+), \text{C}(+)\) \\
\hline 
\(\mathbf{\Gamma}_2 (B_1)\)  & \(\sigma_1(+)\) & $k^2_x-k_y^2(+)$, $k^4_x-k_y^4(+)$ \\ 
\hline
\(\mathbf{\Gamma}_3 (B_2)\) & \(\sigma_2(+)\) & $k_xk_y(+)$ \\ 
\hline
\(\mathbf{\Gamma}_3 (A_2)\)  & $\sigma_3(-)$ &  $k_xk_y(k^2_x-k_y^2)(+)$ \\ 
\hline
\end{tabular}
\caption{\justifying  The classification of the representation matrices and $k$-polynomials for the $C_{4v}$ group. The + or - in the bracket indicates how the matrices transform under TR (even or odd). The symbol $\text{C}$ in the column of $k$-polynomial is for a constant. \label{TAB: symmetry C4v matrix} }
\end{table}

\begin{figure}[h!]
\centering
\includegraphics[width=0.7\textwidth]{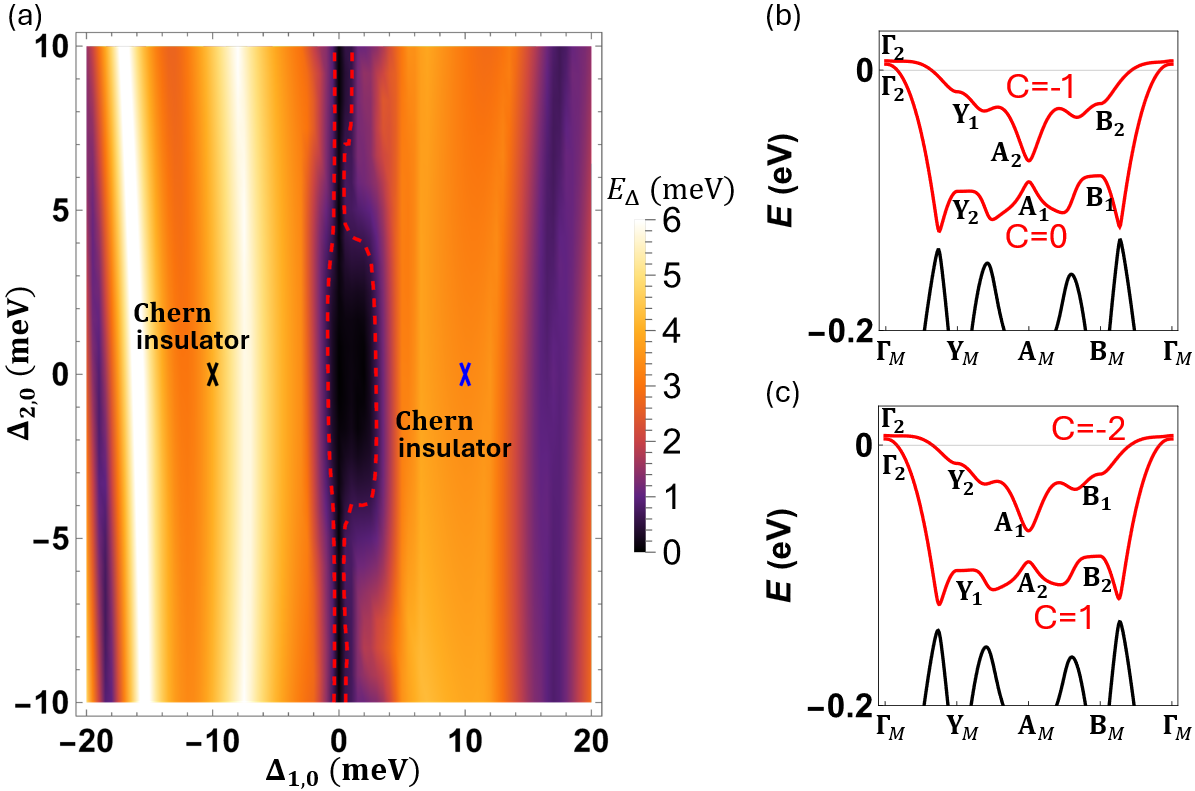}
\caption{\justifying\label{fig:spinless two bands} 
(a) Phase diagram of the low-energy two bands. $E_\Delta$ denotes the direct band gap between the second and third low-energy  bands. The red dashed lines separate the Chern insulator region and semimetal regions. (b) The Band dispersions with parameters labeled by the blue cross in  (a).  (c) The Band dispersions with parameters labeled by the black cross in  (a).  }

\end{figure}

The symmetry group of $\hat{H}_{2}$ in Eq.(\ref{eq: spinless moire}) is described by the space group $P2$, with the little groups at high symmetry momenta $\mathbf{\Gamma}_M$, $\textbf{Y}_M$, $\textbf{A}_M$, and $\textbf{B}_M$ to be the point groups $C_{2}$. In Sec.~\ref{sec: cases withour TR}, we have shown that the irreps of the two low-energy bands are always $(2\mathbf{\Gamma}_{2}, \mathbf{Y}_1 \oplus \mathbf{Y}_2, \mathbf{A}_1 \oplus \mathbf{A}_2, \mathbf{B}_{1} \oplus \mathbf{B}_2)$, and that these two bands carry a total odd Chern number in the weak moiré potential limit, regardless of the specific form of the potential. To verify this, we choose $A=8$  $nm^4\cdot$eV, $B=32$  $nm^4\cdot$eV, $a_{M}$=$11.46$ $nm$ and set the moiré potential strengths as tunable parameters. Here, we choose the parameters as $\Delta_{1,1} = -4\Delta_{1,0}$, $\Delta_{1,2} = \Delta_{1,0}$, $\Delta_{1,3} = 3\Delta_{1,0}$, $\Delta_{2,1} = 3\Delta_{2,0}$, $\Delta_{2,2} = \Delta_{2,0}$, and $\Delta_{2,3} = -4\Delta_{2,0}$, such that the moir\'e potential preserves only $C_{2z}$ symmetry. This choice is made for illustrative purposes. The direct band gap between the second and third topmost bands as a function of $\Delta_{1,0}$ and $\Delta_{2,0}$ is shown in Fig. \ref{fig:spinless two bands}(a). The red dashed lines separate the semimetallic phase, where the low-energy two bands connect to other bands at generic (non–high-symmetry) momenta, from the Chern insulator phase, where the low-energy two bands carry a total Chern number of $-1$. Figure \ref{fig:spinless two bands}(b) and (c) show the band dispersions for $\Delta_{1,0} = 10$\,meV and $\Delta_{1,0} = -10$\,meV, respectively, with $\Delta_{2,0} = 0$\,meV in both cases. We can see that the top two bands have a total Chern number $-1$ for both parameters.

We can go one step further, and determine the irreps and Chern number of each individual band among the low-energy two bands. The two-fold rotation $C_{2z}$ acts on fermion operators as 
\begin{eqnarray}
  C_{2z}c^\dagger_{n, \kk,  \gs}C^{-1}_{2z}=\sum_{m \gs'}c^\dagger_{m, C_{2z} \kk,  \gs'}D_{m\gs',n\gs}(C_{2z}).
\end{eqnarray}
with
\begin{eqnarray}\label{eq: D of C6z}
  D_{m\gs',n\gs}(C_{2z})=-\delta_{\gs',C_{2z}\gs}\delta_{m,n}.
\end{eqnarray}
Under $C_{2z}$, $H_{2}(\kk)$ is invariant
\begin{eqnarray}
 D(C_{2z})H_{2}(\kk) D^{-1}(C_{2z})=H_{2}(C_{2z}\kk).
\end{eqnarray}

The irreps at all high symmetry momenta can be determined from the eigenvalues of $C_{2z}$. In Fig.\ref{fig:spinless two bands}(b), the highest-energy band has the irreps $(\mathbf{\Gamma}_2, \mathbf{Y}_1, \mathbf{A}_2, \mathbf{B}_2)$ and a Chern number $\text{C} = -1$ while the second highest energy band has the irreps $(\mathbf{\Gamma}_2, \mathbf{Y}_2, \mathbf{A}_1, \mathbf{B}_1)$ and a Chern number $\text{C} =0$.   In Fig.\ref{fig:spinless two bands}(c), the top band has the irreps $(\mathbf{\Gamma}_2,\mathbf{Y}_2,\mathbf{A}_1,\mathbf{B}_1)$ and a Chern number $\text{C}=-2$. Meanwhile, the second top band has the irreps $(\mathbf{\Gamma}_2, \mathbf{Y}_1, \mathbf{A}_2, \mathbf{B}_2)$ and a Chern number $\text{C}=1$.

\section{material candidates}

In this section, we will discuss the material candidates.
We first pick out monolayer SOC materials  in the existing non-moiré 2D material databases \cite{vergniory2019complete, haastrup2018computational,jiang20242d}, and these materials satisfy two conditions: (i) VBM or CBM is located at $\Gamma$ and (ii) there is no symmetries that will force the atomic band splitting to  be zero  along any line in the BZ.  We identify 92 monolayer materials in the atomic plane groups $P31m$, $P3m1$, and $P3$, with their atomic symmetry data listed in Table~1 of the main text. Table~\ref{tab: material p3m1} lists all 92 candidates, organized by band extrema location: (a) only the VBM at $\Gamma$, (b) only the CBM at $\Gamma$, and (c) both the VBM and CBM at $\Gamma$.

The easiest way to introduce  moir\'e potential is twisting homobilayer, in which we require the moir\'e space group $G_0$ to be a subgroup of material space group $\mathcal{G}^A_0$. For a small twisted angle, the mBZ rotates $90^\circ$ relative to ABZ. By breaking the  inplane mirrors, we get the   $G_0$ from $\mathcal{G}^A_0$.
 For  space groups $P3m1$, $P3m1$ and $P3$, the corresponding moir\'e space groups is $P3$, as shown in the Table.\ref{Tab:group corresponds}. According to Table.1 in the main text, we immediately know that if the CBM or VBM at $\Gamma$ corresponds to $J_z = \pm \frac{3}{2}$ states, symmetry-enforced topology emerges in the twisted homobilayer. 
Based on the above discussion, we found that for 71 of 92 candidates highlighted by the red color in Table~\ref{tab: material p3m1}, the required moiré symmetry group can be naturally realized by a twisted homobilayer structure. Meanwhile, according to the Table.1 in the main text, the symmetry-enforced topological bands in those 71 materials can be insulating.


\begin{table}[H]
  \centering

  \begin{minipage}[t]{0.49\textwidth}
    \vspace{0pt}           
    \centering
    \begin{subtable}[t]{\textwidth}
      \centering
      \renewcommand{\arraystretch}{1.15}
      \begin{tabular}{lcc}
\toprule
Name      & SG  & Irrep of VBM \\
\hline
 \color{red}AlSeI           & 156   & $ \bar\Gamma_4  \bar\Gamma_5 $ 
\\ \color{red}AsSI            & 156   & $ \bar\Gamma_4  \bar\Gamma_5 $ 
\\ \color{red}AsSeI           & 156   & $ \bar\Gamma_4  \bar\Gamma_5 $ 
\\ \color{red}AsTeBr          & 156   & $ \bar\Gamma_4  \bar\Gamma_5 $ 
\\ \color{red}AsTeI           & 156   & $ \bar\Gamma_4  \bar\Gamma_5 $ 
\\ \color{red}BaIBr           & 156   & $ \bar\Gamma_4  \bar\Gamma_5 $ 
\\ \color{red}BaICl           & 156   & $ \bar\Gamma_4  \bar\Gamma_5 $ 
\\ \color{red}BiSI-1          & 156   & $ \bar\Gamma_4  \bar\Gamma_5 $ 
\\ \color{red}BiSI-2          & 156   & $ \bar\Gamma_4  \bar\Gamma_5 $ 
\\ \color{red}BiSeF           & 156   & $ \bar\Gamma_4  \bar\Gamma_5 $ 
\\ \color{red}BiSeI           & 156   & $ \bar\Gamma_4  \bar\Gamma_5 $ 
\\ \color{red}BiTeBr          & 156   & $ \bar\Gamma_4  \bar\Gamma_5 $ 
\\ \color{red}BiTeCl-1        & 156   & $ \bar\Gamma_4  \bar\Gamma_5 $ 
\\ \color{red}BiTeCl-2        & 156   & $ \bar\Gamma_4  \bar\Gamma_5 $ \\ \color{red}
BiTeF           & 156   & $ \bar\Gamma_4  \bar\Gamma_5 $ \\ \color{red}
BiTeI           & 156   & $ \bar\Gamma_4  \bar\Gamma_5 $ \\ \color{red}
{Cd2SIBr}       & 156   & $ \bar\Gamma_4  \bar\Gamma_5 $ \\ \color{red}
{CdHSI}         & 156   & $ \bar\Gamma_4  \bar\Gamma_5 $ \\ \color{red}
CdIBr           & 156   & $ \bar\Gamma_4  \bar\Gamma_5 $ \\ \color{red}
CoTeI           & 156   & $ \bar\Gamma_4  \bar\Gamma_5 $ \\ \color{red}
{CuGeTeBr}      & 156   & $ \bar\Gamma_4  \bar\Gamma_5 $ \\ \color{red}
GeSe            & 156   & $ \bar\Gamma_4  \bar\Gamma_5 $ \\ \color{red}
GeTe            & 156   & $ \bar\Gamma_4  \bar\Gamma_5 $ \\ \color{red}
HfSeS           & 156   & $ \bar\Gamma_4  \bar\Gamma_5 $ \\ \color{red}
HfTeSe          & 156   & $ \bar\Gamma_4  \bar\Gamma_5 $ \\ \color{red}
InSb            & 156   & $ \bar\Gamma_4  \bar\Gamma_5 $ \\ 
MoSO            & 156   & $ \bar\Gamma_6 $ \\ 
MoSeO           & 156   & $ \bar\Gamma_6 $ \\ \color{red}
MoTeO           & 156   & $ \bar\Gamma_4  \bar\Gamma_5  $ \\ \color{red}
{Nb3OI7}        & 156   & $ \bar\Gamma_4  \bar\Gamma_5 $ \\ \color{red}
{Nb3SI7}        & 156   & $ \bar\Gamma_4  \bar\Gamma_5 $ \\ \color{red}
Nb3SeI7         & 156   & $ \bar\Gamma_4  \bar\Gamma_5 $ \\ \color{red}
{Nb3TeI7}       & 156   & $ \bar\Gamma_4  \bar\Gamma_5 $ \\ \color{red}
PtTeSe          & 156   & $ \bar\Gamma_4  \bar\Gamma_5 $ \\ \color{red}
{SbAs}          & 156   & $ \bar\Gamma_4  \bar\Gamma_5 $ \\ \color{red}
{SbSI}          & 156   & $ \bar\Gamma_4  \bar\Gamma_5 $ \\ \color{red}
SbSeI           & 156   & $ \bar\Gamma_4  \bar\Gamma_5 $ \\ \color{red}
ScTeCl          & 156   & $ \bar\Gamma_4  \bar\Gamma_5 $ \\ \color{red}
SnSeS           & 156   & $ \bar\Gamma_4  \bar\Gamma_5 $ \\ \color{red}
SnTe            & 156   & $ \bar\Gamma_4  \bar\Gamma_5 $ \\ \color{red}
{Ta3SI7}        & 156   & $ \bar\Gamma_4  \bar\Gamma_5 $ \\ \color{red}
{Ta3SeI7}       & 156   & $ \bar\Gamma_4  \bar\Gamma_5 $ \\ \color{red}
TiSeS           & 156   & $ \bar\Gamma_4  \bar\Gamma_5 $ \\ 
WSO             & 156   & $ \bar\Gamma_6 $ \\ \color{red}
YSI             & 156   & $ \bar\Gamma_4  \bar\Gamma_5 $ \\ \color{red}
YSeBr           & 156   & $ \bar\Gamma_4  \bar\Gamma_5 $ \\ \color{red}
{ZrOBr2}        & 156   & $ \bar\Gamma_4  \bar\Gamma_5 $ \\ \color{red}
ZrSeS           & 156   & $ \bar\Gamma_4  \bar\Gamma_5 $ \\ 
ZrTeS           & 156   & $ \bar\Gamma_6 $ \\ \color{red}
ZrTeSe          & 156   & $ \bar\Gamma_4  \bar\Gamma_5 $ \\ \color{red}
{AgI}           & 157   & $ \bar\Gamma_4  \bar\Gamma_5 $ \\ \color{red}
In2Cl3          & 157   & $ \bar\Gamma_4  \bar\Gamma_5 $ \\ \color{red}
{PbO}           & 157   & $ \bar\Gamma_4  \bar\Gamma_5  $ \\ 
\toprule
\end{tabular}
\caption{\justifying Candidate materials with the VBM at $\Gamma$. The symmetry-enforced topology can emerge in moir\'e valence bands. The first and second columns give the material names and the corresponding space groups. The third column indicates the irrep of the VBM.}
    \end{subtable}
  \end{minipage}%
  \hfill
\  \begin{minipage}[t]{0.49\textwidth}
    \vspace{0pt}          
      \begin{subtable}[t]{\textwidth}
      \centering
      \renewcommand{\arraystretch}{1.15}
      \begin{tabular}{lcc}
\toprule
Name      & SG  & Irrep of CBM  \\
\hline
{AgSbP2Se6}       & 143   & $ \bar\Gamma_5  \bar\Gamma_6 $ \\
{AuSbP2Se6}       & 143   & $ \bar\Gamma_5  \bar\Gamma_6 $ \\
{CuSbAs2Se6}      & 143   & $ \bar\Gamma_5  \bar\Gamma_6 $ \\
{CuSbP2Se6}       & 143   & $ \bar\Gamma_5  \bar\Gamma_6 $ \\
AsTeCl            & 156   & $ \bar\Gamma_6 $ \\
BaBrCl            & 156   & $ \bar\Gamma_6 $ \\
Bi2SeS2           & 156   & $ \bar\Gamma_6 $ \\
{CdHSBr}          & 156   & $ \bar\Gamma_6 $ \\
GaAs              & 156   & $ \bar\Gamma_6 $ \\
GaP               & 156   & $ \bar\Gamma_6 $ \\
\color{red}HgTe              & 156   & $ \bar\Gamma_4  \bar\Gamma_5 $ \\
InGaS2            & 156   & $ \bar\Gamma_6 $ \\
SbTeCl            & 156   & $ \bar\Gamma_6 $ \\
AlAsS3            & 157   & $ \bar\Gamma_6 $ \\
CuCl              & 157   & $ \bar\Gamma_6 $ \\
{FeSe2}           & 157   & $ \bar\Gamma_6 $ \\
\toprule
\end{tabular}
      \caption{\justifying Candidate materials with the CBM at $\Gamma$. The symmetry-enforced topology can emerge in both moir\'e valence and conduction bands.
      The first and second columns give the material names and the corresponding space groups. The third column indicates the irrep of the CBM.}
    \end{subtable}

    \vspace{9.6 em}
    \centering
    \begin{subtable}[t]{\textwidth} 
      \centering
      \renewcommand{\arraystretch}{1.15}
      \begin{tabular}{lccc}
\toprule
Name        & SG & Irrep of VBM & Irrep of CBM\\ 
\hline
\color{red}{AgBiAs2Se6}      & 143   & $ \bar\Gamma_4\bar\Gamma_4 $ & $ \bar\Gamma_5  \bar\Gamma_6 $ \\ \color{red}
{AgSbTe6P2}       & 143   & $ \bar\Gamma_4\bar\Gamma_4 $ & $ \bar\Gamma_5  \bar\Gamma_6 $ \\ \color{red}
CuInP2Se6         & 143   & $ \bar\Gamma_4\bar\Gamma_4 $ & $ \bar\Gamma_5  \bar\Gamma_6 $ \\ \color{red}
{CdGaAlS4}        & 156   & $ \bar\Gamma_4  \bar\Gamma_5 $ & $ \bar\Gamma_6 $ \\ \color{red}
{CdHOBr}          & 156   & $ \bar\Gamma_4  \bar\Gamma_5 $ & $ \bar\Gamma_6 $ \\ \color{red}
CdHOCl            & 156   & $ \bar\Gamma_4  \bar\Gamma_5 $ & $ \bar\Gamma_6 $ \\ 
CdHOF             & 156   & $ \bar\Gamma_6 $               & $ \bar\Gamma_6 $ \\ \color{red}
{CdHSCl}          & 156   & $ \bar\Gamma_4  \bar\Gamma_5 $ & $ \bar\Gamma_6 $ \\ \color{red}
{CdInAlS4}        & 156   & $ \bar\Gamma_4  \bar\Gamma_5 $ & $ \bar\Gamma_6 $ \\ \color{red}
{CdInAlSe4}       & 156   & $ \bar\Gamma_4  \bar\Gamma_5 $ & $ \bar\Gamma_6 $ \\ \color{red}
In2Se3            & 156   & $ \bar\Gamma_4  \bar\Gamma_5 $ & $ \bar\Gamma_6 $ \\ \color{red}
MgBrCl            & 156   & $ \bar\Gamma_4  \bar\Gamma_5 $ & $ \bar\Gamma_6 $ \\ \color{red}
MgIBr             & 156   & $ \bar\Gamma_4  \bar\Gamma_5 $ & $ \bar\Gamma_6 $ \\ \color{red}
MgICl             & 156   & $ \bar\Gamma_4  \bar\Gamma_5 $ & $ \bar\Gamma_6 $ \\ \color{red}
PbTe              & 156   & $ \bar\Gamma_4  \bar\Gamma_5 $ & $ \bar\Gamma_6 $ \\ \color{red}
SiCF2             & 156   & $ \bar\Gamma_4  \bar\Gamma_5 $ & $ \bar\Gamma_6 $ \\ \color{red}
{SiCH2}           & 156   & $ \bar\Gamma_4  \bar\Gamma_5 $ & $ \bar\Gamma_6 $ \\ \color{red}
SrHOI             & 156   & $ \bar\Gamma_4  \bar\Gamma_5 $ & $ \bar\Gamma_6 $ \\ \color{red}
SrIBr             & 156   & $ \bar\Gamma_4  \bar\Gamma_5 $ & $ \bar\Gamma_6 $ \\ \color{red}
{Ta3TeI7}         & 156   & $ \bar\Gamma_4  \bar\Gamma_5 $ & $ \bar\Gamma_6 $ \\ \color{red}
TiSO              & 156   & $ \bar\Gamma_4  \bar\Gamma_5 $ & $ \bar\Gamma_6 $ \\ \color{red}
ZnBrCl            & 156   & $ \bar\Gamma_4  \bar\Gamma_5 $ & $ \bar\Gamma_6 $ \\ 
{HgS}             & 157   & $ \bar\Gamma_6 $               & $ \bar\Gamma_6 $ \\ 
\toprule
\end{tabular}
      \caption{\justifying Candidate materials with both the VBM and CBM at $\Gamma$.  The symmetry-enforced topology can emerge in moir\'e conduction bands.
      The first and second columns give the material names and the corresponding space group. The third and fourth columns indicate the irreps of the VBM and CBM, respectively.}
    \end{subtable}

  \end{minipage}
  \caption{\centering \justifying Candidate materials in space group $P3$ (143), $P3m1$ (156) and  $P31m$ (157).  \label{tab: material p3m1} For the materials highlighted in red, the symmetry-enforced moiré bands can be insulating, whereas for the others they can only be semimetallic.} 
 
\end{table}

   \begin{table}[h!]
    \centering
    \renewcommand{\arraystretch}{1.5}
    \begin{tabular}{|c|c|c|c|c|c|c|}
    \hline
    $\mathcal{G}^A_0$  & $P3m1$ (156)  & $P31m$ (157)   & $P3$ (143) \\ 
    \hline
    $G^A_0$ & $C_{3v}$ &  $C_{3v}$    & $C_3$ \\ 
    \hline 
    $G_0$ & $P3$ & $P3$ & $P3$  \\ 
    \hline
    \end{tabular}
    \caption{\justifying The corresponding moir\'e space groups for twisted homobilayers with single-layer plane groups $P3m1$, $P31m$, $P3$. \label{Tab:group corresponds} }
    \end{table}

\clearpage
\bibliographystyle{unsrt}
\bibliography{ref}